\documentclass[review,twocolumn,sort&compress]{elsarticle}
\usepackage{amsmath,graphicx}
\usepackage{siunitx}
\usepackage{braket}
\usepackage{amssymb}
\usepackage[acronym]{glossaries}
\usepackage{subcaption}
\usepackage[modulo]{lineno}
\usepackage[breaklinks, colorlinks=true]{hyperref}
\usepackage[capitalise]{cleveref}
\usepackage{xfrac}
\usepackage{xcolor}
\usepackage[normalem]{ulem}
\usepackage{booktabs}
\usepackage{float}
\journal{Nuclear Fusion}

\newcommand{\rtxt}[1]{{\color{black}#1}}

\makeglossaries
\newacronym{div}{DIV}{divertor}
\newacronym{bkfw}{BKFW}{blanket first wall}
\newacronym{ccfw}{CCFW}{center column first wall}
\newacronym{vv}{VV}{vacuum vessel}
\newacronym{stm}{STM}{Stochastic Tools Module}
\newacronym{v44}{V44}{V-4Cr-4Ti}
\newacronym{ss}{SS}{stainless steel}
\newacronym{w}{W}{tungsten}
\newacronym{nqa}{NQA-1}{Nuclear Quality Assurance, Level 1}
\newacronym{rmspe}{RMSPE}{Root Mean Squared Percentage Error}
\newacronym{tmap8}{TMAP8}{Tritium Migration Analysis Program, Version 8}
\newacronym{ode}{ODE}{ordinary differential equation}

\begin{document}

\begin{frontmatter}
\onecolumn

\title{Multiscale Assessment of Tritium Behavior in Preliminary Fusion Pilot Plant Design Using Surrogate Models in TMAP8}

\author[inladdress]{Lin Yang\corref{cor1}}
\ead{lin.yang@inl.gov}
\author[inladdress,psuaddress]{Pierre-Cl\'{e}ment A. Simon\corref{cor1}}
\ead{pierreclement.simon@inl.gov}
\author[teaddress]{Emre Yildirim}
\author[teaddress]{Jos\'{e} Trueba}
\author[teaddress]{Matthew Robinson}
\author[inladdress]{Masashi Shimada}

\address[inladdress]{Idaho National Laboratory, Idaho Falls, ID, 83415, USA}
\address[psuaddress]{The Pennsylvania State University, University Park, PA, 16802, USA}
\address[teaddress]{Tokamak Energy \rtxt{Ltd.}, Milton Park, Abingdon, Oxfordshire, OX14 4SD, United Kingdom}

\cortext[cor1]{Corresponding author}

\begin{abstract}

The complexity and significance of multiscale phenomena in fusion energy systems make advanced modeling necessary for designing, optimizing, and safely deploying fusion plants. Tritium accountancy is one of those challenges for deuterium–tritium fusion systems. Its availability is constrained by its short half-life (12.33 years) and limited natural abundance, which require fusion plants to breed tritium onsite. Therefore, accurate tritium accountancy is essential for effective resource management, safety, and economics in fusion plants. Through the U.S. Department of Energy milestone program, Tokamak Energy Ltd. is developing a fusion pilot plant design and evaluating tritium retention and loss in key components and their effect on the fuel cycle. 
To rapidly explore design trade-offs and quantify design decisions on tritium management, this study presents a multiscale analysis to investigate tritium diffusion, trapping, and recovery in key plasma-facing components. To enhance computational efficiency, we integrate surrogate models at the component-level within a fuel cycle model at the system-level, enabling rapid evaluation of tritium recycling dynamics and inventory under various operational scenarios. The goal of this study is twofold: (1) demonstrate the feasibility of utilizing surrogate models to increase the accuracy of fuel cycle modeling, and (2) rapidly evaluate the performance of fusion technologies to accelerate design iterations. This multiscale model provides the tritium transport and retention behavior and supports the plasma-facing components design optimization in normal and bake-out operations. The work is implemented using the \acrfull{tmap8}, an open-source application for tritium transport analysis in fusion systems.

\end{abstract}


\end{frontmatter}
\newpage
\section*{Highlights}
\begin{itemize}
\item Develop transport model with implantation, diffusion, trapping, and thermal effects
\item Develop surrogate model to predict tritium release and retention behaviors
\item Develop two-parameter residence time fuel cycle model to increase the model accuracy
\item Integrated capabilities accelerate and optimize rapid preliminary fusion plant design
\end{itemize}
\newpage


\section{Introduction}
\label{sec:introduction}

Most fusion power plant designs rely on deuterium and tritium as fusion fuel. These fusion systems must carefully account for tritium accountancy challenges because tritium is radioactive, naturally scarce due to its short half-life (12.33 years), and must be carefully managed to ensure safety, fuel self-sufficiency, and advantageous economics \cite{tanabe2009tritium,pearson2018tritium,ying2020recent}. Only a fraction of injected tritium participates in fusion reactions, while the remainder is either exhausted or retained in plasma-facing components \cite{causey2002hydrogen,pajuste2021tritium,widdowson2021evaluation}. Thus, robust tritium accountancy and recycling strategies are critical to minimize losses, ensure regulatory compliance, and support sustainable reactor operation. The accountancy effort is relevant at several length scales, i.e., at the scale of individual components, and at the scale of the full system, which encompasses the full fuel cycle.

Tokamak Energy is developing an upgradable fusion power plant, ST-E1, to support the refinement and optimization of critical fusion components. Because ST-E1 is not intended to supply tritium to additional plants, it need only replenish tritium consumed in fusion reactions, lost through radioactive decay, and lost through permeation in the plant's components. The primary objectives of ST-E1 include understanding tritium production, processing, and transport behavior throughout the entire fuel cycle in order to better achieve tritium requirements and tritium self-sufficiency.

The tritium fuel cycle in fusion energy plants is typically divided into two primary parts: the inner fuel cycle and the outer fuel cycle \cite{abdou1986deuterium,abdou2020physics,meschini2023modeling}. The inner fuel cycle pumps fusion products and unburned fuels out of the vacuum chamber, although part of the fuels are retained in the first wall. These products are then processed to separate the unburned fuels from impurities, and the purified deuterium and tritium are re-injected into the vacuum chamber for subsequent fusion reactions. Shortening the cycle time of the inner fuel cycle can accelerate progress toward tritium self-sufficiency. Due to inevitable tritium losses from the deuterium--tritium reaction, radioactive decay, and permeation through materials, the outer fuel cycle is designed to breed tritium and efficiently extract it from the breeding blanket and the inner fuel cycle exhaust. In the outer fuel cycle, lithium-containing breeding blanket materials utilize the neutrons produced by the deuterium--tritium fusion reaction,
\begin{equation}
    \label{eqn:Intro:neutrons}
    D + T \rightarrow He + neutron + 17.59\ MeV,
\end{equation}
to produce tritium. The generated tritium is then extracted from the breeder, goes through the isotope separation system, and is processed and stored until re-injection. Accurate modeling of both the inner and outer cycles is critical for evaluating tritium inventories, identifying loss mechanisms, and optimizing system design.

Traditional fuel cycle models use a constant residence time to describe the tritium transport and retention in each subsystem (e.g., blanket, extraction system, etc.) \cite{abdou2020physics,meschini2023modeling}. However, it is not always accurate for two reasons: first, tritium behavior is not always best described by such constant residence-time-based models, and second, these models can be highly dependent on the selected residence time value, which can be highly uncertain. For example, in components such as the \acrfull{div}, \acrfull{ccfw}, \acrfull{bkfw}, and \acrfull{vv}, which operate under extreme thermal, neutron, and tritium fluxes, high-fidelity component-level models are essential for capturing the complex physics of tritium diffusion, trapping, and desorption. 
The challenge then becomes to capture the intricacies of tritium behavior at the component scale in the system-level model. 
To perform an effective parametric study of design options and system-level fuel cycle analysis for iterative design, simulations need to be fast and automated. Having to perform new component-level simulations and then input the results into the system-level model every time the design changes would prohibitively limit design iteration speed.
To that end, large combinations of component-level tritium transport and heat conduction simulations with different parameter configurations are performed to develop a surrogate model for each component using the \acrfull{stm} in the Multiphysics Object-Oriented Simulation Environment (MOOSE) framework \cite{audet2000surrogate,alizadeh2020managing,dhulipala2025moose,slaughter2023moose}. The surrogate models are then used in the system-level fuel cycle analysis, which now can enable rapid design iterations while capturing the fidelity of the component-level calculations.

In this study, we develop component-level models for tritium transport in key plasma-facing and vessel components, including the \acrshort{div}, \acrshort{ccfw}, \acrshort{bkfw}, and \acrshort{vv}.
Based on the simulation data from these models, we construct corresponding surrogate models using the Gaussian process method, implemented via the \acrshort{stm} in the TMAP8, which is based on the MOOSE framework \cite{williams2006gaussian,adam2014method,slaughter2023moose,dhulipala2025moose}.
These surrogate models are then coupled with a system-level fuel cycle model to achieve an optimal balance between computational efficiency and predictive accuracy.
The goal of this study is twofold: (1) to demonstrate the feasibility and impact of utilizing surrogate models of component-scale simulations to increase the accuracy of fuel cycle modeling, and (2) to rapidly derive key insights to assess the design and performance of fusion technologies.
The component structures and operating conditions are presented in \cref{sec:components_information}.
The modeling methodology is detailed in \cref{sec:methods}. The results from the component-level simulations, surrogate model performance, and integrated fuel cycle analysis are presented in \cref{sec:results} and discussed in \cref{sec:discussion}. The main conclusions are summarized in \cref{sec:conclusion}.

\section{Component Configurations}
\label{sec:components_information}

\subsection{Component Geometries}
\label{sec:components_information:components}

\begin{figure}[h!]
\centering
\begin{subfigure}{.45\linewidth}
  \centering
	\includegraphics[width = \linewidth]{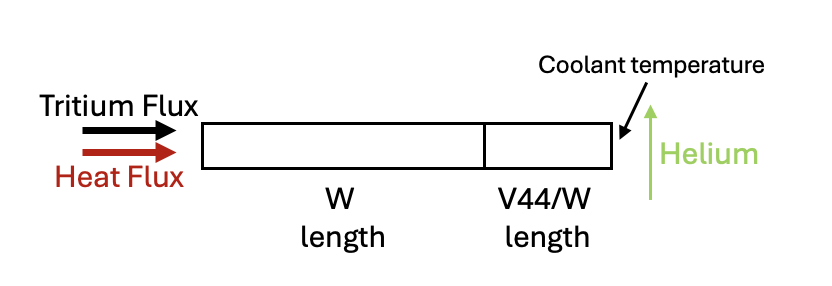}
    \caption{}
    \label{fig:schematics_all:schematics}
\end{subfigure}
\begin{subfigure}{.45\linewidth}
  \centering
	\includegraphics[width = \linewidth]{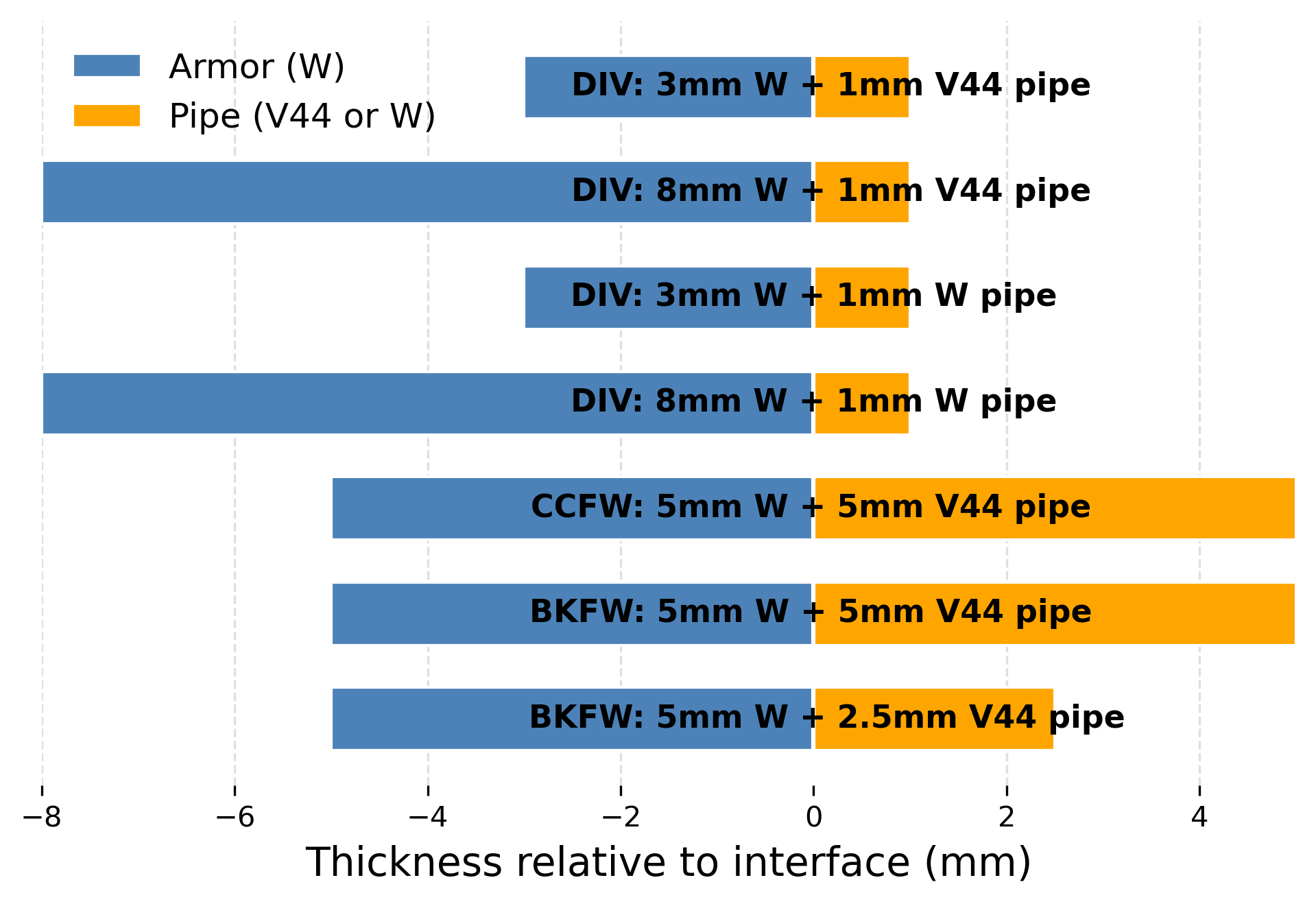}
    \caption{}
    \label{fig:schematics_all:schematics_multiple}
\end{subfigure}
\caption{(a) General schematic and (b) detailed component thicknesses of all plasma-facing components.}
\label{fig:schematics_all}
\end{figure}

The study only considers the tritium transport, motivated by its short half-life and limited natural abundance, across major plasma-facing components of a fusion pilot plant, including the \acrshort{div}, \acrshort{ccfw}, \acrshort{bkfw}, and \acrshort{vv}. Except for the \acrshort{vv}, all components incorporate a \acrfull{w} armor layer on the plasma-facing surface to withstand high heat and particle fluxes. The \acrshort{vv} does not have a \acrshort{w} armor due to the relatively low heat and tritium fluxes on its surface. The general schematic for all plasma-facing components is shown in \cref{fig:schematics_all:schematics}.
%

The \acrshort{div} has four structural configurations: 8 mm \acrshort{w} armor with a 1 mm \acrfull{v44} pipe, 3 mm \acrshort{w} armor with a 1 mm \acrshort{v44} pipe, 8 mm \acrshort{w} armor with a 1 mm \acrshort{w} pipe, and 3 mm \acrshort{w} armor with a 1 mm \acrshort{w} pipe.
The \acrshort{ccfw} configuration consists of 5 mm \acrshort{w} armor and a 5 mm \acrshort{v44} pipe.
For the \acrshort{bkfw}, two configurations are considered: 5 mm \acrshort{w} armor with either a 2.5 mm or 5 mm \acrshort{v44} pipe. 
The \acrshort{vv} is considered as a 50 mm \acrfull{ss} structure. All components are connected to high-flow helium coolant channels, which remove both deposited heat and released tritium from the coolant surfaces. The geometric parameters of these plasma-facing components are summarized in \cref{tab:operation_parameters}.  The detailed configurations are presented in \cref{fig:schematics_all:schematics_multiple}.


\subsection{Operating Conditions}
\label{sec:components_information:conditions}

All plasma-facing components are studied under pulsed operation relevant to the plant designs considered by Tokamak Energy. Each operation consists of 50 tritium injection pulses with a total operation time of 80,000 s, which ensures the system approaches a steady-state condition in both temperature and tritium inventory. \rtxt{Note that in future version of the model when microstructural evolution and irradiation effects on critical material properties and tritium transport behavior are accounted for, the steady state condition might be different and require more time to be reached}.
Each pulse consists of a 600 s tritium injection phase and a 1,000 s interval phase with no plasma exposure.

\begin{figure}[htb]
\centering
\includegraphics[width=0.5\textwidth]{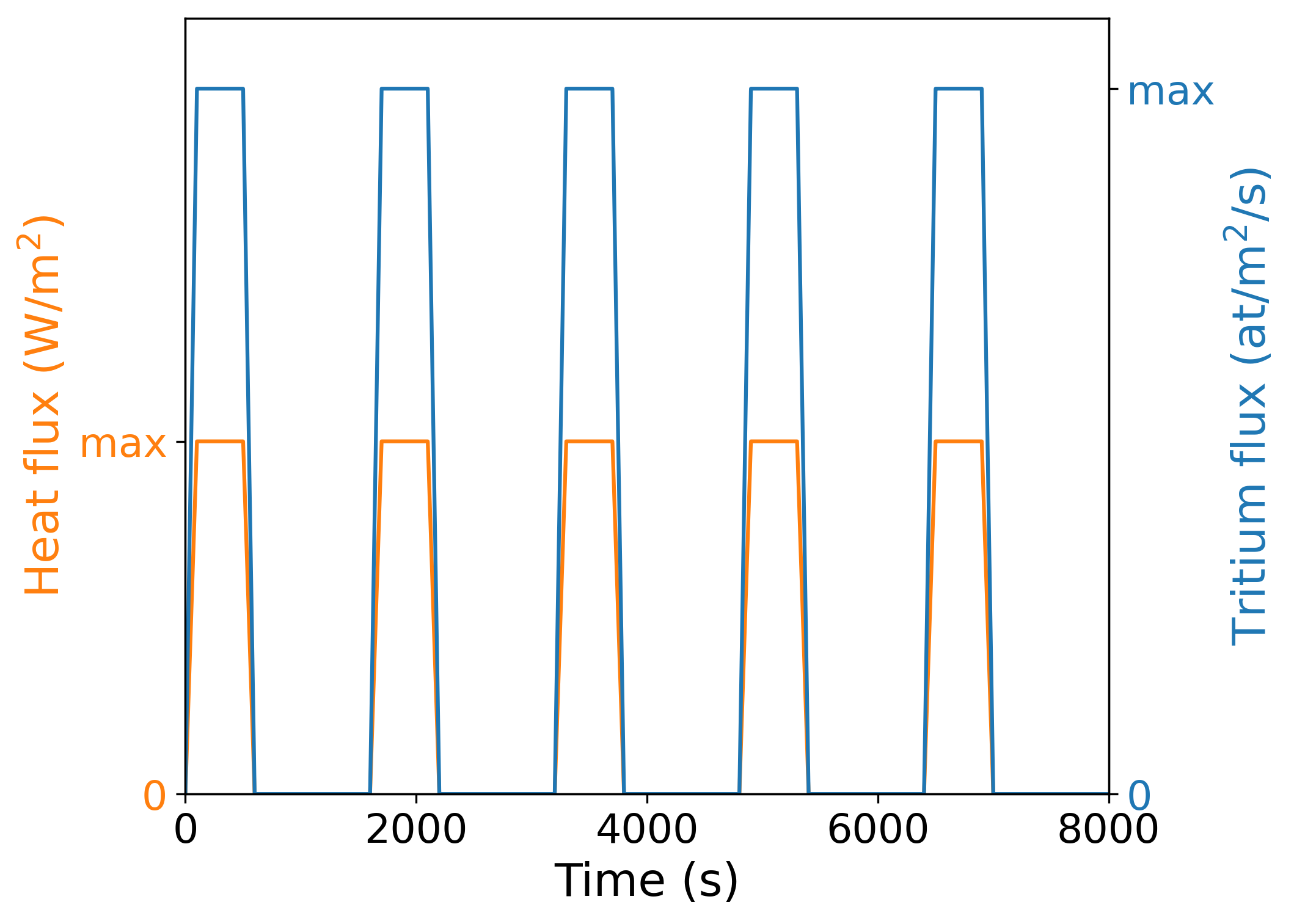}
\caption{\label{fig:tritium_and_heat_flux_evolution} The tritium flux and heat flux evolutions during the first five pulses of operation. Following pulses follow the same pattern.}
\end{figure}

The peak tritium fluxes at plasma-facing surface are $1\times 10^{25}$ at/m$^2$/s for the \acrshort{div}, $3\times 10 ^{23}$ at/m$^2$/s for the \acrshort{ccfw} and \acrshort{bkfw}, and $2.48\times 10 ^{18}$ at/m$^2$/s for the \acrshort{vv}. These fluxes correspond to a conservative assumption and likely overestimate the realistic fluxes \rtxt{in a worse case scenario}, since re-emission processes are not included in the present estimates. \rtxt{In practice, only a small fraction of the tritium flux is retained within plasma-facing components, while the majority is pumped through the vacuum pumping system.} The corresponding peak heat fluxes are $1\times 10 ^{7}$ W/m$^2$ for the \acrshort{div}, $5\times 10 ^{5}$ W/m$^2$ for the \acrshort{ccfw} and \acrshort{bkfw}, and no external heat flux for the \acrshort{vv}. The time evolution of these applied tritium and heat fluxes during the initial five pulses is shown in \cref{fig:tritium_and_heat_flux_evolution}.

In this study, as a first approximation, a constant temperature is maintained for each component at the coolant surface during each pulse, representing the high heat transfer efficiency of helium-cooling channels. For components with \acrshort{v44} cooling pipes (i.e., \acrshort{div}, \acrshort{ccfw}, and \acrshort{bkfw}), the helium coolant temperature is set to 673 K. Although the coolant temperature is adjustable, a constant value is used in this study for simplicity. For \acrshort{div} configurations with \acrshort{w} pipes, a higher coolant temperature of 873 K is adopted because tritium release from \acrshort{w} is more difficult. The \acrshort{vv} is maintained at 300 K. All parameters describing the operational conditions are summarized in \cref{tab:operation_parameters}.

\begin{table}[!ht]
\centering
\caption{The parameters of operation conditions of interest for the \acrshort{div}, \acrshort{ccfw}, \acrshort{bkfw}, and \acrshort{vv}.}
\label{tab:operation_parameters}
\begin{tabular}{|c|c|c|c|c|}
\hline
\textbf{Parameter} & \textbf{Description} & \textbf{Value} & \textbf{Units} & \textbf{Source} \\ \hline
$l_\text{W}^\text{DIV}$ & thickness of \acrshort{w} armor in \acrshort{div} & 3 or 8 & mm & Tokamak Energy \\ \hline
$l_\text{W}^\text{FW}$ & thickness of \acrshort{w} armor in first wall & 5 & mm & Tokamak Energy \\ \hline
$l_\text{V44}^\text{DIV}$ & thickness of \acrshort{v44} pipe in \acrshort{div} & 1 & mm & Tokamak Energy \\ \hline
$l_\text{V44}^\text{FW}$ & thickness of \acrshort{v44} pipe in first wall & 2.5 or 5 & mm & Tokamak Energy \\ \hline
$l_\text{SS}^\text{VV}$ & thickness of \acrshort{ss} in \acrshort{vv} & 50 & mm & Tokamak Energy \\ \hline
$J_\text{plasma}^\text{DIV}$ & \rtxt{peak} tritium flux in \acrshort{div} & $1\times10^{25}$ & at/m$^2$ & Tokamak Energy  \\ \hline
$J_\text{plasma}^\text{FW}$ & \rtxt{peak} tritium flux in first wall & $3\times10^{23}$ & at/m$^2$ & Tokamak Energy \\ \hline
$J_\text{plasma}^\text{VV}$ & \rtxt{peak} tritium flux in \acrshort{vv} & $2.48\times10^{18}$ & at/m$^2$ & Tokamak Energy  \\ \hline
$\Phi_\text{plasma}^\text{DIV}$ & \rtxt{peak} heat flux in \acrshort{div} & $1\times10^{7}$ & W/m$^2$ & Tokamak Energy  \\ \hline
$\Phi_\text{plasma}^\text{FW}$ & \rtxt{peak} heat flux in first wall & $5\times10^{5}$ & W/m$^2$ & Tokamak Energy  \\ \hline
$\Phi_\text{plasma}^\text{VV}$ & \rtxt{peak} heat flux in \acrshort{vv} & 0 & W/m$^2$ & Tokamak Energy  \\ \hline
$T_\text{coolant}^\text{DIV-V44}$ & coolant temperature in V44-pipe \acrshort{div} & 873 & K & Tokamak Energy  \\ \hline
$T_\text{coolant}^\text{DIV-W}$ & coolant temperature in W-pipe \acrshort{div} & 673 & K & Tokamak Energy  \\ \hline
$T_\text{coolant}^\text{FW}$ & coolant temperature in first wall & 673 & K & Tokamak Energy  \\ \hline
$T_\text{coolant}^\text{VV}$ & coolant temperature in \acrshort{vv} & 300 & K & Tokamak Energy  \\ \hline
\end{tabular}
\end{table}

\subsection{Material Properties}
\label{sec:components_information:parameters}

To accurately predict tritium transport behavior in components of interest, we collect material properties from the literature. Current components consist of three major material systems: \acrshort{w}, \acrshort{v44}, and \acrshort{ss}. The associated material parameters are summarized in \cref{tab:material_parameters_diff_solu,tab:material_parameters_thermal,tab:material_parameters_trapping}. \Cref{tab:material_parameters_diff_solu} lists the diffusivity and solubility parameters used to predict tritium transport through each material. \Cref{tab:material_parameters_thermal} includes the specific heat capacity and thermal conductivity to couple heat transfer behavior for each material. \Cref{tab:material_parameters_trapping} provides the trapping and release parameters for each material. Note that while a plethora of other material property models exist in the literature, the ones listed in \cref{tab:material_parameters_diff_solu,tab:material_parameters_thermal,tab:material_parameters_trapping} are selected for being either the most established or representative of average measurements. Quantifying the uncertainty of these model parameters and including more complex models for irradiation effects, multi-traps, and complex surface effects are left for future work. As the goal of this study is to enable a rapid design parameter space exploration, the main priority is to perform fast-running simulations and deploy surrogate models for multiscale analysis, as opposed to high-fidelity component-level tritium simulations \cite{shimada2024toward}.

\begin{table}[!ht]
\centering
\caption{Diffusivity and solubility parameters in tritium transport models for plasma-facing components.}
\label{tab:material_parameters_diff_solu}
\begin{tabular}{|c|c|c|c|c|}
\hline
\textbf{Parameter} & \textbf{Description} & \textbf{Value} & \textbf{Units} & \textbf{Reference} \\ \hline
$D^\text{W}$ & diffusivity in \acrshort{w} & $2.4\times10^{-7} \exp{\left(-\frac{37627.5}{RT}\right)}$ & m$^2$/s & \cite{shimada2024toward,hodille2021modelling,simon2025moose} \\ \hline
$D^\text{V44}$ & diffusivity in \acrshort{v44} & $7.5\times10^{-8} \exp{\left(-\frac{12548.9}{RT}\right)}$ & m$^2$/s & \cite{hashizume2007diffusional} \\ \hline
$D^\text{SS}$ & diffusivity in \acrshort{ss} & $9.18\times10^{-7} \exp{\left(-\frac{56500}{RT}\right)}$ & m$^2$/s & \cite{byeon2020transport} \\ \hline
$S^\text{W}$ & solubility in \acrshort{w} &
\begin{tabular}{@{}c@{}}
$1.87\times10^{24} \exp\left(-\frac{140341}{RT}\right)$ \\
$+ 3.14\times10^{21} \exp\left(-\frac{54993}{RT}\right)$
\end{tabular}
& at/m$^3$/Pa$^{0.5}$ & \cite{shimada2024toward,hodille2021modelling,simon2025moose} \\ \hline
$S^\text{V44}$ & solubility of \acrshort{v44} & $5.97\times10^{17} \exp\left(\frac{29099}{RT}\right)$ & at/m$^3$/Pa$^{0.5}$ & \cite{buxbaum2002hydrogen} \\ \hline
$S^\text{SS}$ & solubility of \acrshort{ss} & $1.32\times10^{23} \exp\left(-\frac{10100}{RT}\right)$ & at/m$^3$/Pa$^{0.5}$ & \cite{byeon2020transport} \\ \hline
\end{tabular}
\end{table}

\begin{table}[!ht]
\centering
\caption{Thermal parameters in tritium transport models for plasma-facing components.}
\label{tab:material_parameters_thermal}
\begin{tabular}{|c|c|c|c|c|}
\hline
\textbf{Parameter} & \textbf{Description} & \textbf{Value} & \textbf{Units} & \textbf{Reference} \\ \hline
$c_p^\text{W}$ & specific heat capacity of \acrshort{w} &
\begin{tabular}{@{}c@{}c@{}c@{}}
$0.116 + 7.11\times10^{-5} T$ \\
$- 6.58\times10^{-8} T^2$ \\
$+ 3.24\times10^{-11} T^3$ \\
$- 5.45\times10^{-15} T^4 $
\end{tabular}
& J/g/K & \cite{shimada2024toward,simon2025moose}  \\ \hline
$k^\text{W}$ & thermal conductivity of \acrshort{w} &
\begin{tabular}{@{}c@{}c@{}c@{}}
$0.0241 - 2.9\times10^{-1} T$ \\
$+ 2.54\times10^{-4} T^2$ \\
$- 1.03\times10^{-7} T^3$ \\
$+ 1.52\times10^{-11} T^4 $
\end{tabular}
& W/m/K & \cite{shimada2024toward,simon2025moose}  \\ \hline
$c_p^\text{V44}$ & specific heat capacity of \acrshort{v44} & $0.58 - \frac{21.1}{T}$ & J/g/K & \cite{porter1995subtask}  \\ \hline
$k^\text{V44}$ & thermal conductivity of \acrshort{v44} & $27.827 + 8.6\times10^{-3} T$ & W/m/K & \cite{porter1995subtask}  \\ \hline
$c_p^\text{SS}$ & specific heat capacity of \acrshort{ss} & $0.459 + 1.328\times10^{-4} T$ & J/g/K & \cite{byeon2020transport}  \\ \hline
$k^\text{SS}$ & thermal conductivity of \acrshort{ss} & $9.248 - 1.571\times10^{-2} T$ & W/m/K & \cite{byeon2020transport}  \\ \hline
\end{tabular}
\end{table}

\begin{table}[!ht]
\centering
\caption{Trapping parameters in tritium transport models for plasma-facing components.}
\label{tab:material_parameters_trapping}
\begin{tabular}{|c|c|c|c|c|}
\hline
\textbf{Parameter} & \textbf{Description} & \textbf{Value} & \textbf{Units} & \textbf{Reference} \\ \hline
$\epsilon_t^\text{W}$ & trapping energy of \acrshort{w} & 27010.8 & J/mol & \cite{hodille2021modelling,dark2024modelling}  \\ \hline
$\epsilon_r^\text{W}$ & release energy of \acrshort{w} & 100300.98 & J/mol & \cite{hodille2021modelling,dark2024modelling}  \\ \hline
$\chi^\text{W}$ & trapping fraction of \acrshort{w} & $3.8\times10^{-7}$ & - & \cite{hodille2021modelling,dark2024modelling}  \\ \hline
$\tau_t^\text{W}$ & trapping pre-factor of \acrshort{w} & $3.3\times10^{12}$ & 1/s & \cite{hodille2021modelling,dark2024modelling,shimada2024toward}  \\ \hline
$\tau_r^\text{W}$ & release pre-factor of \acrshort{w} & $1\times10^{13}$ & 1/s & \cite{hodille2021modelling,dark2024modelling,shimada2024toward}  \\ \hline

$\epsilon_t^\text{V44}$ & trapping energy of \acrshort{v44} & 0 & J/mol & \cite{hodille2021modelling,shimada2024toward}  \\ \hline
$\epsilon_r^\text{V44}$ & release energy of \acrshort{v44} & 17174.3 & J/mol & \cite{hodille2021modelling,hashizume2007diffusional}  \\ \hline
$\chi^\text{V44}$ & trapping fraction of \acrshort{v44} & $4.0\times10^{-2}$ & - & \cite{hodille2021modelling,hashizume2007diffusional}  \\ \hline
$\tau_t^\text{V44}$ & trapping pre-factor of \acrshort{v44} & $2.75\times10^{11}$ & 1/s & \cite{shimada2024toward}  \\ \hline
$\tau_r^\text{V44}$ & release pre-factor of \acrshort{v44} & $8.4\times10^{12}$ & 1/s & \cite{shimada2024toward}  \\ \hline

$\epsilon_t^\text{SS}$ & trapping energy of \acrshort{ss} & 0 & J/mol & \cite{hodille2021modelling,shimada2024toward}  \\ \hline
$\epsilon_r^\text{SS}$ & release energy of \acrshort{ss} & 44900 & J/mol & \cite{esteban2007hydrogen}  \\ \hline
$\chi^\text{SS}$ & trapping fraction of \acrshort{ss} & $3.85\times10^{-5}$ & - & \cite{esteban2007hydrogen}  \\ \hline
$\tau_t^\text{SS}$ & trapping pre-factor of \acrshort{ss} & $2.75\times10^{11}$ & 1/s & \cite{shimada2024toward}  \\ \hline
$\tau_r^\text{SS}$ & release pre-factor of \acrshort{ss} & $8.4\times10^{12}$ & 1/s & \cite{shimada2024toward}  \\ \hline
\end{tabular}
\end{table}

\subsection{Critical Input Parameters for Surrogate Model}
\label{sec:components_information:surrogate}

To ensure the surrogate models discussed in \cref{sec:methods:STM} can simulate all plasma-facing component configurations, seven key input parameters are selected from \cref{tab:operation_parameters,tab:material_parameters_trapping}. The corresponding parameter ranges used for training and validation are summarized in \cref{tab:input_parameters}.
\begin{table}[!ht]
\centering
\caption{The selected input parameters for tritium transport surrogate model training.}
\label{tab:input_parameters}
\begin{tabular}{|c|c|c|c|c|}
\hline
\textbf{Parameter} & \textbf{Description} & \textbf{Range} & \textbf{Units} \\ \hline
$J_\text{plasma}$ & tritium flux on plasma-facing surface & [$2\times10^{23}$, $2\times10^{25}$] & at/m$^2$ \\ \hline
$\Phi_\text{plasma}$ & heat flux on plasma-facing surface & [$2\times10^{5}$, $2\times10^{7}$] & W/m$^2$ \\ \hline
$T_\text{coolant}$ & coolant temperature & [573, 673] & K \\ \hline
$l_\text{W}$ & thickness of \acrshort{w} armor & [4, 8] & mm \\ \hline
$l_\text{V44}$ & thickness of \acrshort{v44} pipe & [1, 2] & mm \\ \hline
$\chi^\text{W}$ & trapping fraction of tritium in \acrshort{w} & [$5.0\times10^{-5}$, $5.0\times10^{-4}$] & -  \\ \hline
$\epsilon_r^\text{W}$ & release energy of tritium in \acrshort{w} & [0.95, 1.05] & eV  \\ \hline
\end{tabular}
\end{table}

\subsection{Critical Component Properties for Fuel Cycle Model}
\label{sec:components_information:fuel_cycle}

As shown in \cref{tab:fuel_cycle_parameters}, we use the critical component parameters from Tokamak Energy except for the residence time for \acrshort{div}, \acrshort{ccfw}, and \acrshort{bkfw}; the residence time is obtained from our surrogate models discussed in \cref{sec:results:engineering}. This fuel cycle model includes 17 components: Tokamak Exhaust Processing (TEP), Secondary Pump (SP), Gas Detritiation System (GDS), Water Detritiation System (WDS), Stack, Isotope Separation System (ISS), Isotope Adjustment System (IAS), Tritium Extraction System (TES), Heat Exchanger (HX), Blanket (BK), Helium Loop (HL), Air Detritiation System (ADS), Fusion Chamber Pump (FCP), Storage System (STO), and three plasma-facing components---\acrshort{div}, \acrshort{ccfw}, and \acrshort{bkfw}. \rtxt{The residence times of these components do not exactly match the values reported in Ref.~\cite{TE_lead_paper}, as those values represent a single example configuration from Tokamak Energy. However, they are sufficient to demonstrate the goals of this study.}

\begin{table}[!ht]
\centering
\caption{Fuel cycle modeling parameters from Tokamak Energy \cite{TE_lead_paper}.}
\label{tab:fuel_cycle_parameters}
\begin{tabular}{|c|c|c|c|}
\hline
\textbf{Parameter} & \textbf{Description} & \textbf{Value} & \textbf{Unit} \\ \hline
TBR & tritium breeding ratio & 1.113 & -  \\ \hline
TBE & tritium burn efficiency & 0.02 & -  \\ \hline
AF & availability factor & 0.375 & -  \\ \hline
$\varepsilon_i$ & non-radioactive tritium losses & 
\begin{tabular}{l}
$\varepsilon_{SP}=2.78e-8$, $\varepsilon_{Stack}=9.98e-8$, \\ $\varepsilon_{TES}=1.67e-7$, $\varepsilon_{BK}=1.62e-8$, \\ $\varepsilon_{FW}=3.33e-7$, $\varepsilon_{ADS}=2.78e-8$, \\ $\varepsilon_{FCP}=5.56e-8$, \\ other components with $\varepsilon_i=0$
\end{tabular}
& - \\ \hline
$\dot{N}_{T,burn}$ & tritium burn rate & 2.7938$\times 10^{-\rtxt{6}}$ & kg/s \\ \hline
$\lambda$ & tritium decay rate & 1.78$\times 10^{-9}$ & s$^{-1}$  \\ \hline
$\tau_i$ & tritium residence time& 
\begin{tabular}{l}
$\tau_{TEP}=950.04$, $\tau_{SP}=3600$, $\tau_{GDS}=7197.12$,\\$\tau_{WDS}=172800$, $\tau_{Stack}=1002.24$, $\tau_{ISS}=21600$, \\$\tau_{IAS}=404.23981$, $\tau_{TES}=600$, $\tau_{HX}=95.04$, \\$\tau_{BK}=6164.38$, $\tau_{HL}=1002.24$, $\tau_{ADS}=3602.88$, \\$\tau_{FCP}=1800$, $\tau_{Plasma}=1800$
\end{tabular}
& s \\ \hline
\end{tabular}
\end{table}

\section{Methods}
\label{sec:methods}

\subsection{TMAP8}
\label{sec:methods:TMAP8}

The models developed in this study are implemented in \acrshort{tmap8} to simulate tritium transport in fusion components and tritium recycling in the tritium fuel cycle \cite{simon2025moose}. \acrshort{tmap8} is a MOOSE-based, \acrfull{nqa} compliant, open-source application that has advanced multiscale capabilities for tritium transport and fusion fuel cycle modeling \cite{simon2025moose}. \acrshort{tmap8} has been verified and validated using verification and validation cases from TMAP4, TMAP7, and beyond \cite{longhurst1992verification,ambrosek2008verification, simon2025moose,TMAP8}, which demonstrate that \acrshort{tmap8} retains the capabilities of previous versions while extending to advanced functionalities. Currently, TMAP8 includes capabilities for modeling tritium diffusion, kinetic surface chemical reactions, and trapping effects, although these capabilities are not fully exploited in this model, they are key to predicting tritium transport in the components of interest in this study.
In addition, \acrshort{tmap8} can leverage the \acrshort{stm} within the MOOSE framework to construct surrogate models \cite{slaughter2023moose,dhulipala2025moose}. These surrogate models can be trained using component-level results to approximate tritium transport behavior with an order-of-magnitude lower computational cost. These capabilities enable detailed component-level simulations and system-level fuel cycle simulations to be efficiently coupled, which facilitates optimization and uncertainty quantification of tritium inventory and recovery analysis.

\subsection{Governing Equations for Component-Level Models}
\label{sec:methods:governing_components}

Tritium transport in each component is governed by the tritium diffusion and trapping kinetics within the armor and pipe materials. The diffusion of mobile tritium is described as:
\begin{equation}
    \label{eqn:governing:transport}
    \frac{\partial C}{\partial t} = \nabla \cdot D \nabla C - \frac{\partial C_{trap}}{\partial t},
\end{equation}
where $C$ is the concentration of mobile tritium in the material (e.g., \acrshort{w}, \acrshort{v44}, \acrshort{ss}) in mol/m$^3$, $t$ is the time in s, $D$ is the tritium diffusivity in m$^2$/s, and $C_{trap}$ is the concentration of trapped tritium in mol/m$^3$. Trapping sites represent features in a material that can capture tritium and hence slow its transport, e.g., interstitial sites, vacancies, dislocations, pores, grain boundaries, etc. \cite{lee1983hydrogen}. 

The evolution of trapped tritium concentration is governed by:
\begin{equation}
    \label{eqn:governing:trapping}
    \frac{\partial C_{trap}}{\partial t} = \alpha_t \frac{\chi N - C_{trap}}{N} C - \alpha_r C_{trap},
\end{equation}
where $N$ is the density of the host material in mol/m$^3$, $\chi$ is the atomic fraction of the trapping site in the host material, and $\alpha_t$ and $\alpha_r$ are the trapping and release rates in 1/s, respectively. In the proposed model, both rates are functions of the temperature $T$ and follow the Arrhenius equation:
\begin{equation}\label{eqn:governing:trapping_rate}
    \alpha_t = \tau_{t0} \exp \left( \frac{-\epsilon_t}{k_B T} \right),
\end{equation}
and
\begin{equation}
\label{eqn:governing:release_rate}
    \alpha_r = \tau_{r0} \exp \left( \frac{-\epsilon_r}{k_B T} \right) ,
\end{equation}
where $\tau_{t0}$ and $\tau_{r0}$ are the trapping and release rate coefficients in 1/s, respectively, $\epsilon_t$ and $\epsilon_r$ are the trapping and release energy in eV, respectively, $k_B$ is the Boltzmann constant in eV/K, and $T$ is the temperature in K.

Heat transfer is coupled with tritium transport in the models to simulate the temperature evolution during pulsed plasma operation. The transient heat conduction is governed by the heat diffusion equation:
\begin{equation}
\label{eqn:governing:heat_diffusion}
    \rho c_p \frac{\partial T}{\partial t} = \nabla \cdot k \nabla T,
\end{equation}
where $\rho$ is the corresponding material density in g/m$^3$, $c_p$ is the corresponding specific heat capacity in J/g/K, and $k$ is the corresponding thermal conductivity in W/m/K.

At the plasma-facing surface of each component, the injected tritium and heat fluxes are applied as Neumann boundary conditions:
\begin{equation}
\label{eqn:governing:neumann_tritium}
    \left. \frac{\partial C}{\partial x} \right|_{x=0}  = f_T (t) - 2\left(K_r C^2 - K_d P\right),
\end{equation}
and
\begin{equation}
\label{eqn:governing:neumann_heat}
    \left. \frac{\partial T}{\partial x} \right|_{x=0}  = f_H(t) ,
\end{equation}
where $f_T$ and $f_H$ are functions of applied tritium and heat fluxes evolving with time $t$ (as illustrated in \cref{fig:tritium_and_heat_flux_evolution}), respectively, $x$ is the distance from the plasma-facing surface ($x=0$) into the material, $K_r$ and $K_d$ are the recombination and dissociation rates for tritium on the \acrshort{w} surface, respectively, and $P$ is the tritium partial pressure at the plasma-facing surface, with a value of 0 for simplicity. Tritium release from the plasma-facing surface to the vacuum chamber occurs continuously through recombination during both the injection and interval phases.

At the coolant surface, Dirichlet boundary conditions are imposed to represent the strong convective removal of tritium and heat by the helium coolant, which is described as
\begin{equation}
\label{eqn:governing:dirchilet_tritium}
    C = 0,
\end{equation}
and
\begin{equation}
\label{eqn:governing:dirchilet_heat}
    T = T_{\text{coolant}},
\end{equation}
where $T_{\text{coolant}}$ is the prescribed coolant temperature for each component. These boundary conditions assume that the high helium flow rate effectively maintains zero tritium concentration and a constant temperature at the coolant interface. This conservatively overestimates the flux of tritium to the coolant. This assumption can be refined in future work with a more accurate description of heat and mass convection conditions at the surface.

\subsection{Surrogate Tritium Transport Model Using STM in MOOSE}
\label{sec:methods:STM}


The surrogate tritium transport model is used for fast and automated system-level fuel cycle analysis. The Gaussian process surrogate is utilized in this study because of its superior performance and robustness, which is demonstrated in \ref{sec:appendix:surrogates}, where it is compared to other surrogate model methods.
We perform 200, 400, 800, 1,600, 3,200, and 6,400 component-level tritium transport simulations as six datasets. Each dataset includes seven input parameters: tritium flux in the plasma-facing surface, heat flux in the plasma-facing surface, \acrshort{w} armor thickness, pipe thickness, coolant temperature, trapping site fraction in the \acrshort{w} armor, and release energy in the \acrshort{w} armor. The outputs are (1) the steady-state tritium flux at the coolant surface, denoted as $J_{\infty}$, and (2) the residence time, represented by the parameters $\tau_0$ and $\tau_1$. The residence time is extracted by fitting the time-dependent coolant-side tritium flux $J_{\text{back}}$ using the following exponential form:
\begin{equation}
\label{eqn:surrogate:resident_time_fitting}
J_\text{back} = 
\begin{cases}
0, & t \le \tau_0, \\
J_{\infty} \left(1 - \exp\left(-\frac{t-\tau_0}{\tau_1}\right)\right), & t > \tau_0,
\end{cases},
\end{equation}
where $\tau_0$ is the delay time before tritium flux reaches the coolant surface, and $\tau_1$ is the time constant of coolant-side flux reaching steady state. Both $\tau_0$ and $\tau_1$ are used in the system-level fuel cycle analysis described in \cref{sec:results:engineering}. 

The original one-parameter residence time equation is described as \cite{abdou1986deuterium,abdou2020physics}:
\begin{equation}
\label{eqn:surrogate:one_parameter_resident_time_fitting}
J_\text{back} = 
J_{\infty} \left(1 - \exp\left(-\frac{t}{\tau}\right)\right),
\end{equation}
where $\tau$ is the time constant of coolant-side flux reaching steady state. This one-parameter residence time is not used in this work since it cannot consider the delay time observed in tritium losses to the coolant in plasma-facing components.

For each entry in the dataset, the tritium flux evolution from the component-level simulation is recorded over 80,000 s. The resulting flux is then averaged using a 1,600 s moving window (i.e., pulse cycle length) to reduce fluctuations caused by pulsed loading and to facilitate subsequent curve fitting. During the fitting process, the steady-state tritium flux at the coolant surface and the corresponding residence-time parameters are extracted for use in further surrogate model training.

Some component-level model simulation results exhibit numerical instabilities in some pulse loops. These cases are removed from the dataset to prevent potential adverse effects on the surrogate model training, as discussed in \ref{sec:appendix:flaw}. 
Furthermore, due to fitting-related uncertainties, some calculated values of $\tau_0$ and $\tau_1$ exceed the maximum simulation time of 80,000 s. To maintain data consistency, all simulations with $\tau_0$ or $\tau_1$ greater than 80,000 s are excluded from the training datasets. Representative examples of the type of simulation results excluded from the datasets are shown in \ref{sec:appendix:flaw}. It is also important to note that 20 \% or less of all the simulations are excluded from the datasets. In addition, to avoid nonphysical negative residence times, we use the logarithmic values of $\tau_0$ and $\tau_1$ as input parameters during model training.

The model accuracy is quantified using the \acrfull{rmspe}, which is defined as:
\begin{equation}
\label{eqn:surrogate:RMSPE}
    \text{RMSPE}(x) = \frac{\sqrt{\sum{(x_{\text{surrogate}} - x_{\text{TMAP}})^2} / n}}{\sum{x_\text{TMAP}} / n},
\end{equation}
where $x_\text{TMAP}$ represents the reference results calculated from the \acrshort{tmap8} simulations, $x_\text{surrogate}$ represents the surrogate model predictions, and $n$ is the number of data points. A lower \acrshort{rmspe} value indicates higher predictive accuracy. To ensure statistical robustness, the \acrshort{rmspe}s for both the training and test datasets are compared to evaluate and prevent potential underfitting or overfitting of the surrogate model.

\subsection{Governing Equations for System-Level Models}
\label{sec:methods:governing_system}

To evaluate the tritium evolution across the entire fusion fuel cycle, each major component can be represented by a zero-dimensional model. The time-dependent tritium retention in each component $i$ is described by an \acrfull{ode}:
\begin{equation}
\label{eqn:goerning:fuel_cycle}
\frac{dI_i}{dt} = \sum_{j\neq i}{\frac{I_j}{\tau_j}}  - \left(1+\epsilon_i\right)\frac{I_i}{\tau_i} - \lambda I_i + S_i,
\end{equation}
where $I_i$ and $\tau_i$ are the tritium inventory and tritium residence time of component $i$ in kg and s, respectively, $I_j$ and $\tau_j$ are the tritium inventory and tritium residence time in other systems feeding into system $i$, $\epsilon_i$ is the non-radioactive tritium losses in system $i$ (e.g., flux to other subsystems), $\lambda$ is the radioactive decay in 1/s, and $S_i$ is the tritium source term in kg/s, which will only appear in plasma chamber components due to injected fusion fuel. 

\begin{figure}[htb]
\centering
\includegraphics[width=0.65\textwidth]{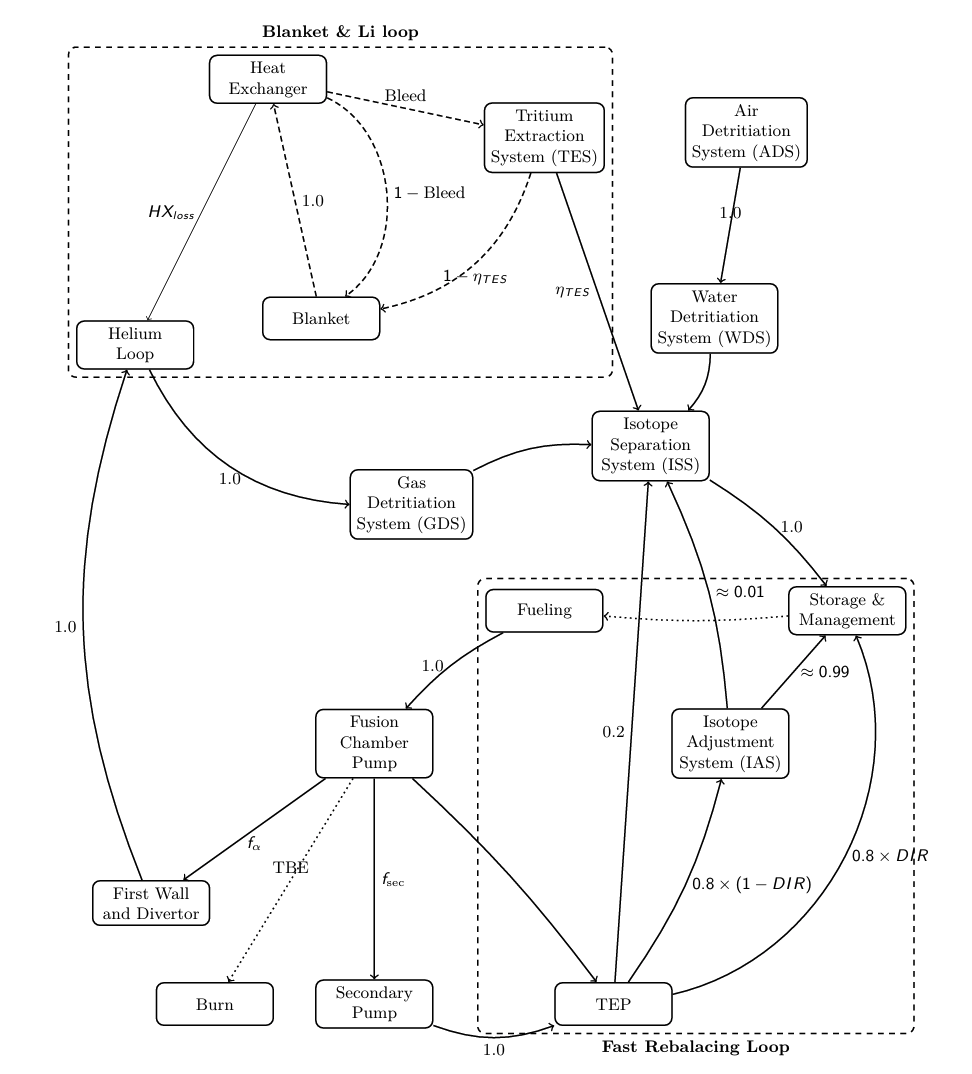}
\caption{\label{fig:TE_fuel_cycle_configuration} The fuel cycle configuration with all critical components for ST-E1 from Ref.~\cite{TE_lead_paper}.}
\end{figure}

The primary difference between the fuel cycle model used in this work and the conventional one in TMAP8 \cite{abdou2020physics,simon2025moose} is the configuration of the fusion plant components. The fuel cycle configuration adopted from Ref. \cite{TE_lead_paper} includes a more detailed representation of subsystem boundaries, which enables a more comprehensive and design-oriented analysis, as shown in \cref{fig:TE_fuel_cycle_configuration}. In addition, the First Wall and Divertor component is divided into the \acrshort{ccfw}, \acrshort{bkfw}, and \acrshort{div} so the tritium behavior in each plasma-facing component can be comprehensively investigated.

\subsection{Numerical Method}
\label{sec:methods:numerical_method}

All component-level tritium transport simulations are conducted using the finite-element implementation in \acrshort{tmap8}. A uniform mesh size of 0.1 mm is used to balance spatial resolution and computational cost. 

During the simulations, an adaptive time stepping is implemented to dynamically adjust the time step size based on the converge efficiency, minimizing computational costs. The second-order backward differentiation formula time integration scheme is used.

All the components in the system-level fuel cycle model are built on a zero-dimensional mesh because their structures are simplified as residence time to reduce the computational cost \cite{simon2025moose}. A similar adaptive time stepping and the second-order backward differentiation formula time integration scheme are implemented in the system-level simulations.

All simulations are conducted on a high-performance computing cluster with nodes of 2 Intel Xeon Platinum 8480+ CPUs. Each CPU has 56 cores and 256 GB of memory.

\section{Results}
\label{sec:results}

This section presents the results from the component-scale simulations for the base design and operation conditions (see \cref{sec:results:components}), the performance of the surrogate models based on the component-scale simulations on a wide range of design parameters and operational conditions (see \cref{sec:results:surrogate}), and predictions from the system-level fuel cycle model (see \cref{sec:results:engineering}). 

\subsection{One-Dimensional Tritium Diffusion from Component-Level Model}
\label{sec:results:components}

\begin{figure}[H]
\centering
\begin{subfigure}{.43\linewidth}
  \centering
	\includegraphics[width = \linewidth]{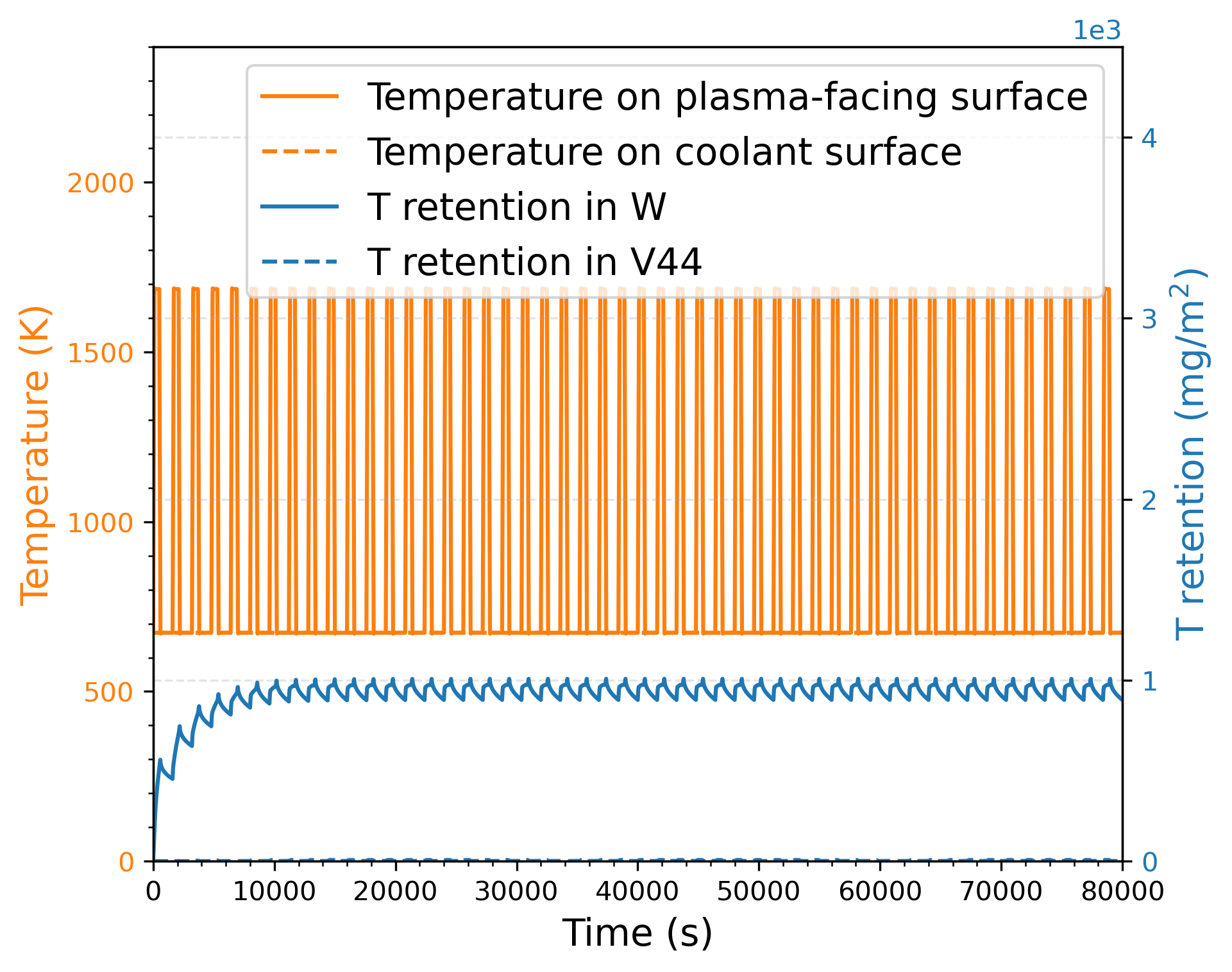}
    \caption{}
    \label{fig:1D_results:DIV_W_V44_v1_comp:tritium_rentension}
\end{subfigure}
\begin{subfigure}{.43\linewidth}
  \centering
	\includegraphics[width = \linewidth]{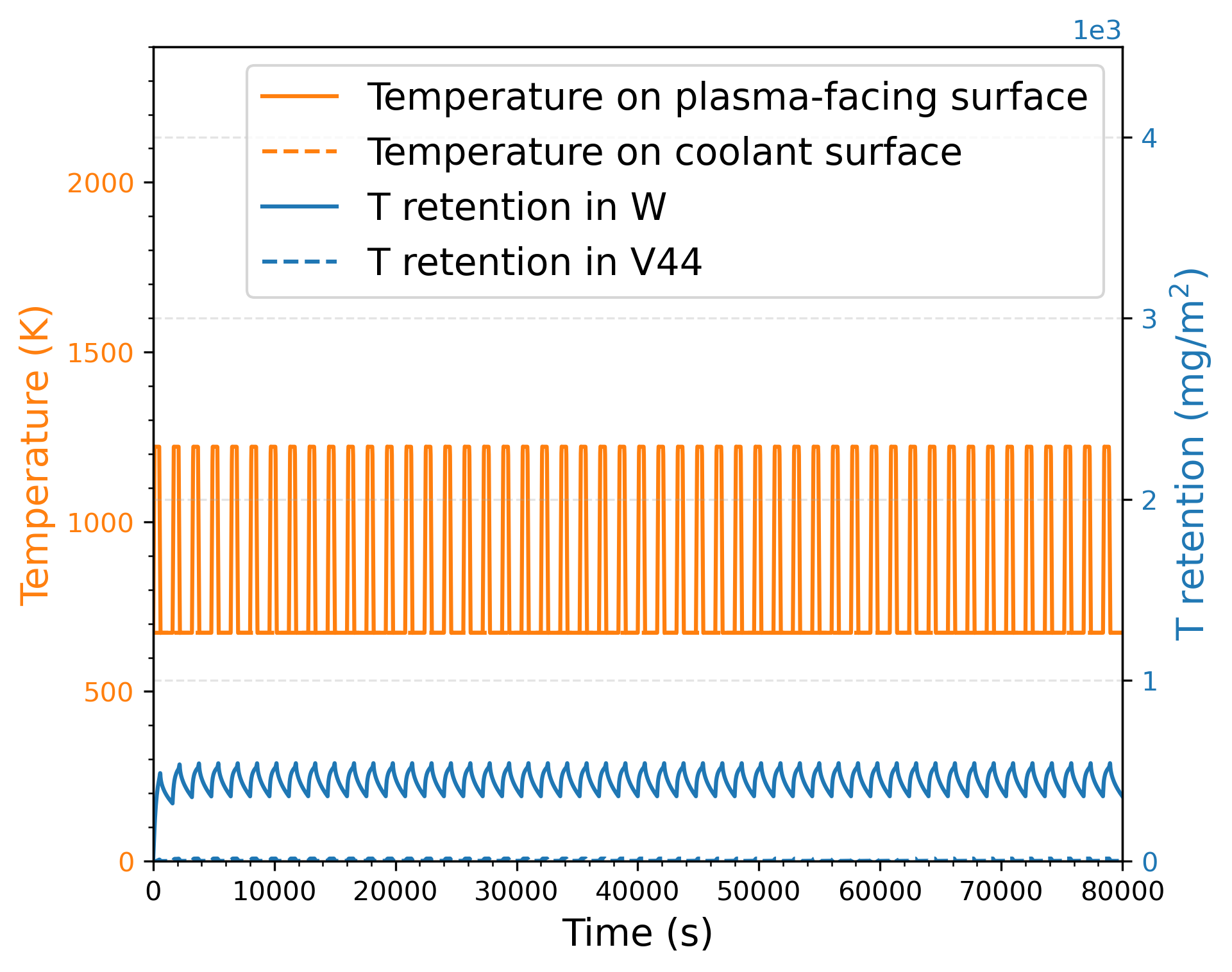}
    \caption{}
    \label{fig:1D_results:DIV_W_V44_v2_comp:tritium_rentension}
\end{subfigure}
\begin{subfigure}{.43\linewidth}
  \centering
	\includegraphics[width = \linewidth]{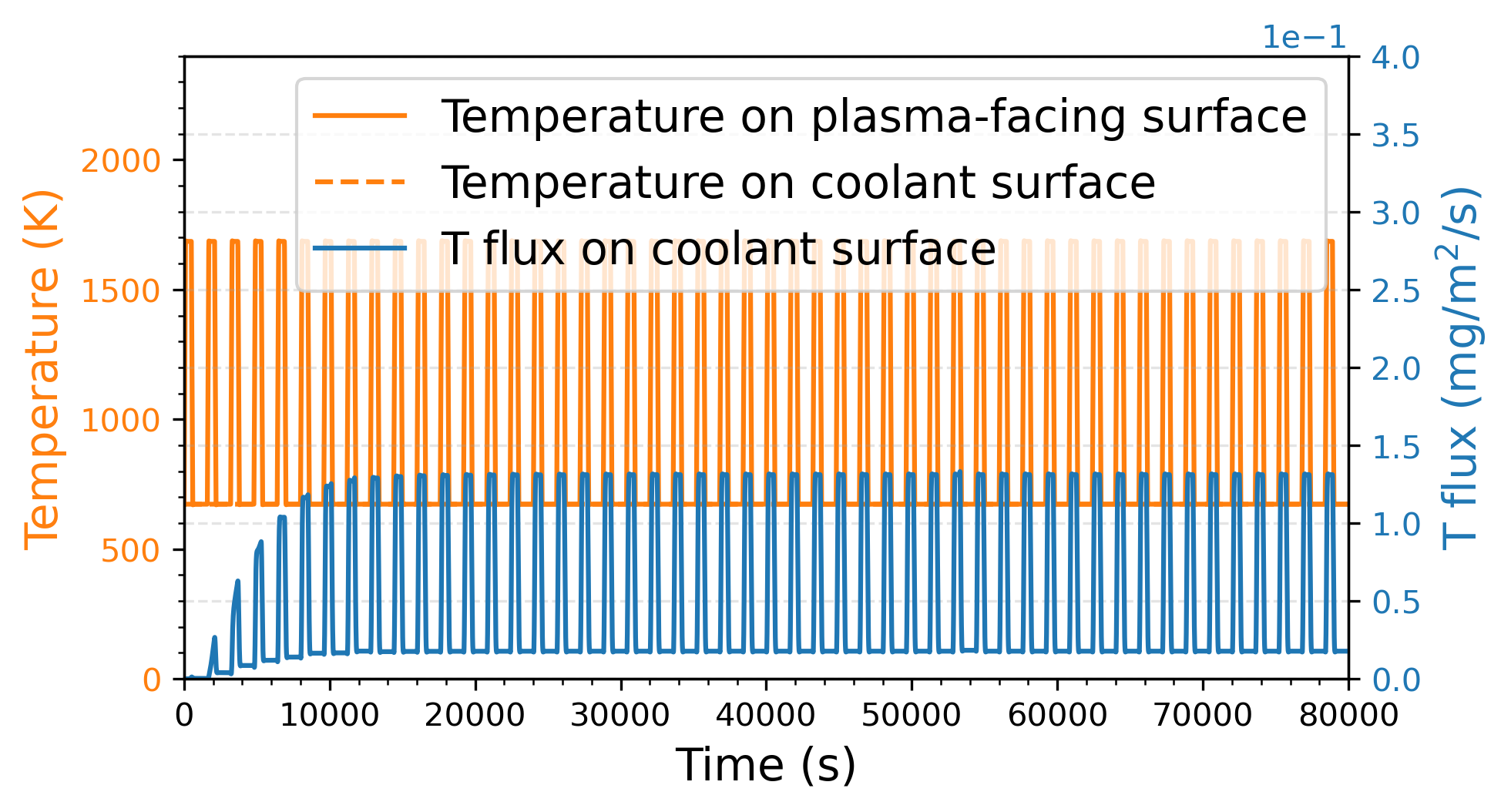}
    \caption{}
    \label{fig:1D_results:DIV_W_V44_v1_comp:tritium_flux}
\end{subfigure}
\begin{subfigure}{.43\linewidth}
  \centering
	\includegraphics[width = \linewidth]{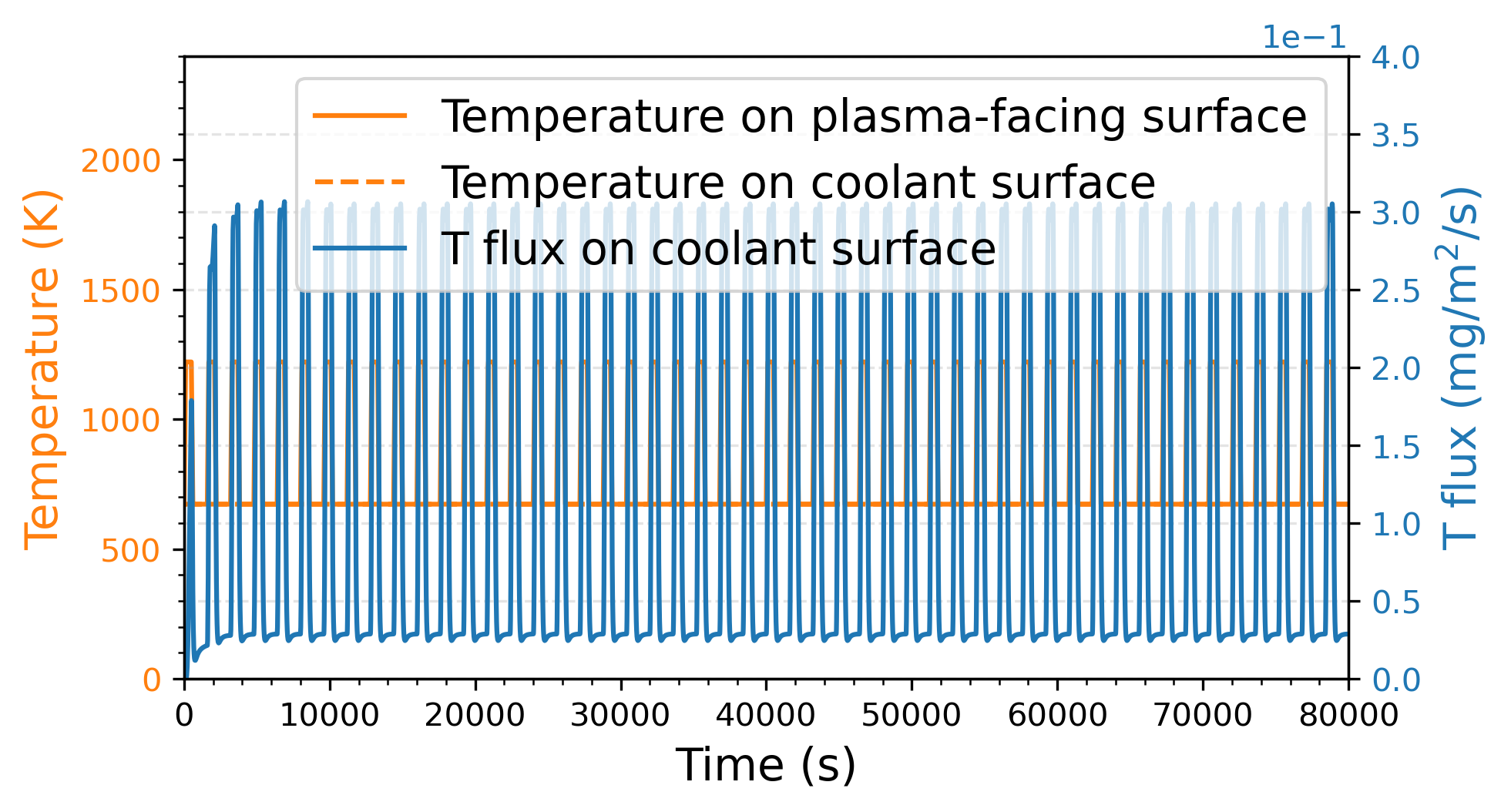}
    \caption{}
    \label{fig:1D_results:DIV_W_V44_v2_comp:tritium_flux}
\end{subfigure}
\begin{subfigure}{.43\linewidth}
  \centering
	\includegraphics[width = \linewidth]{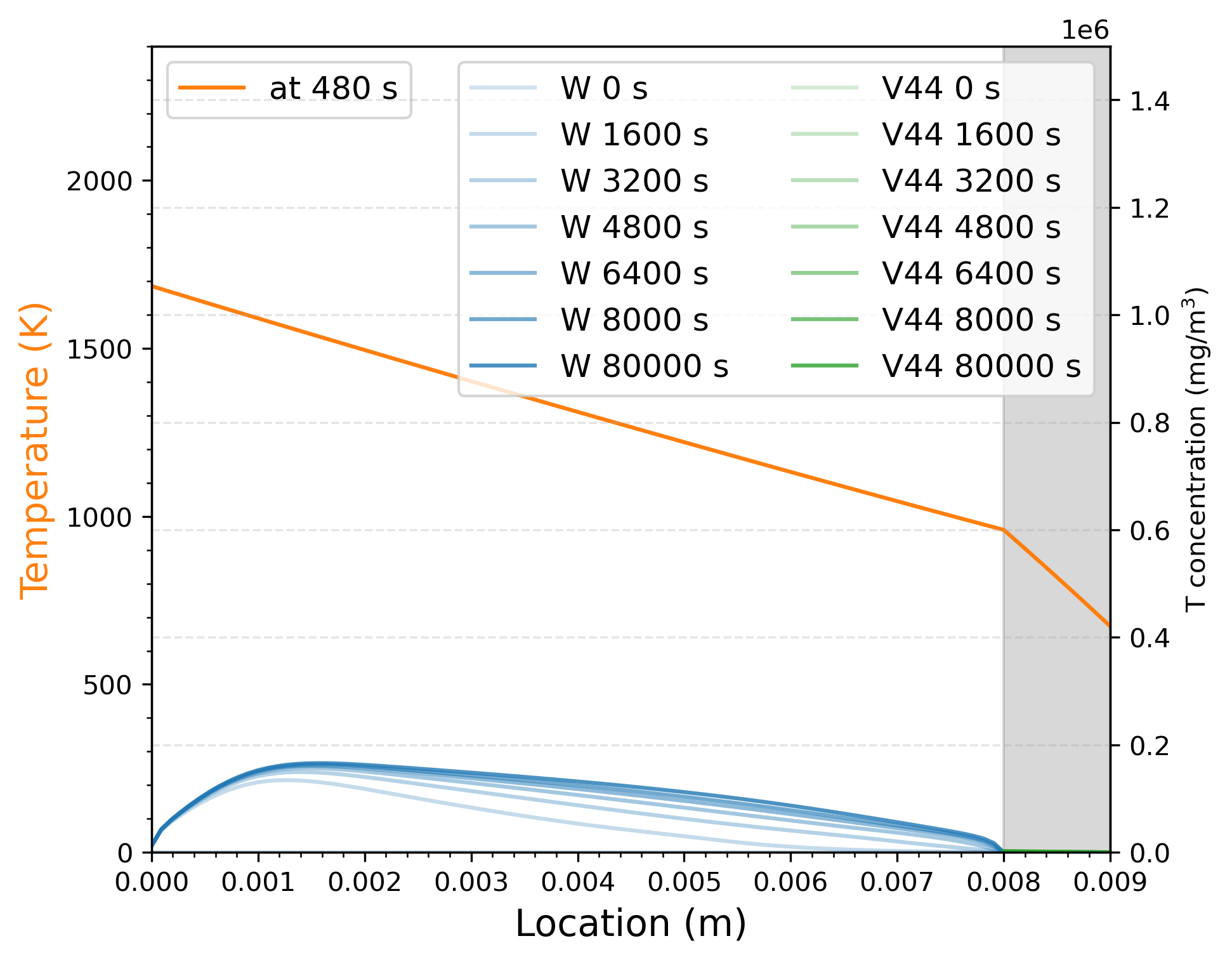}
    \caption{}
    \label{fig:1D_results:DIV_W_V44_v1_comp:tritium_profile}
\end{subfigure}
\begin{subfigure}{.43\linewidth}
  \centering
	\includegraphics[width = \linewidth]{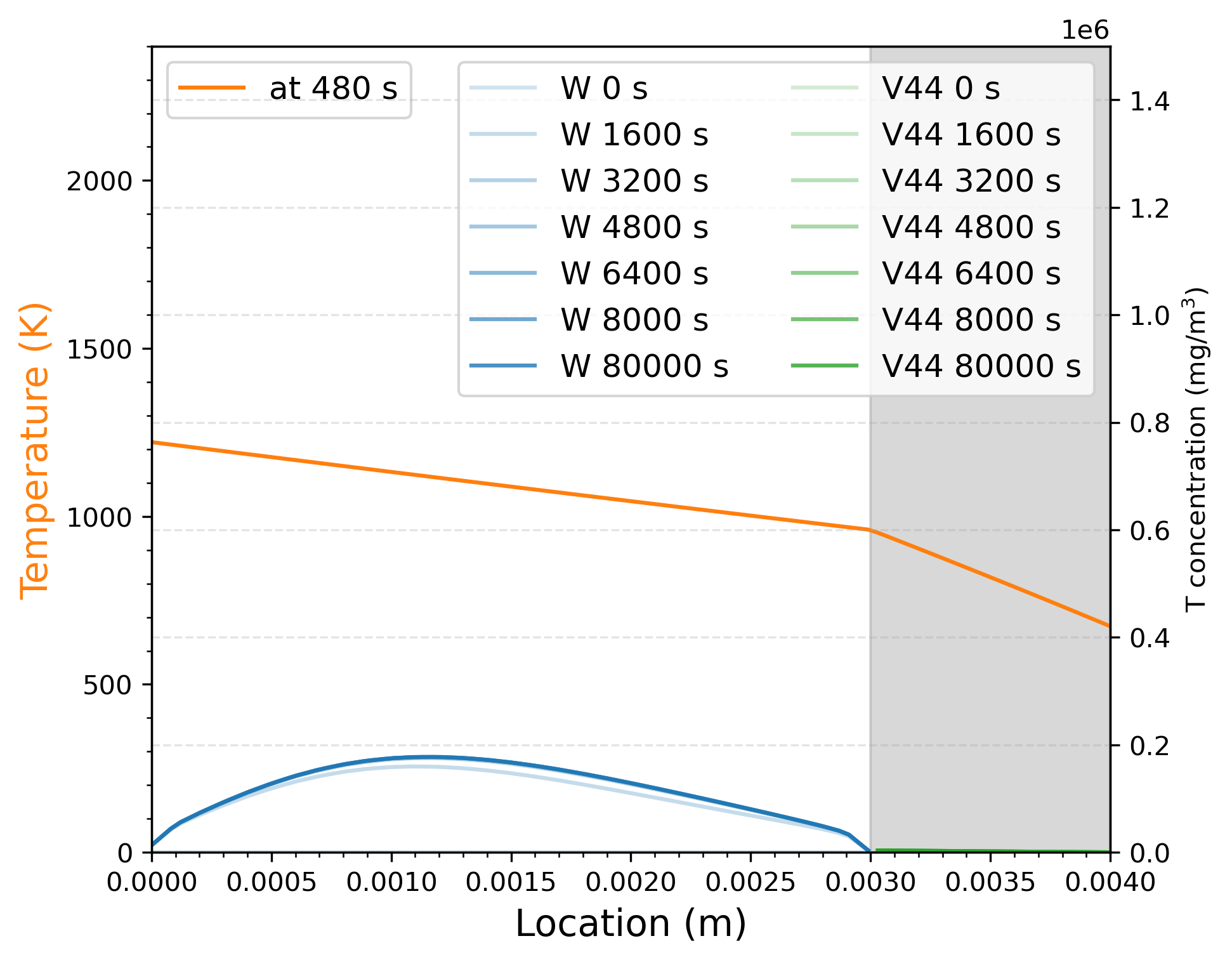}
    \caption{}
    \label{fig:1D_results:DIV_W_V44_v2_comp:tritium_profile}
\end{subfigure}
\caption{Temperature and tritium evolution and corresponding profiles for \acrshort{div} with 8 mm or 3 mm \acrshort{w} armor and \acrshort{v44} pipe during 50 pulses. Panels (a, c, e) and (b, d, f) are the temperature and tritium retention evolution, the temperature and tritium flux evolution at the coolant side, and the temperature and tritium profiles after the first five pulses and after the final pulse for the 8 mm or 3 mm armor, respectively. The orange curves represent temperature and the blue curves the tritium fluxes.}
\label{fig:1D_results:DIV_W_V44_compare}
\end{figure}

\begin{figure}[htb]
\centering
\begin{subfigure}{.43\linewidth}
  \centering
	\includegraphics[width = \linewidth]{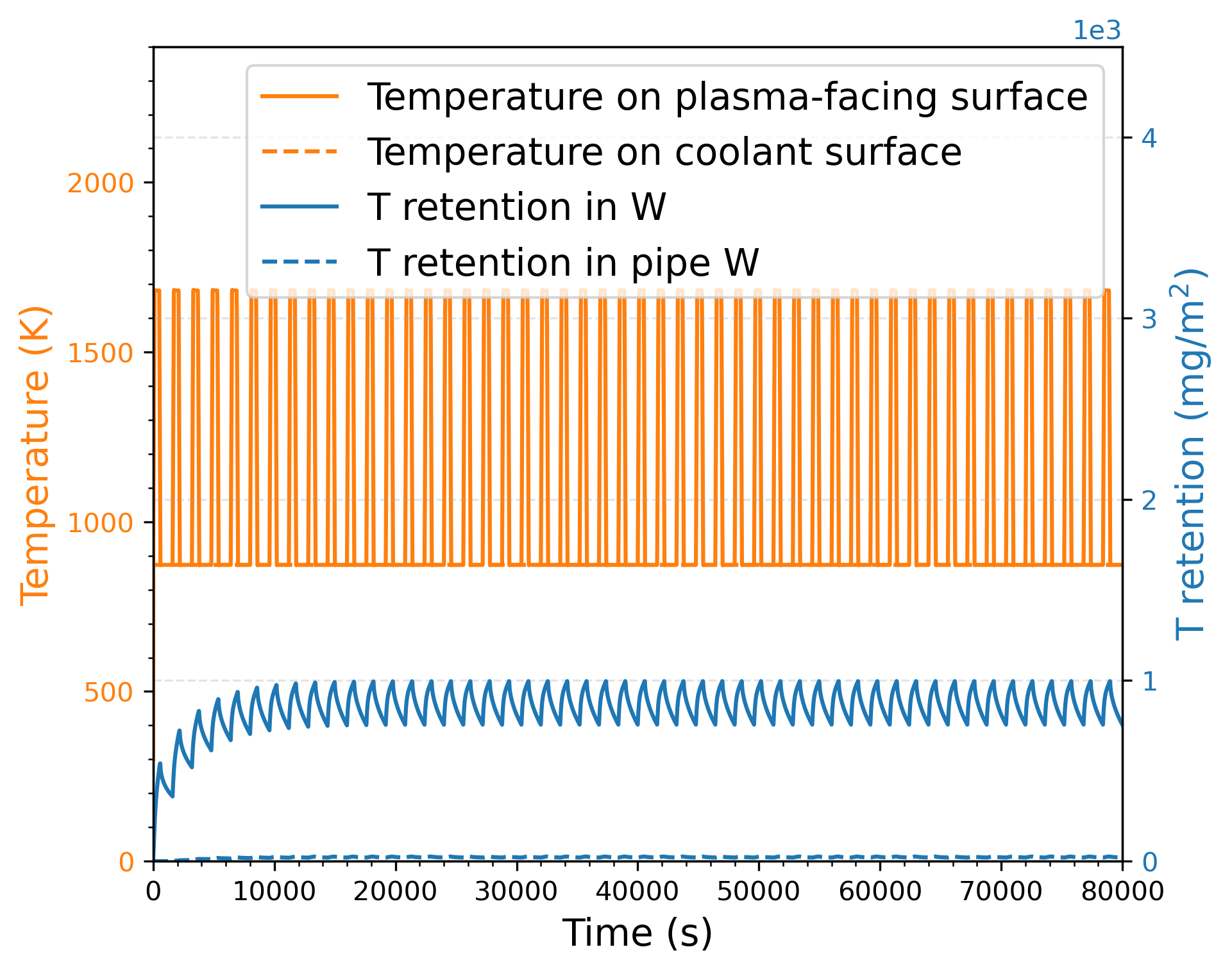}
    \caption{}
    \label{fig:1D_results:DIV_W_W_v1_comp:tritium_rentension}
\end{subfigure}
\begin{subfigure}{.43\linewidth}
  \centering
	\includegraphics[width = \linewidth]{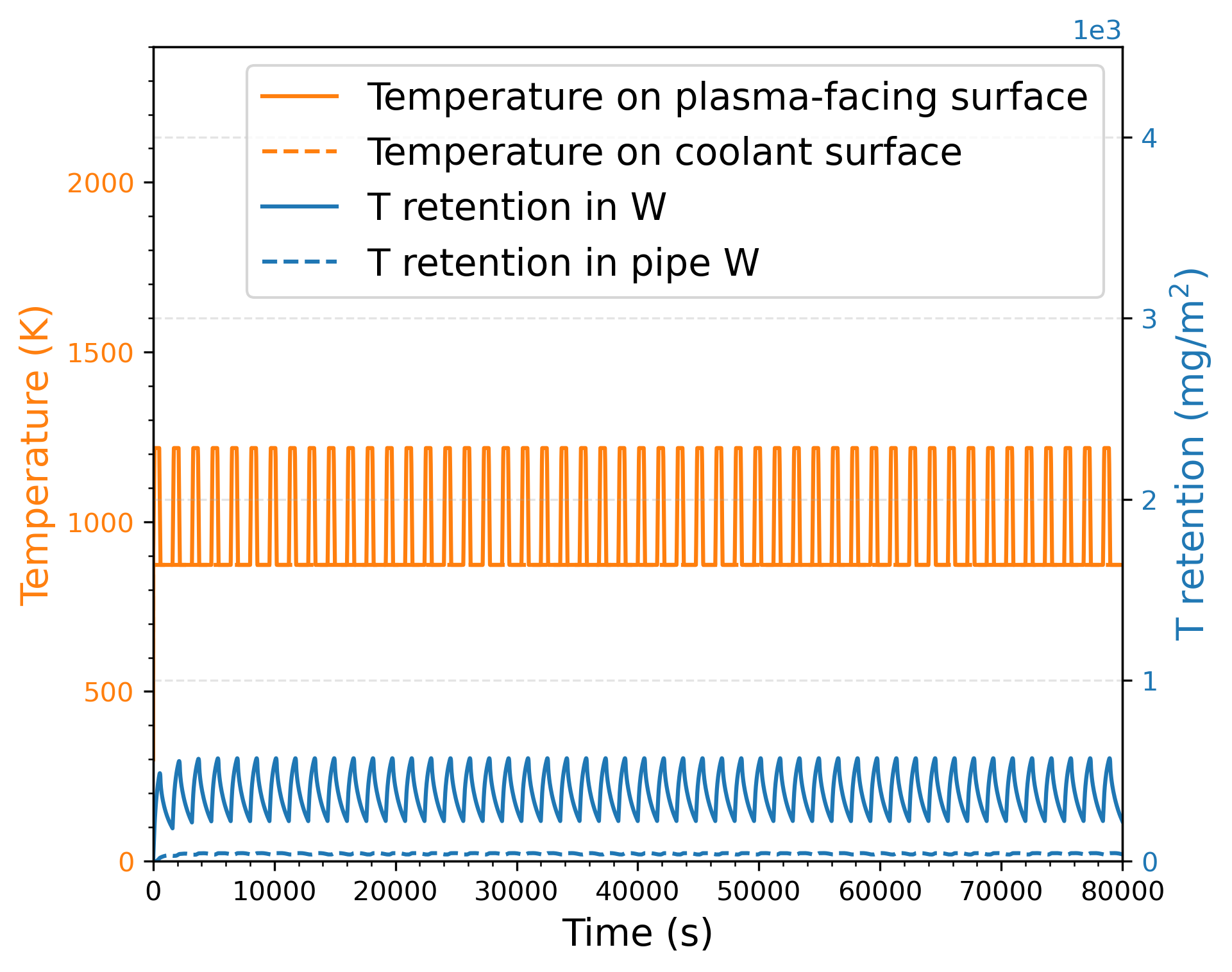}
    \caption{}
    \label{fig:1D_results:DIV_W_W_v2_comp:tritium_rentension}
\end{subfigure}
\begin{subfigure}{.43\linewidth}
  \centering
	\includegraphics[width = \linewidth]{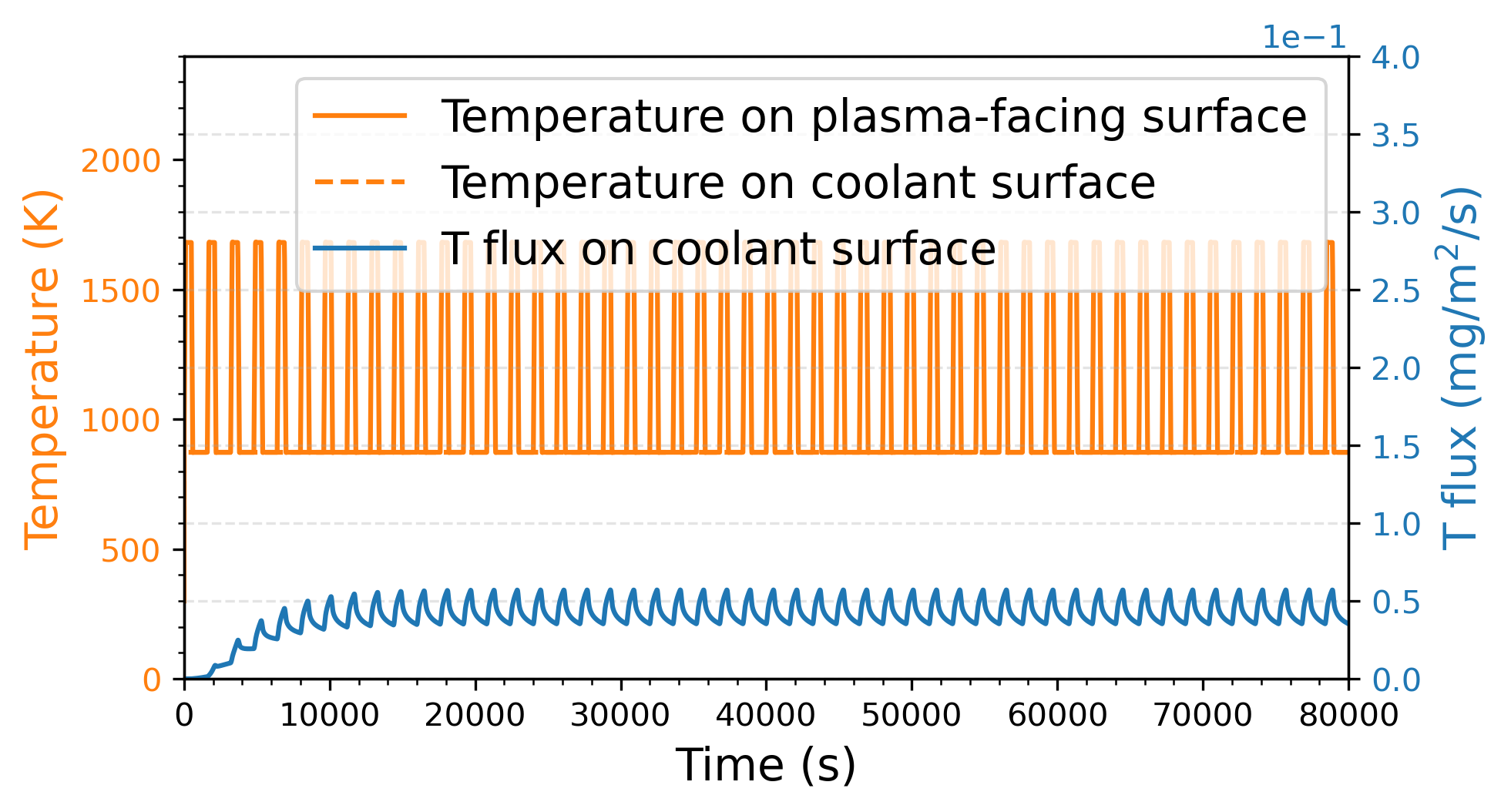}
    \caption{}
    \label{fig:1D_results:DIV_W_W_v1_comp:tritium_flux}
\end{subfigure}
\begin{subfigure}{.43\linewidth}
  \centering
	\includegraphics[width = \linewidth]{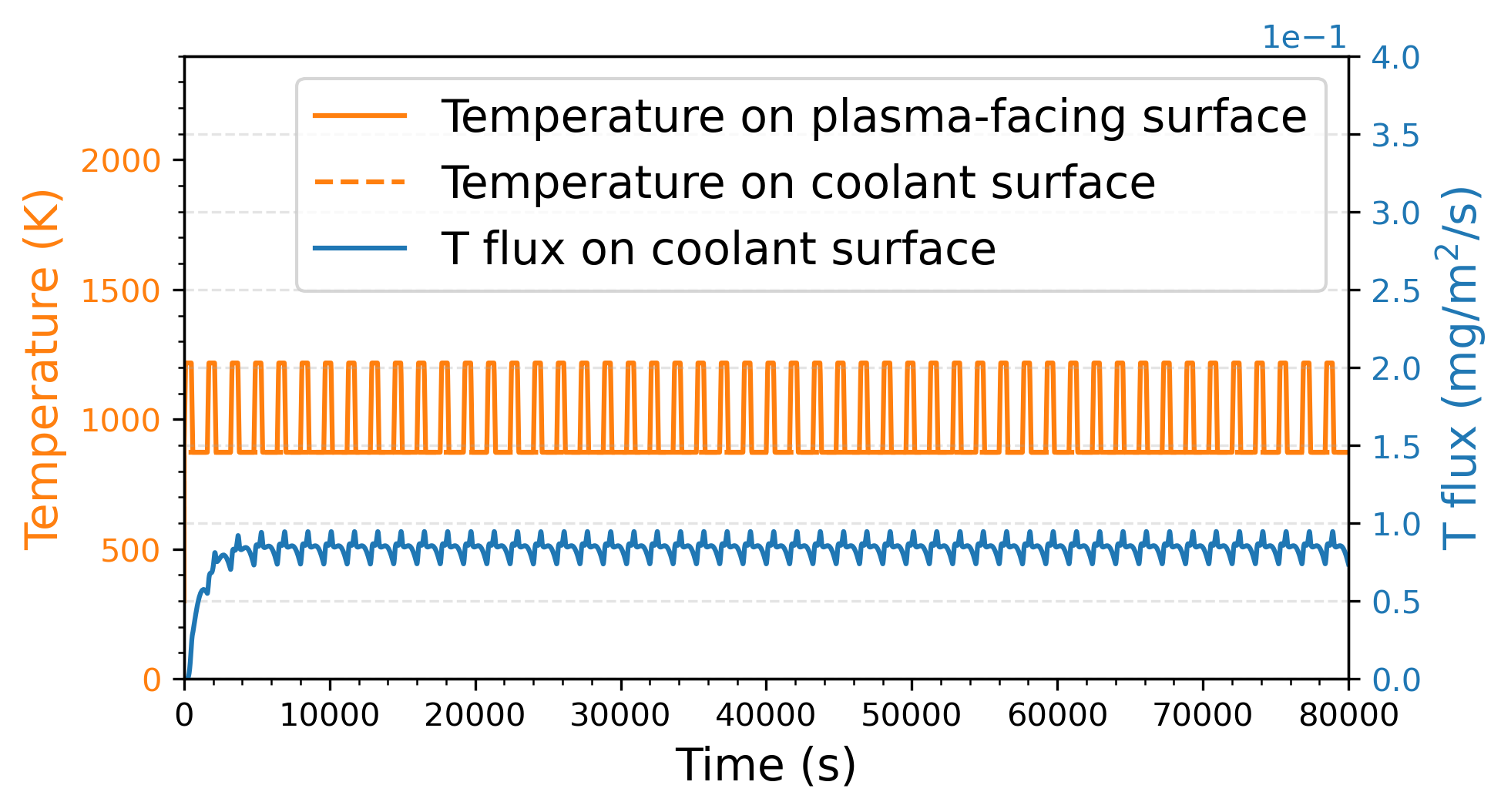}
    \caption{}
    \label{fig:1D_results:DIV_W_W_v2_comp:tritium_flux}
\end{subfigure}
\caption{The temperature and tritium evolution and corresponding profiles for the \acrshort{div} with 8 mm or 3 mm \acrshort{w} armor and \acrshort{w} pipe during 50 pulses. Panels (a, c) and (b, d) are the temperature and tritium retention evolution and the temperature and tritium flux evolution at the coolant side for the 8 mm and 3 mm armor, respectively. The orange curves represent temperature and the blue curves the tritium fluxes.}
\label{fig:1D_results:DIV_W_W_compare}
\end{figure}

We develop one-dimensional component-scale models for the \acrshort{div}, \acrshort{ccfw}, \acrshort{bkfw}, and \acrshort{vv} to investigate coupled thermal and tritium transport behavior. The simulations capture the evolution of temperature and tritium retention under cyclic plasma loading, as shown in \cref{fig:1D_results:DIV_W_V44_compare,fig:1D_results:DIV_W_W_compare} for the \acrshort{div} components. Each configuration reaches a quasi-steady-state condition before 50 pulses are completed, indicating saturation of tritium retention. Due to their similarity, the results for the remaining components (i.e, \acrshort{ccfw}, \acrshort{bkfw}, and \acrshort{vv}) are presented in \ref{sec:appendix:tritium_behavior}.

As shown in \cref{fig:1D_results:DIV_W_V44_v1_comp:tritium_rentension,fig:1D_results:DIV_W_V44_v2_comp:tritium_rentension,fig:1D_results:DIV_W_W_v1_comp:tritium_rentension,fig:1D_results:DIV_W_W_v2_comp:tritium_rentension}, the tritium inventories in all the \acrshort{div} saturate after approximately 10,000 s under the periodic input heat and tritium fluxes. The steady-state condition exhibits a small fluctuation from the pulse frequency, which reflects a dynamic equilibrium between the injection and release processes. The corresponding tritium flux on the coolant surface (\cref{fig:1D_results:DIV_W_V44_v1:tritium_flux,fig:1D_results:DIV_W_V44_v2:tritium_flux,fig:1D_results:DIV_W_W_v1:tritium_flux,fig:1D_results:DIV_W_W_v2:tritium_flux}) displays similar behavior, but with more pronounced fluctuations on the \acrshort{div} with \acrshort{v44} pipe because of the fast exhaustion of tritium in the \acrshort{v44} at the corresponding coolant temperature. \rtxt{The high tritium retention observed in \cref{fig:1D_results:DIV_W_V44_v1_comp:tritium_rentension,fig:1D_results:DIV_W_V44_v2_comp:tritium_rentension} is attributed to the conservative high peak tritium flux without re-emission considered in this study and the presence of additional trapping sites in tungsten.
Future work will refine these assumptions to provide best estimate predictions rather than overly conservative ones.}

As shown in the comparisons of the 8 mm (\cref{fig:1D_results:DIV_W_V44_v1_comp:tritium_rentension,fig:1D_results:DIV_W_V44_v1_comp:tritium_flux}) and 3 mm (\cref{fig:1D_results:DIV_W_V44_v2_comp:tritium_rentension,fig:1D_results:DIV_W_V44_v2_comp:tritium_flux}) armor thickness performance, reducing the thickness significantly decreases the overall tritium retention and increases tritium losses to the coolant due to the lower material volume (and therefore lower number of traps) and reduced diffusion length (and related shortened release time). In addition, the plasma-facing surface temperature is slightly lower in the thinner armor case because the reduced distance between the plasma\rtxt{-facing} and coolant surfaces enhances thermal conduction. A thinner armor, however, might have disadvantages not studied here related to component lifespan \rtxt{and could only be used in a fully detached \acrshort{div} regime with a low erosion.} 

The tritium retention and coolant surface fluxes for the \acrshort{div} configuration with a \acrshort{w} coolant pipe (\cref{fig:1D_results:DIV_W_W_compare}) follow trends that are comparable to those obtained for the \acrshort{v44} pipe configuration. Also, there are no apparent differences in either tritium inventory or coolant-side flux for the two pipe materials, suggesting that the geometric configuration plays a larger role than the material of the coolant pipe in this regime.

The current component-level model can easily extend to other conditions during fusion operations, such as the bake-out process after plasma operations. During operations, tritium accumulates within the plasma-facing components, \acrshort{vv}, and exposed \acrshort{v44} pipe due to continuous exposure to the deuterium--tritium plasma environment. To recover the retained tritium inventory, these components undergo a thermal baking, also called the bake-out process, to facilitate the tritium desorption. \rtxt{In particular, exposed \acrshort{v44} pipes in the breeder blanket and on the backside of plasma-facing components can retain a large amount of tritium due to prolonged exposure to high tritium concentrations. Bake-out of this component is therefore essential to release the accumulated tritium.}

During the bake-out process, the \acrshort{div} is heated from room temperature to a target temperature of 873 K, while the \acrshort{ccfw}, \acrshort{bkfw}, and exposed \acrshort{v44} pipe are heated to 673 K. The \acrshort{vv} is heated to a lowest target temperature of 473 K. In all cases, the temperature ramping phase is completed over 30 hours. After that the components are kept at corresponding target temperatures for an additional 170 hours to enable tritium diffusion and release into the helium coolant flow \rtxt{from plasma-facing components}.

\begin{figure}[h!]
\centering
\begin{subfigure}{.45\linewidth}
  \centering
	\includegraphics[width = \linewidth]{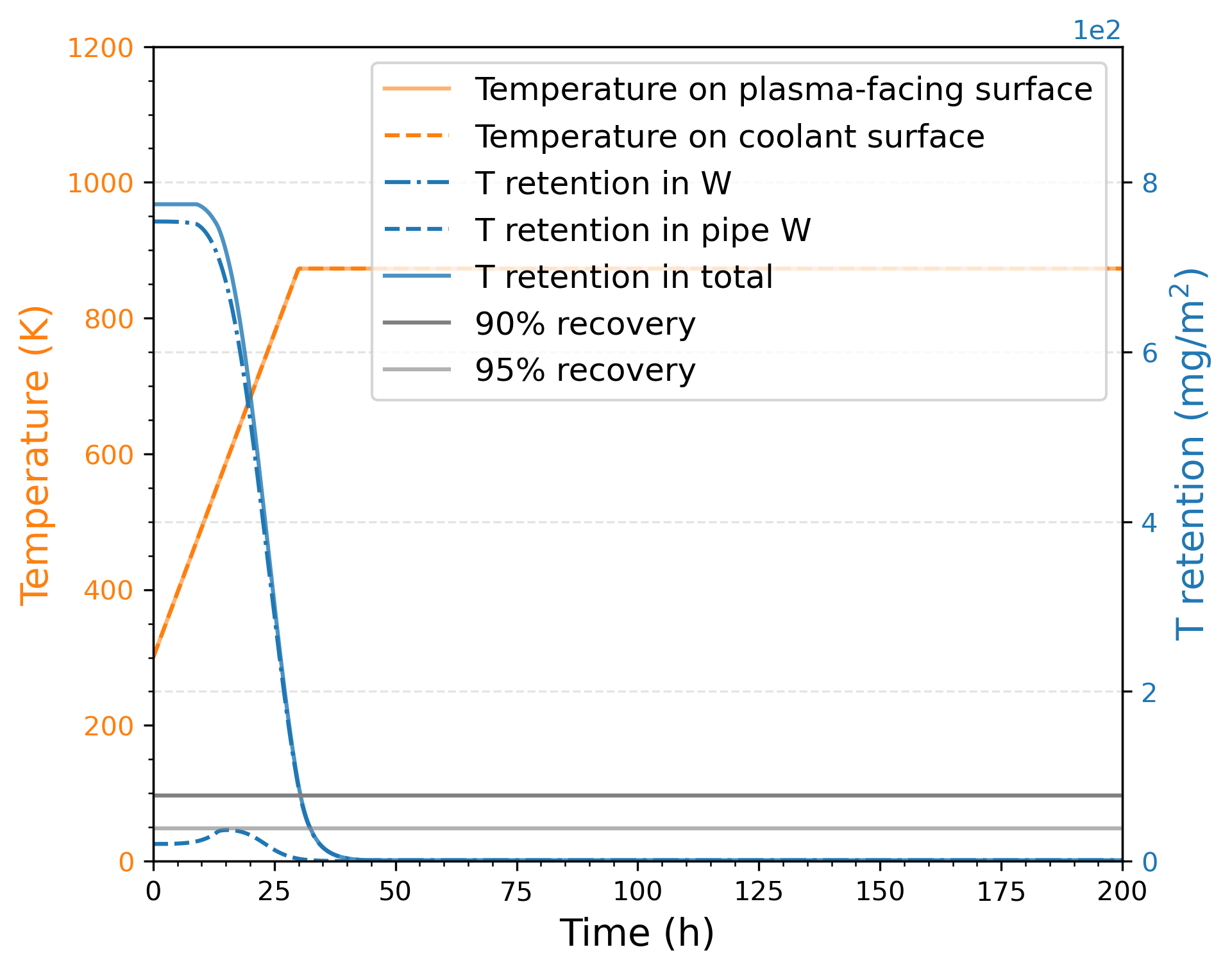}
    \label{fig:1D_results_baking:DIV_W_W_v1:tritium_rentension}
\end{subfigure}
\begin{subfigure}{.45\linewidth}
  \centering
	\includegraphics[width = \linewidth]{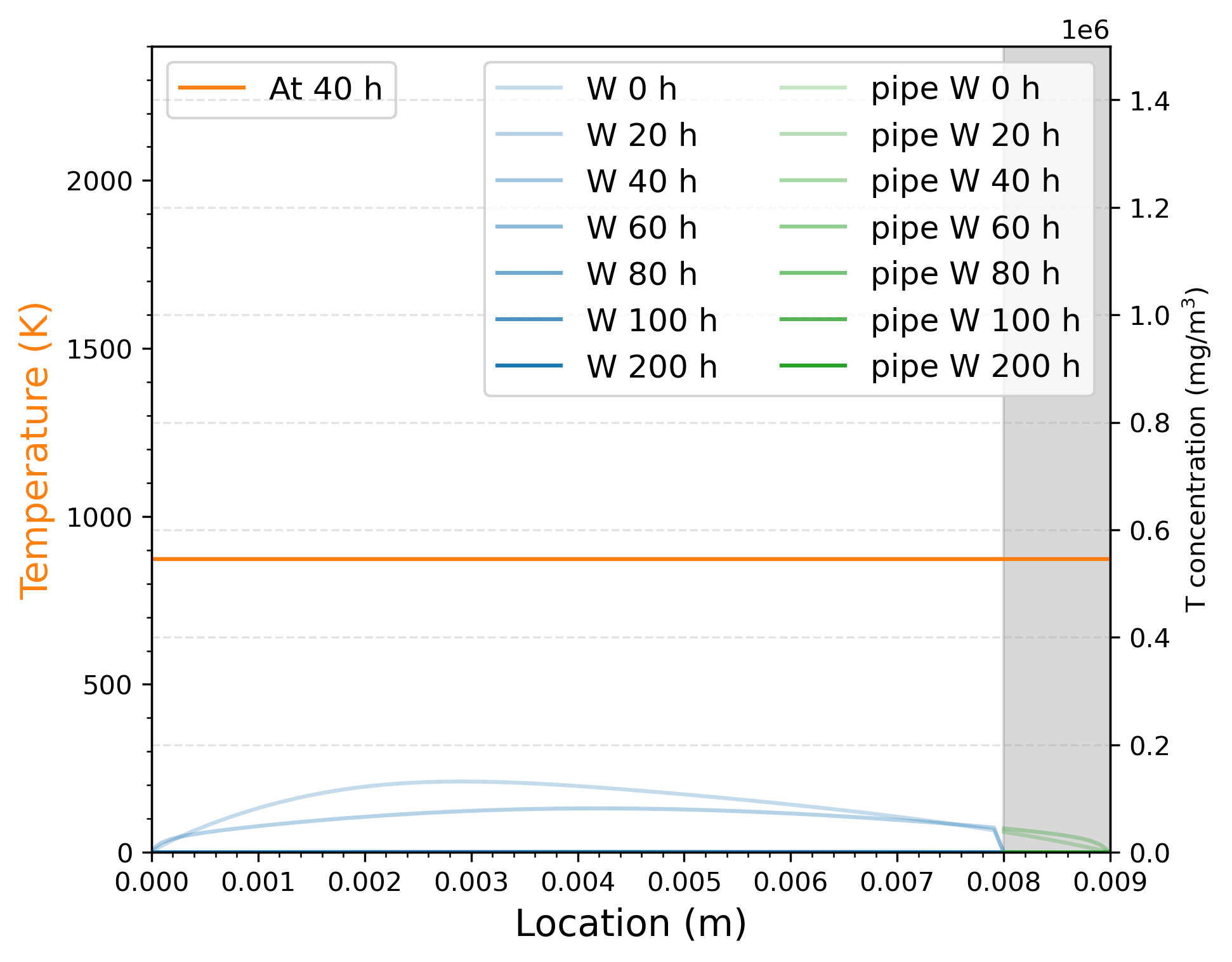}
    \label{fig:1D_results_baking:DIV_W_W_v1:tritium_profile}
\end{subfigure}
\begin{subfigure}{.45\linewidth}
  \centering
	\includegraphics[width = \linewidth]{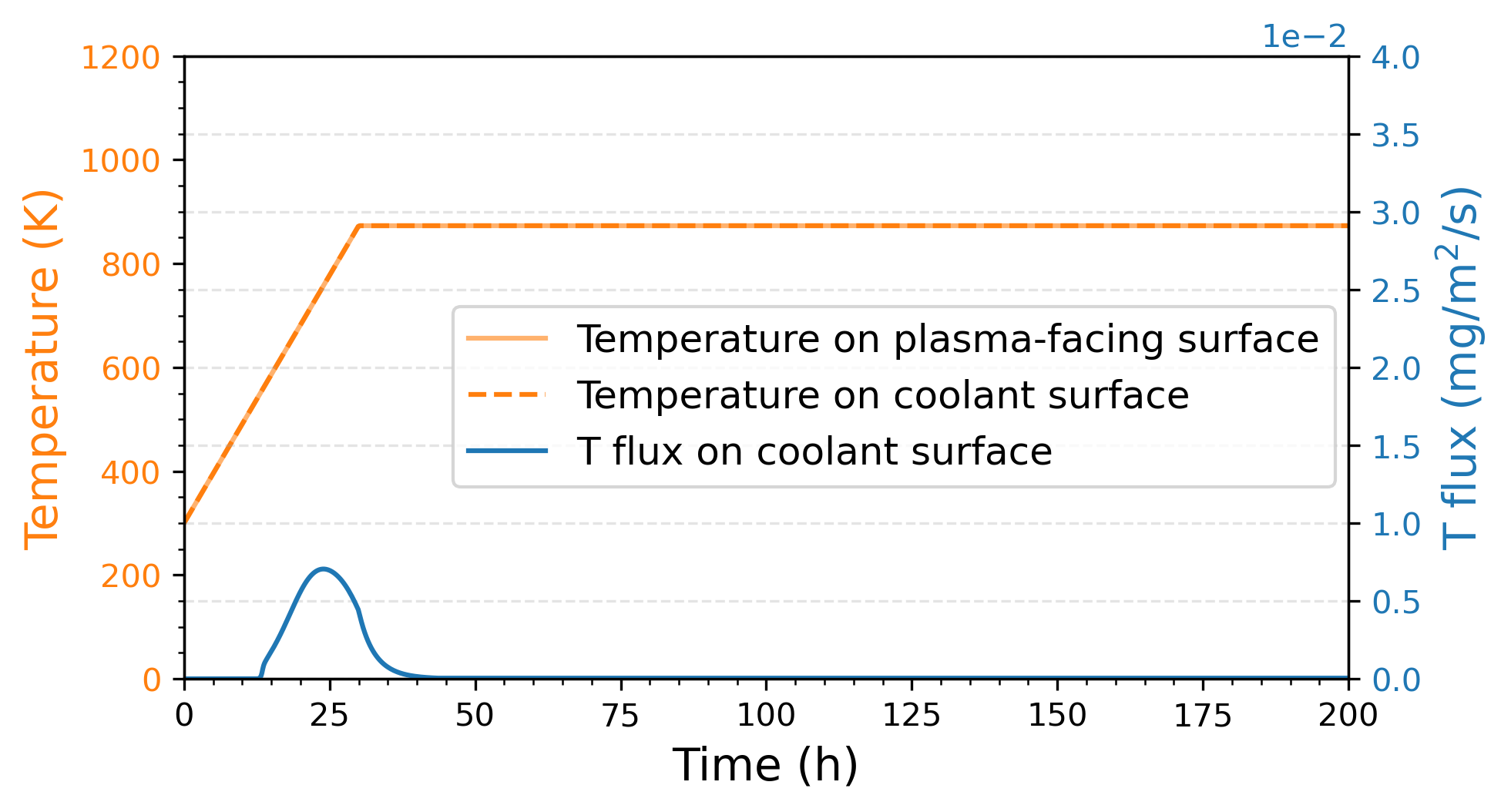}
    \label{fig:1D_results_baking:DIV_W_W_v1:tritium_flux}
\end{subfigure}
\caption{The temperature and tritium evolution and corresponding profiles for the \acrshort{div} with an 8 mm \acrshort{w} armor and a 1 mm \acrshort{w} pipe during the baking operation. Panel (a) shows the temperature and tritium retention evolution during baking. Panel (b) shows the temperature and tritium profiles at 0, 20, 40, 60, 80, 100, and 200 hours. Panel (c) shows the temperature and tritium flux evolution at the coolant surface during baking. The orange curves represent temperature and the blue curves the tritium behavior.}
\label{fig:1D_results_baking:DIV_W_W_v1}
\end{figure}

The tritium release behavior from the \acrshort{div} with an 8 mm \acrshort{w} armor and a 1 mm \acrshort{w} pipe under the corresponding baking temperature profile is shown in \cref{fig:1D_results_baking:DIV_W_W_v1} as a representative case. Similar behaviors are observed in other components. As shown in \cref{fig:1D_results_baking:DIV_W_W_v1:tritium_rentension}, more than 95 \% of the retained tritium is released after approximately 33 hours of baking. 

\begin{figure}[h!]
\centering
\includegraphics[width=0.5\textwidth]{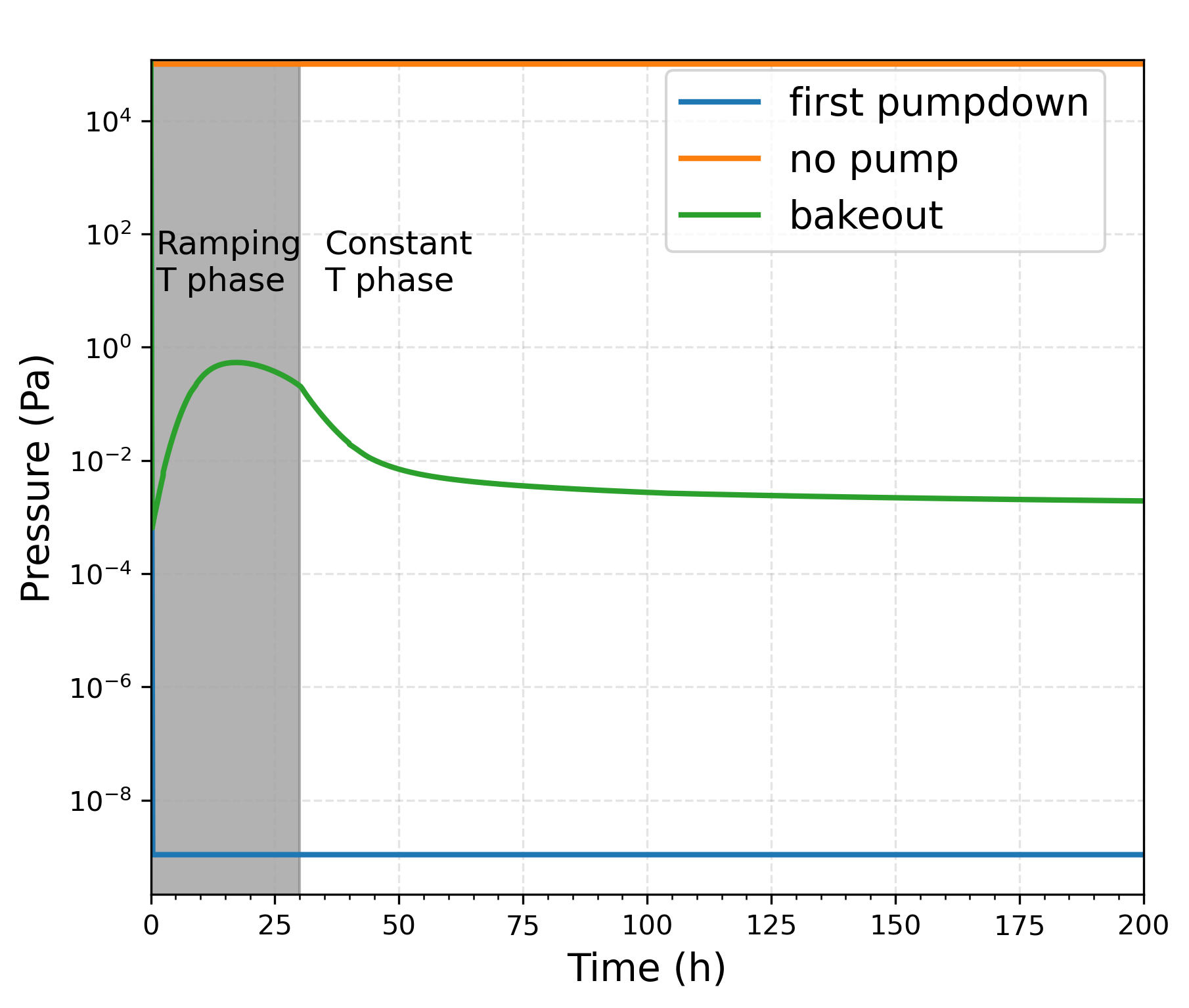}
\caption{\label{fig:1D_results_baking:pressure} The tritium pressure evolution during baking. The blue curve shows the first pump-down considering pumping only, the orange curve the results considering tritium release from components without pumping, and the green curve the pressure prediction when considering both tritium release and pumping.}
\end{figure}

Due to the initially high tritium concentration in the \acrshort{w} armor, a large amount of tritium is released from the plasma-facing surface during the early stages of heating. To evaluate the effect of this release on the vacuum conditions, we developed a pressure model in the vacuum chamber based on the performance of all the components. The model considers the tritium release from all components and a continuous pumping during the bake-out process. The corresponding pressure in the vacuum chamber is shown by the green curve in \cref{fig:1D_results_baking:pressure}, which reaches a minimal pressure of 1.93$\times10^{-3}$ Pa after 33 hours of baking.

\subsection{Tritium Transport from Surrogate Component-Level Model}
\label{sec:results:surrogate}

We construct and train Gaussian process surrogate models using the data from the component-level models with the \acrshort{v44} pipe presented in \cref{sec:results:components}. We use seven input parameters for training, including (1) tritium flux in the plasma-facing surface, (2) heat flux in the plasma-facing surface, (3) \acrshort{w} armor thickness, (4) pipe thickness, (5) coolant temperature, (6) trapping site fraction in the \acrshort{w} armor, and (7) release energy in the \acrshort{w} armor. The output parameters we evaluated are (1) steady-state tritium flux at the coolant interface, $J_\infty$, and (2) the residence times $\tau_0$ and $\tau_1$.

\begin{figure}[h!]
\centering
\begin{subfigure}{.43\linewidth}
  \centering
	\includegraphics[width = \linewidth]{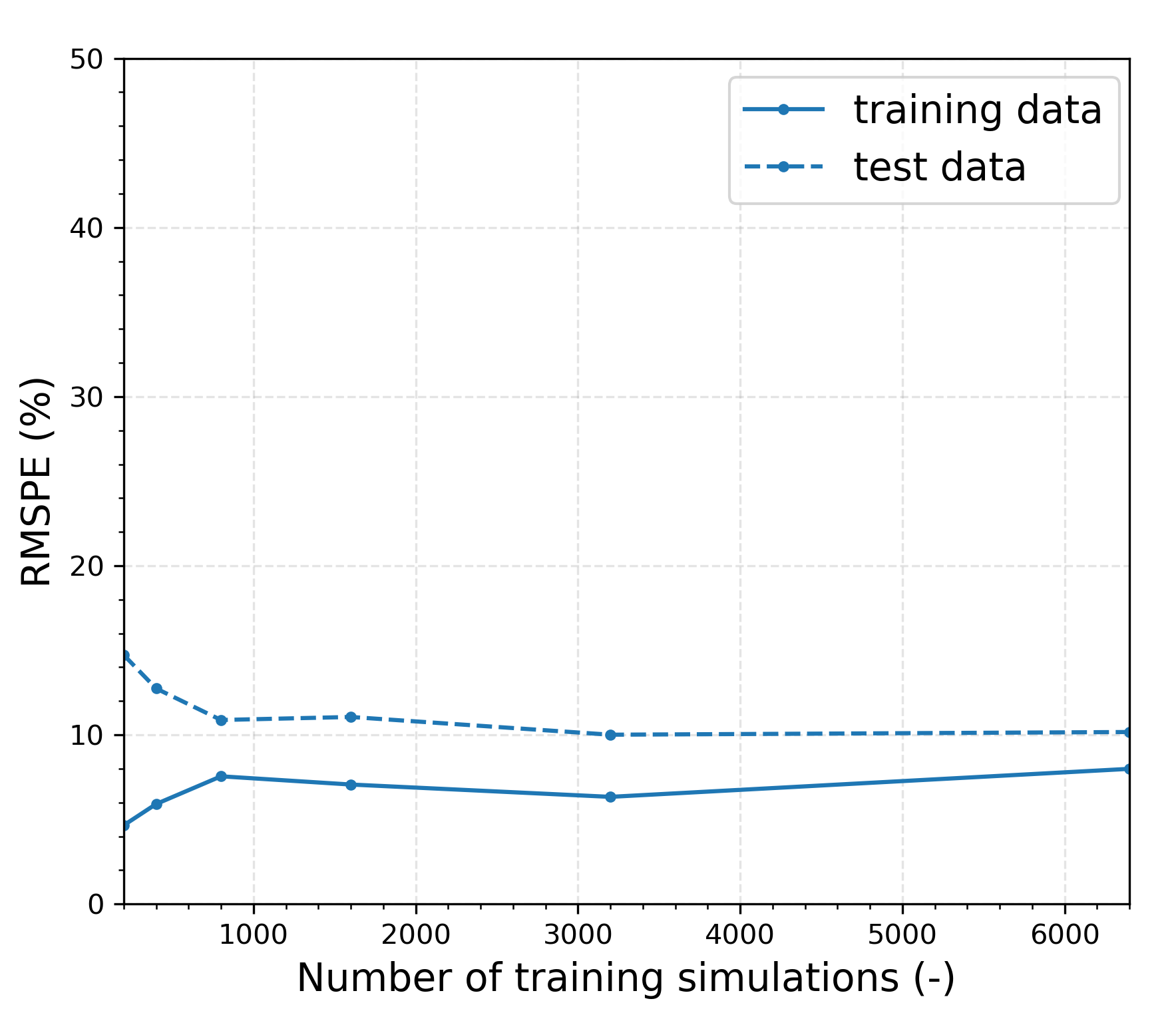}
    \caption{}
    \label{fig:surrogate:fitting:flux}
\end{subfigure}
\begin{subfigure}{.43\linewidth}
  \centering
	\includegraphics[width = \linewidth]{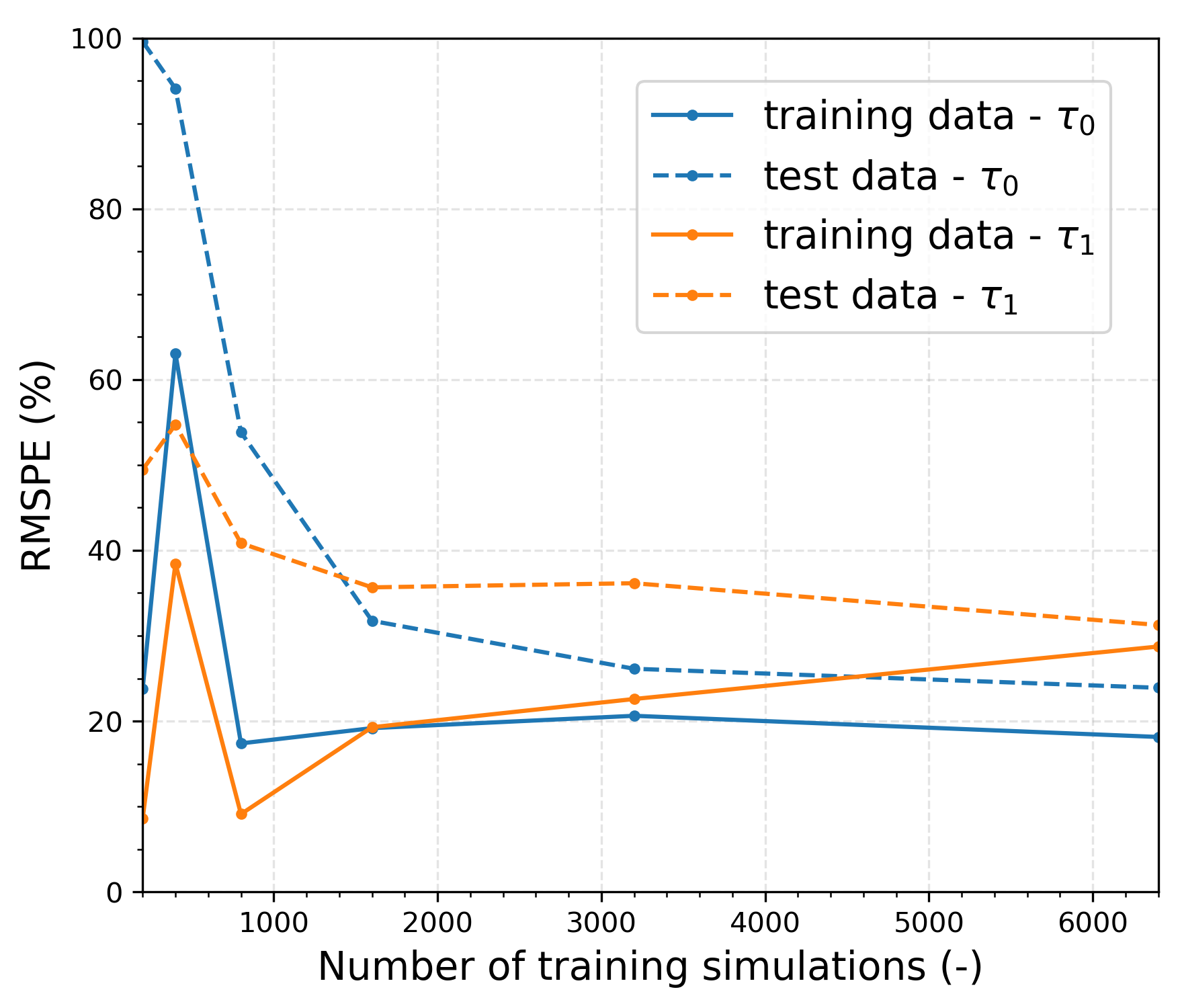}
    \caption{}
    \label{fig:surrogate:fitting:tau}
\end{subfigure}
\begin{subfigure}{.43\linewidth}
  \centering
	\includegraphics[width = \linewidth]{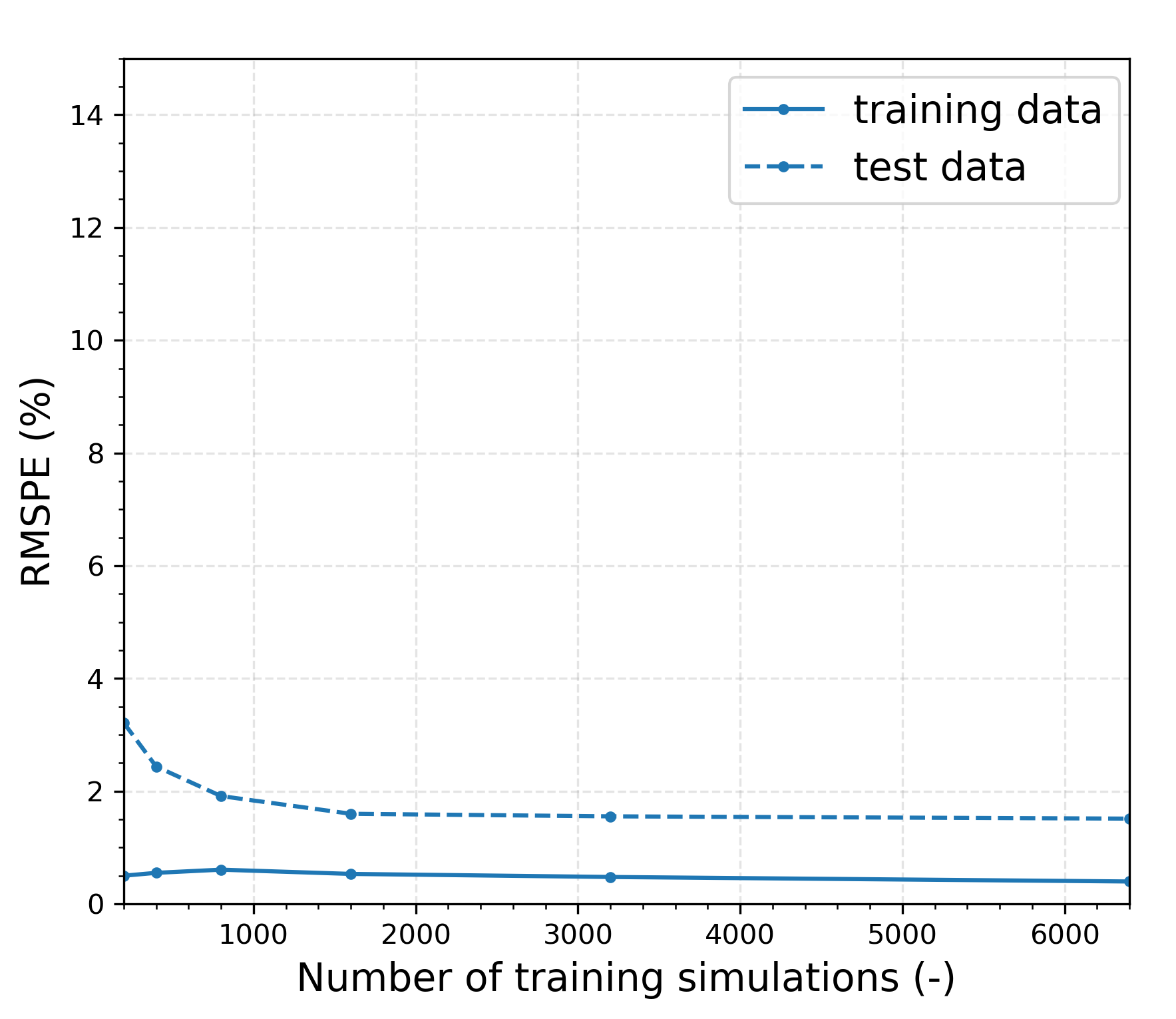}
    \caption{}
    \label{fig:surrogate_W_W:fitting:flux}
\end{subfigure}
\begin{subfigure}{.43\linewidth}
  \centering
	\includegraphics[width = \linewidth]{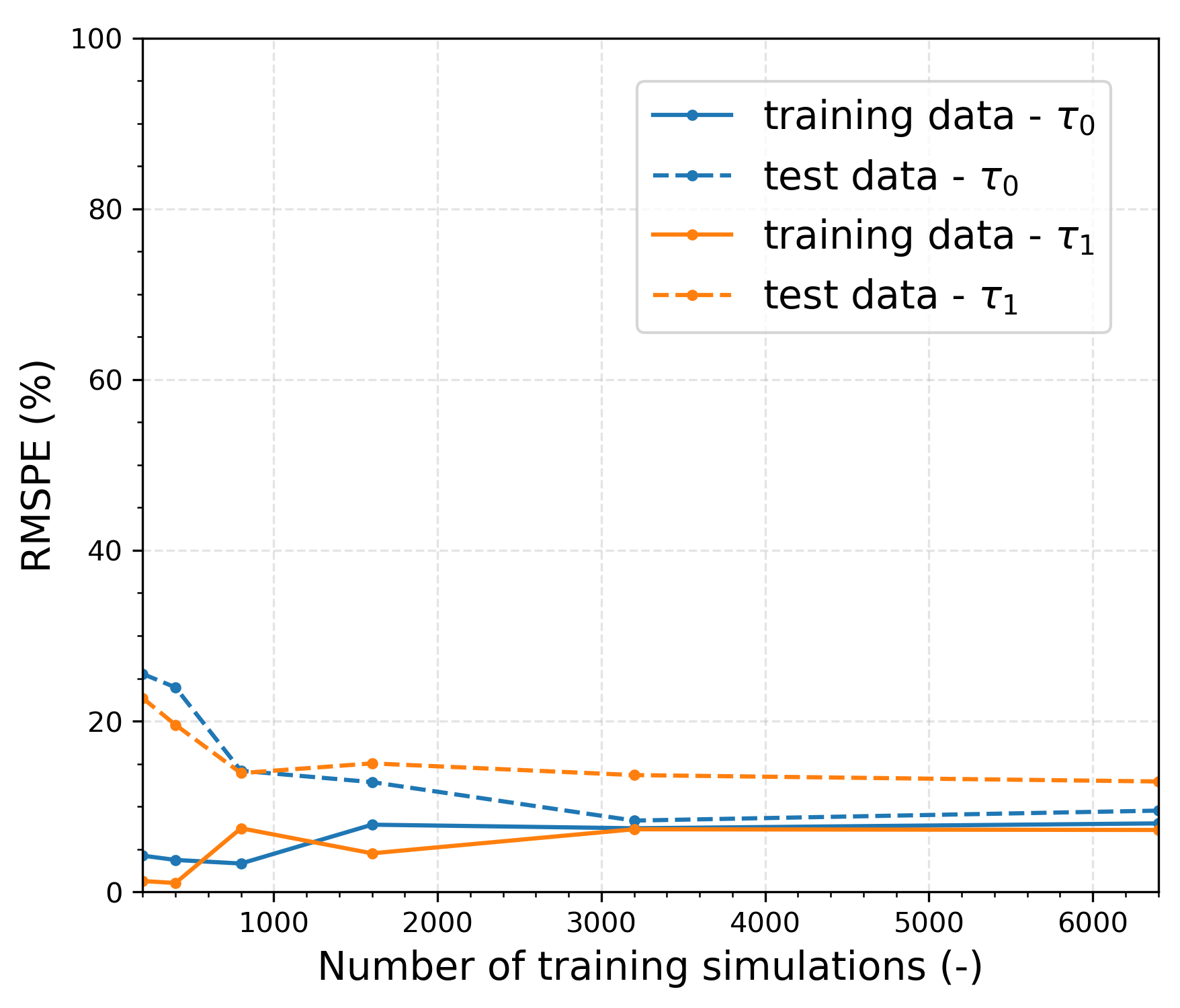}
    \caption{}
    \label{fig:surrogate_W_W:fitting:tau}
\end{subfigure}
\caption{The RMSPEs for the surrogate models on the training and test datasets from components with \acrshort{v44} and \acrshort{w} pipe to predict (a, c) steady-state tritium flux at the coolant surface for the \acrshort{v44} and \acrshort{w} pipes and (b, d) residence time for the \acrshort{v44} and \acrshort{w} pipes.}
\label{fig:surrogate:fitting}
\end{figure}

To evaluate the performance of the Gaussian process surrogate models, we compute the \acrshort{rmspe}s on both the training and test datasets for models trained with varying dataset sizes. As shown in \cref{fig:surrogate:fitting:flux,fig:surrogate:fitting:tau}, the surrogate model has the lowest RMSPE of 9.92\% for $J_\infty$ when trained with 3,200 simulations, which indicates a strong predictive capability for coolant-side tritium flux. For the two-parameter residence time, the minimum RMSPEs are 23.90\% and 31.26\% for $\tau_0$ and $\tau_1$, respectively, using 6,400 training samples, which indicates an increased complexity and sensitivity of residence time fitting. The hyperparameters used to characterize the two Gaussian process models are provided in \ref{sec:appendix:hyperparameters}.

\begin{figure}[h!]
\centering
\begin{subfigure}{.41\linewidth}
  \centering
	\includegraphics[width = \linewidth]{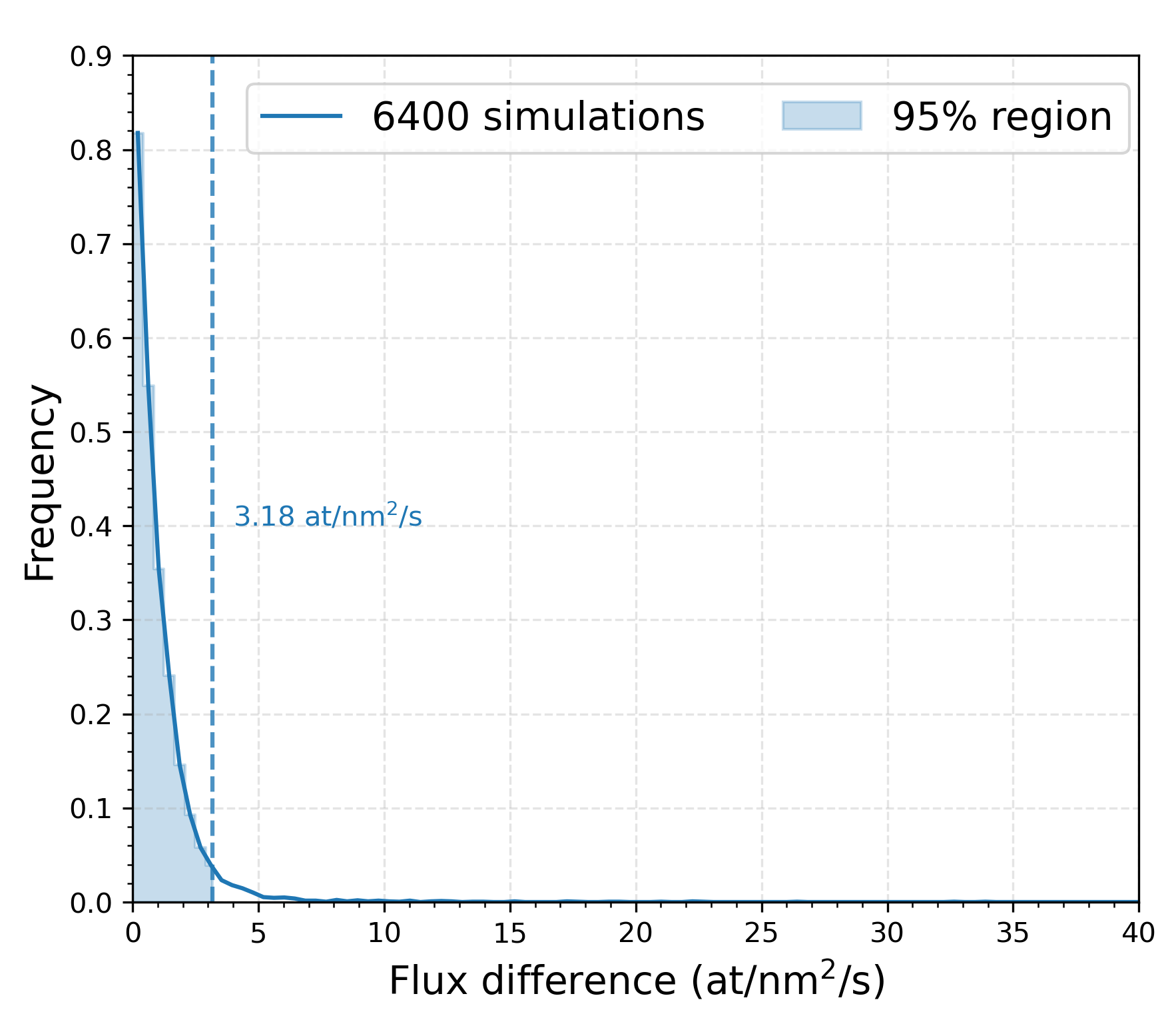}
    \caption{}
    \label{fig:surrogate:error_difference:flux}
\end{subfigure}
\begin{subfigure}{.43\linewidth}
  \centering
	\includegraphics[width = \linewidth]{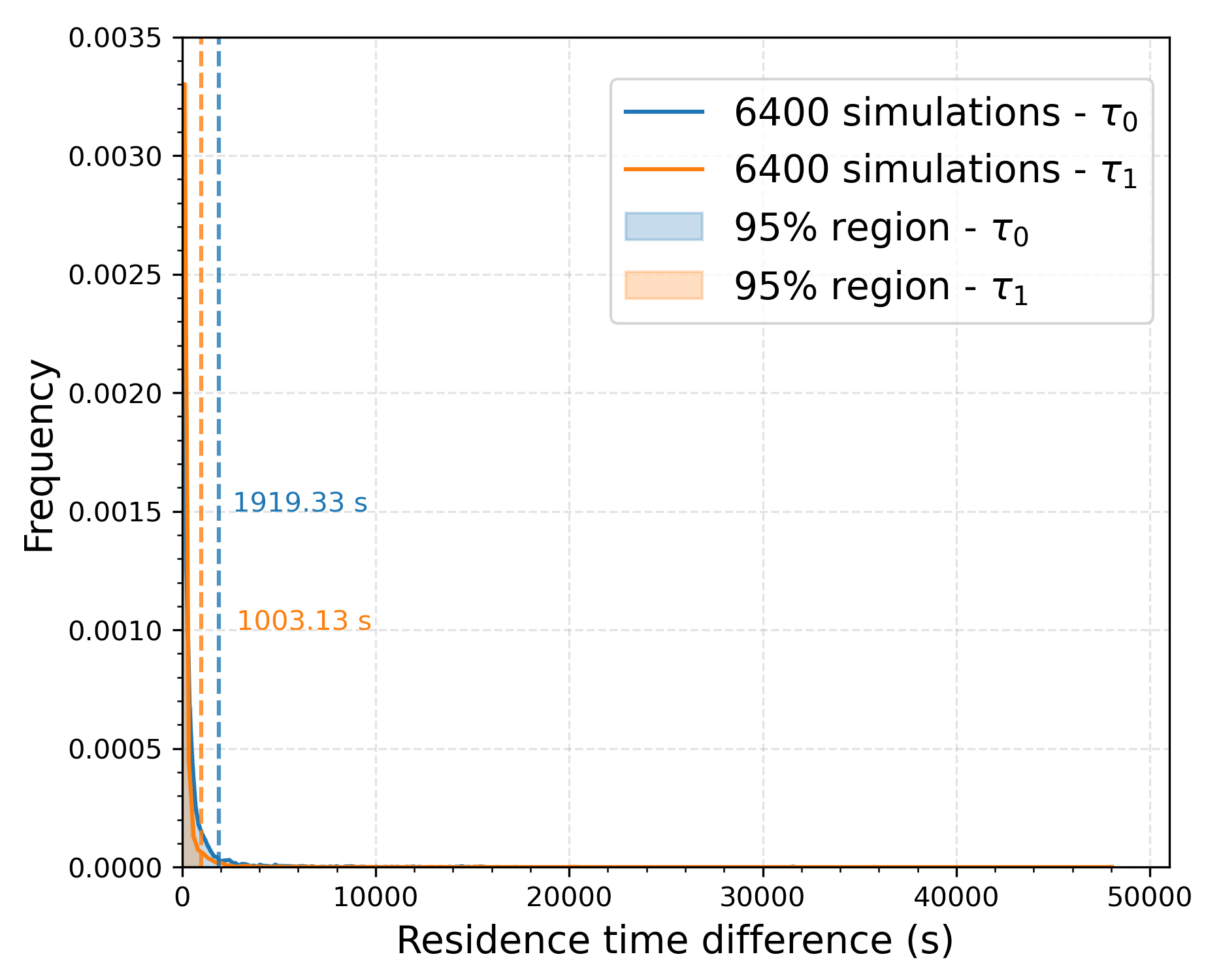}
    \caption{}
    \label{fig:surrogate:error_difference:tau}
\end{subfigure}
\begin{subfigure}{.40\linewidth}
  \centering
	\includegraphics[width = \linewidth]{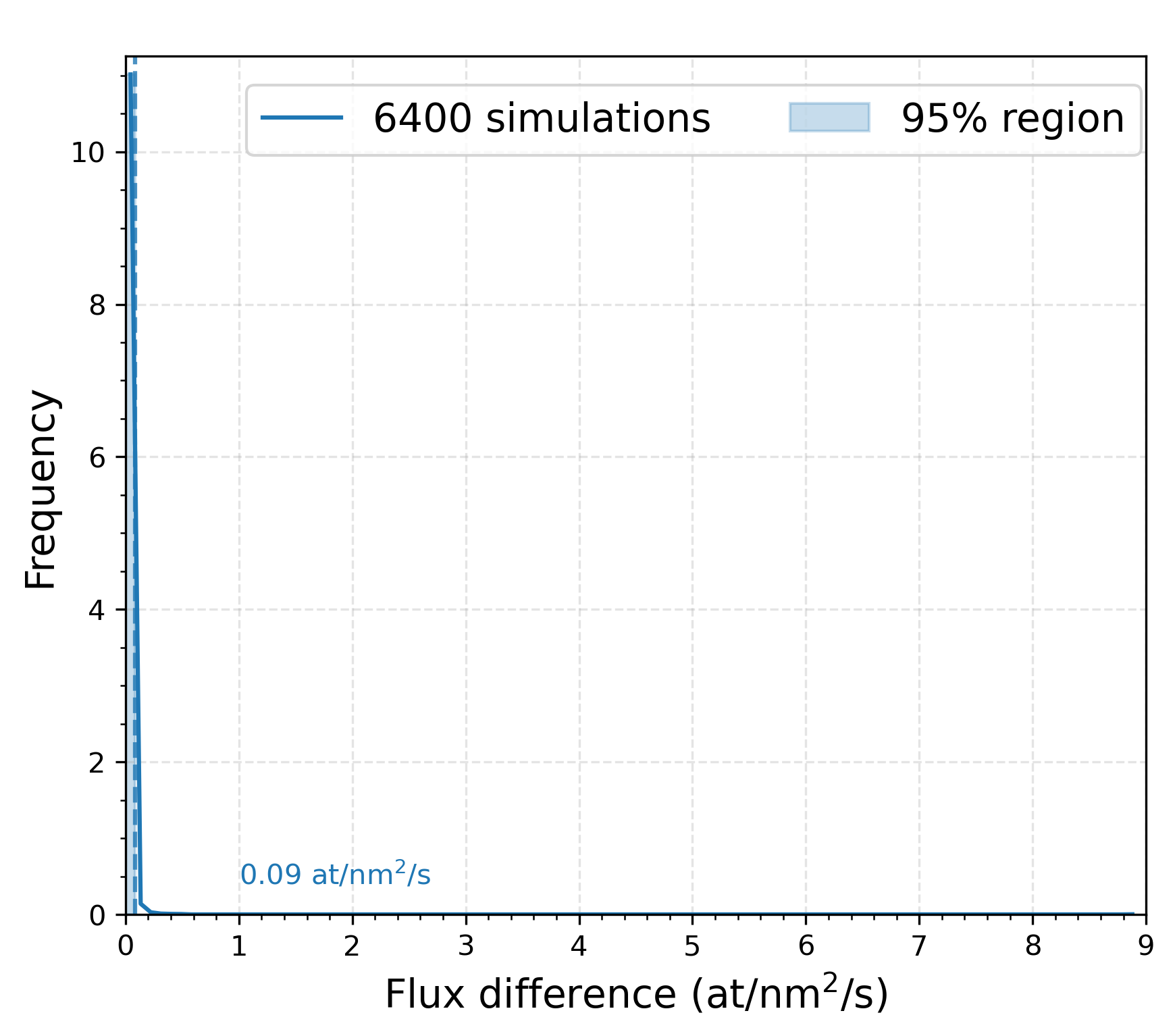}
    \caption{}
    \label{fig:surrogate_W_W:error_difference:flux}
\end{subfigure}
\begin{subfigure}{.43\linewidth}
  \centering
	\includegraphics[width = \linewidth]{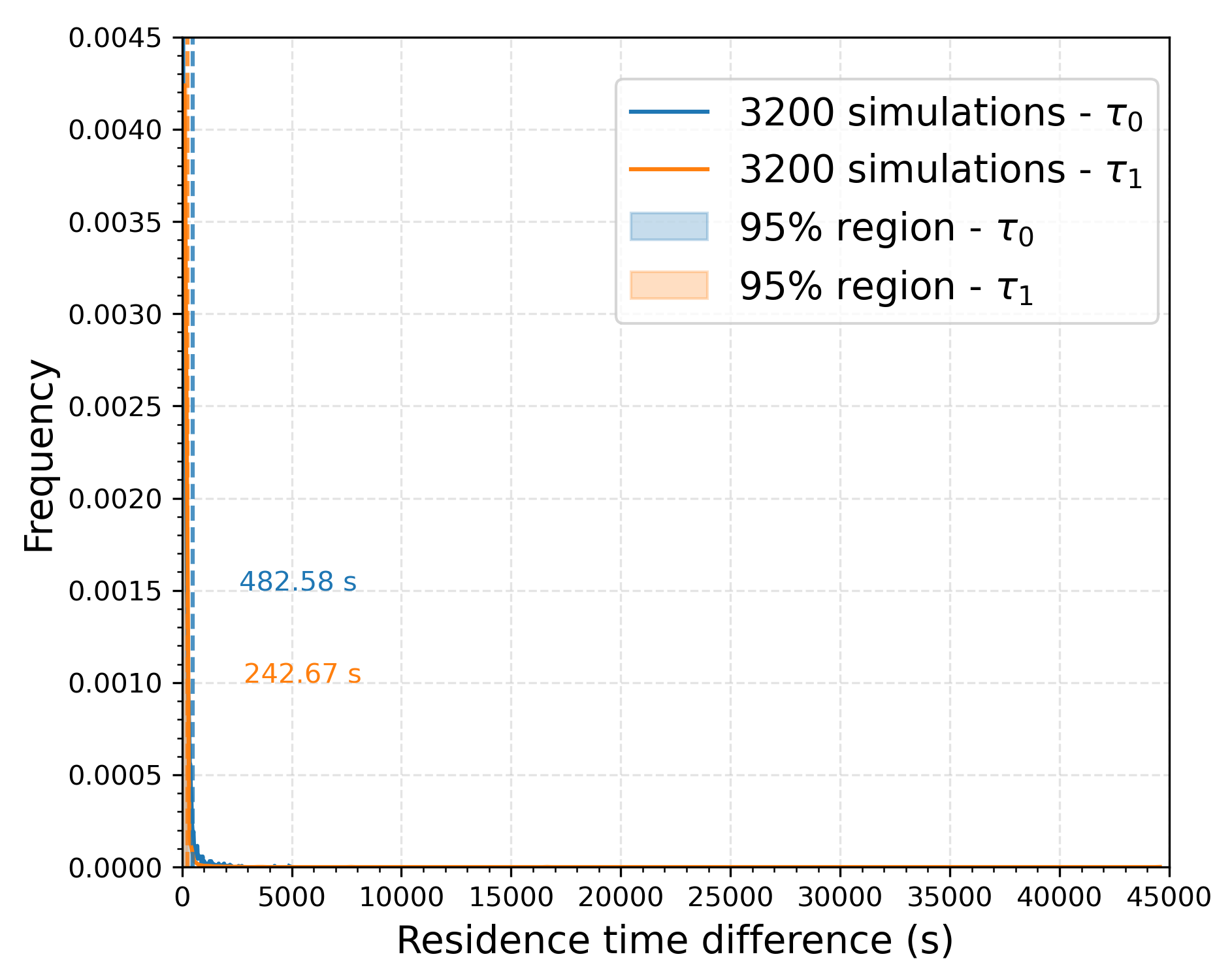}
    \caption{}
    \label{fig:surrogate_W_W:error_difference:tau}
\end{subfigure}
\caption{The distribution of error between results from surrogate model and simulation for components with \acrshort{v44} and \acrshort{w} pipe on predicting (a, c) steady-state tritium flux at coolant surface for the \acrshort{v44} and \acrshort{w} pipes and (b, d) residence time for the \acrshort{v44} and \acrshort{w} pipes.}
\label{fig:surrogate:error_difference}
\end{figure}

The simulation results of the surrogate and component-level models are compared in \cref{fig:surrogate:error_difference:flux,fig:surrogate:error_difference:tau}. The 95\% bounds of the absolute differences indicate a good agreement. The deviation in $J_\infty$ is below 3.18 at/nm$^2$/s, while the deviations in $\tau_0$ and $\tau_1$ are less than 1,919.33 s and 1,003.13 s, respectively. These results confirm that the surrogate model accurately reproduces the steady-state tritium transport behavior while providing significant computational savings compared to the component-level model. The hyperparameters used to characterize the two Gaussian process models are also provided in \ref{sec:appendix:hyperparameters}.

We also train the Gaussian process surrogate models using the data from the component-level simulations on plasma-facing components with \acrshort{w} pipe. As shown in \cref{fig:surrogate_W_W:fitting:flux,fig:surrogate_W_W:fitting:tau}, the surrogate model has the lowest RMSPE of 1.60\% for $J_\infty$ when trained with 1,600 simulations. For the two-parameter residence time, the minimum RMSPEs are 8.35\% and 13.68\% for $\tau_0$ and $\tau_1$, respectively, using 3,200 training samples. The surrogate and component-level simulation results are shown in \cref{fig:surrogate_W_W:error_difference:flux,fig:surrogate_W_W:error_difference:tau}. The 95\% bounds of the absolute differences indicate a good agreement. The deviation in $J_\infty$ is below 0.09 at/nm$^2$/s, while the deviations in $\tau_0$ and $\tau_1$ are less than 482.58 s and 242.67 s, respectively. 
The lower RMSPEs compared to those of the surrogate model on components with \acrshort{v44} pipe indicate the significant impact of the \acrshort{w}-\acrshort{v44} interface on the surrogate model's accuracy. The details of the hyperparameters for the Gaussian process models for components with \acrshort{w} pipe are provided in \ref{sec:appendix:hyperparameters}.

\subsection{Tritium Fuel Cycle Model}
\label{sec:results:engineering}

The results obtained from the surrogate model are integrated into a system-level tritium fuel cycle model to assess their impact on overall tritium inventory. The fuel cycle model is adapted from the model implemented in TMAP8 \cite{meschini2023modeling,simon2025moose} to match the planned design and operation conditions from \rtxt{ST-E1}. The fuel cycle parameters are provided in \cref{tab:fuel_cycle_parameters}. In this work, the model incorporates the surrogate models developed based on the \acrshort{bkfw}, \acrshort{ccfw}, and \acrshort{div} simulations to ensure consistency with the plasma-facing components analyzed in this study. \rtxt{In addition, this model incorporate the residence time from the surrogate models from the three first wall components of interest in this study. The high tritium retention values reported in \cref{sec:results:components} are not used directly, since they result from conservative assumptions, including high peak tritium fluxes without accounting for the re-emission processes and the presence of additional trapping sites in tungsten.}

\begin{figure}[h!]
\centering
\begin{subfigure}{.43\linewidth}
  \centering
	\includegraphics[width = \linewidth]{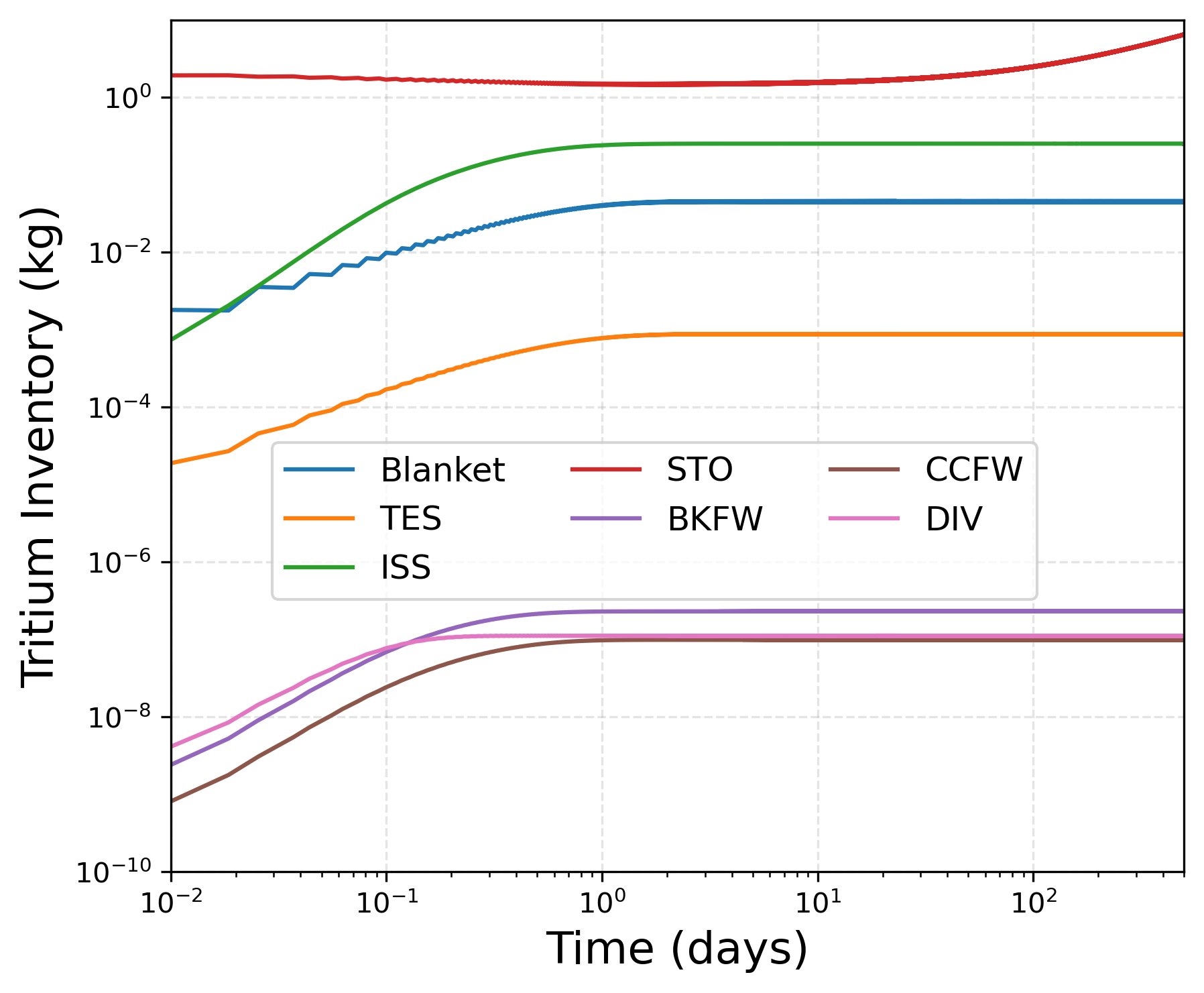}
    \caption{}
    \label{fig:fuel_cycle:compare:original}
\end{subfigure}
\begin{subfigure}{.43\linewidth}
  \centering
	\includegraphics[width = \linewidth]{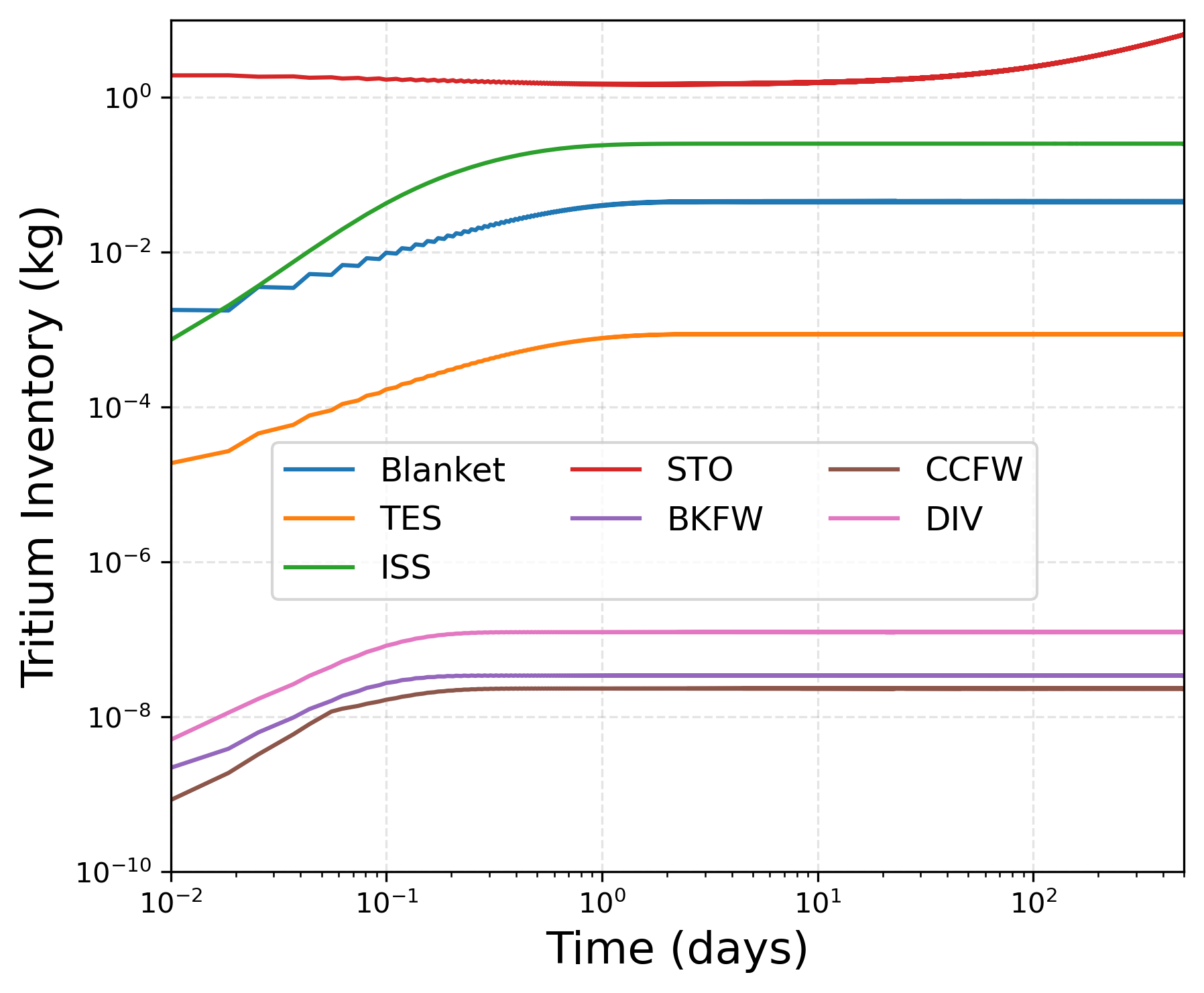}
    \caption{}
    \label{fig:fuel_cycle:compare:two_tau}
\end{subfigure}
\caption{The evolution of tritium inventory in critical fusion components and plasma-facing components during the fuel cycle. Panel (a) shows the fuel cycle results from the one-parameter residence time surrogate model in plasma-facing components. Panel (b) shows the fuel cycle results from the two-parameter residence time surrogate model in plasma-facing components.}
\label{fig:fuel_cycle:compare}
\end{figure}

The residence time parameters for the \acrshort{bkfw}, \acrshort{ccfw}, and \acrshort{div} are updated using the values predicted by the Gaussian process surrogate model trained with 6,400 simulations (described in \cref{sec:results:surrogate}). The residence times for all other components are defined based on  parameters from Tokamak Energy. Two simulation cases are performed: one using results from the one-parameter residence time model shown in \cref{eqn:surrogate:one_parameter_resident_time_fitting} and one using the two-parameter residence time model shown in \cref{eqn:surrogate:resident_time_fitting}. The tritium inventory evolution within key components is shown in \cref{fig:fuel_cycle:compare}. In both cases, as expected, the tritium inventories in the plasma-facing components remain significantly lower than those in the primary tritium-managing components, such as the breeding blanket and tritium extraction systems. In addition, the new two-residence model, developed thanks to the multiscale modeling study, reduce the predicted inventory in plasma-facing components,  therefore alleviating the constraints that would come from an expected higher inventory.
\rtxt{These inventory values differ from those reported in Ref.~\cite{TE_lead_paper} for three reasons: (1) the present fuel cycle model explicitly accounts for both pulsed operation and intervening dwell periods; (2) residence times for the three first-wall components are treated independently and derived from the surrogate models developed in this work; and (3) residence times for other subsystems differ slightly from those adopted in the Tokamak Energy reference, as discussed in \cref{sec:components_information:fuel_cycle}.}
In addition, as shown in \cref{fig:fuel_cycle:compare:two_tau}, the tritium inventory for the \acrshort{ccfw} component, which has the largest delay time $\tau_0$, increases rapidly before residence time $\tau_0$ (i.e, at the elbow around 0.053 day) because no tritium is released from this component during the delay period. 
This behavior reflects a more realistic transient behavior in tritium inventories and release during the fuel cycle.

\begin{figure}[h!]
\centering
\begin{subfigure}{.43\linewidth}
  \centering
	\includegraphics[width = \linewidth]{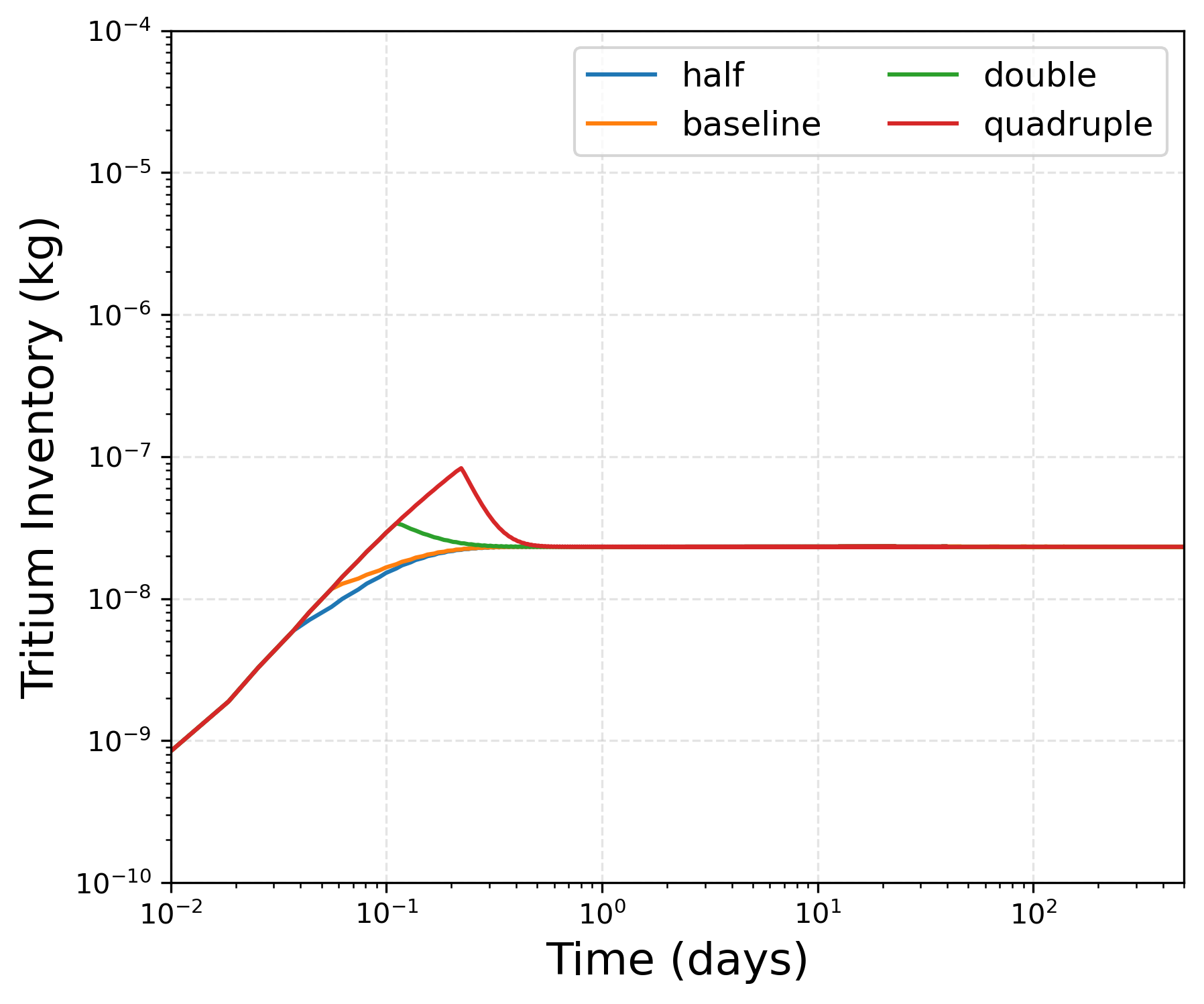}
    \caption{}
    \label{fig:fuel_cycle:change_CCFW_tau:tau0}
\end{subfigure}
\begin{subfigure}{.43\linewidth}
  \centering
	\includegraphics[width = \linewidth]{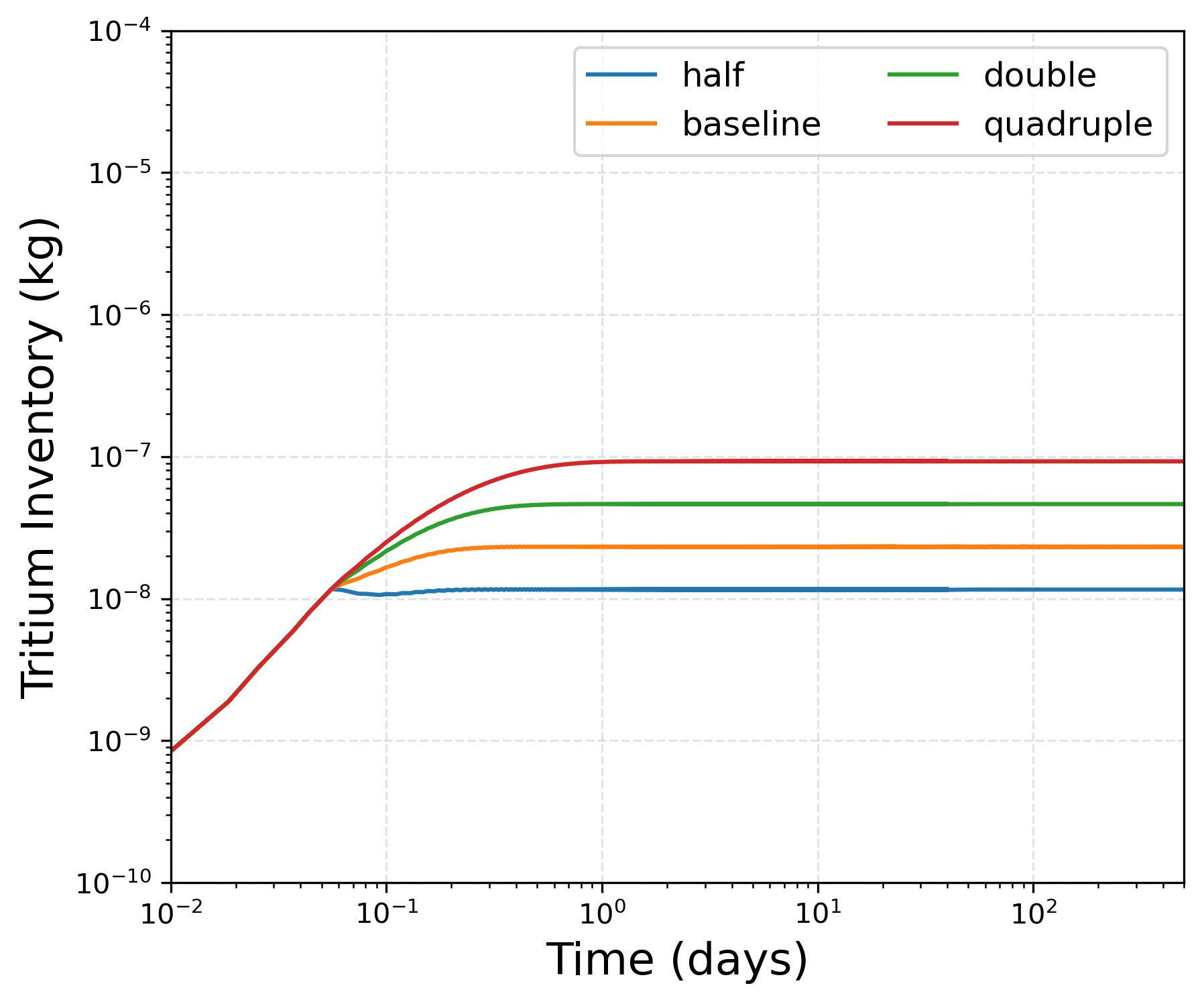}
    \caption{}
    \label{fig:fuel_cycle:change_DIV_tau1:tau1}
\end{subfigure}
\caption{The evolution of tritium inventory in the \acrshort{ccfw} with half of the baseline, baseline, double, and quadruple (a) residence time $\tau_0$ and (b) residence time $\tau_1$}
\label{fig:fuel_cycle:change_CCFW_tau}
\end{figure}

We then vary the residence-time $\tau_0$ and $\tau_1$ of the \acrshort{ccfw} component in the fuel cycle model to investigate its impact on the tritium inventory within the corresponding component. As shown in \cref{fig:fuel_cycle:change_CCFW_tau}, the tritium inventory evolution is compared for cases in which $\tau_0$ and $\tau_1$ in the \acrshort{ccfw} are set to half, equal to, double, and quadruple its original value. The results show that the final tritium inventory increases with increasing residence time $\tau_1$, while the inventory profile increases with similar speed before the residence time delay $\tau_0$. Similar trends are observed when varying $\tau_0$ and $\tau_1$ in other plasma-facing components.

\begin{figure}[h!]
\centering
\includegraphics[width = 0.8\linewidth]{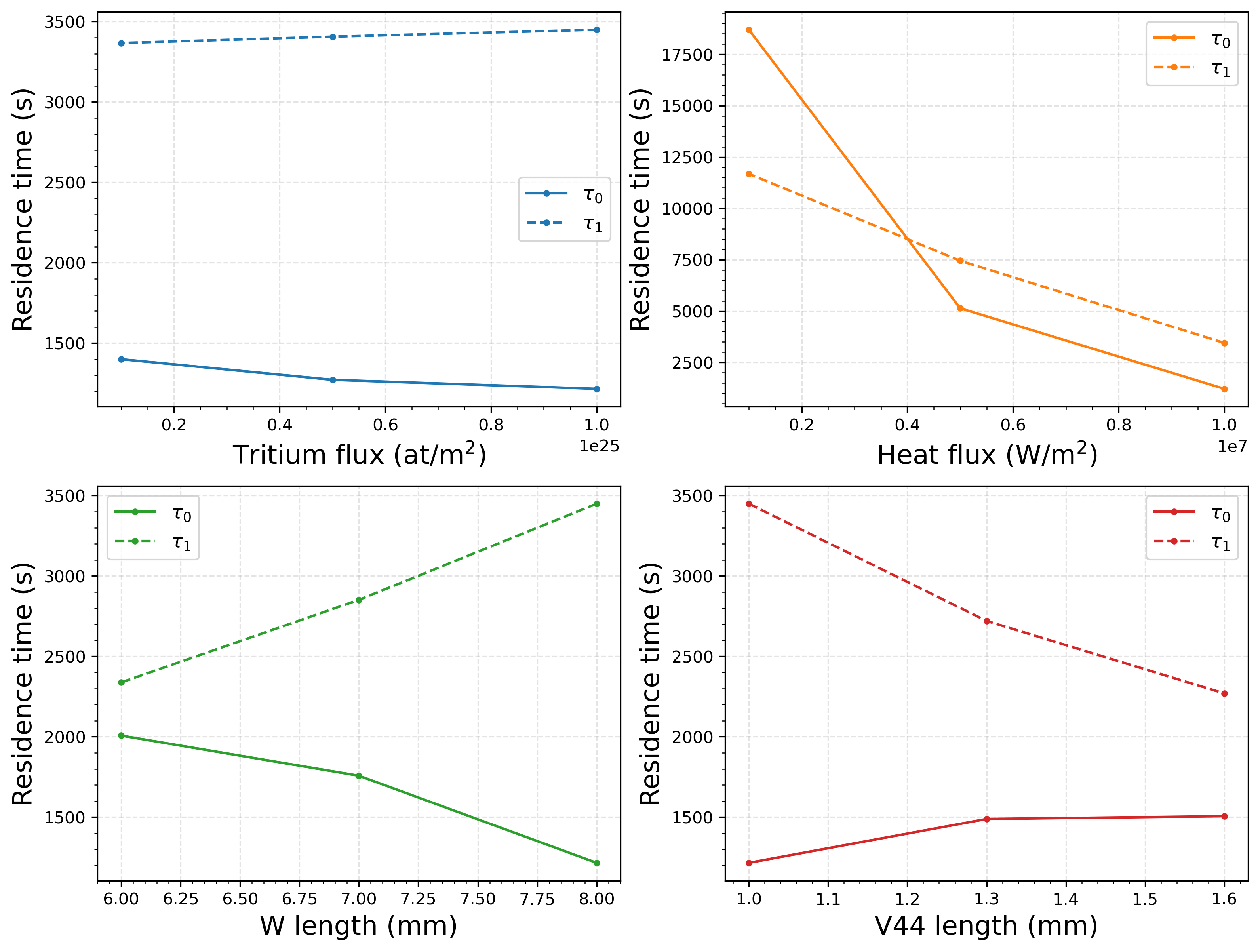}
\caption{\label{fig:fuel_cycle:multiple_changes_input} The residence times on the \acrshort{ccfw} with varying tritium flux, heat flux, \acrshort{w} length, and \acrshort{v44} length (only one parameter is changed at a time). The default values for tritium flux, heat flux, \acrshort{w} length, and \acrshort{v44} length are 1 $\times10^{25}$ at/m$^2$, 1 $\times10^{7}$ W/m$^2$, 8 mm, and 1 mm, respectively.}
\end{figure}

\begin{figure}[h!]
\centering
\begin{subfigure}{.43\linewidth}
  \centering
	\includegraphics[width = \linewidth]{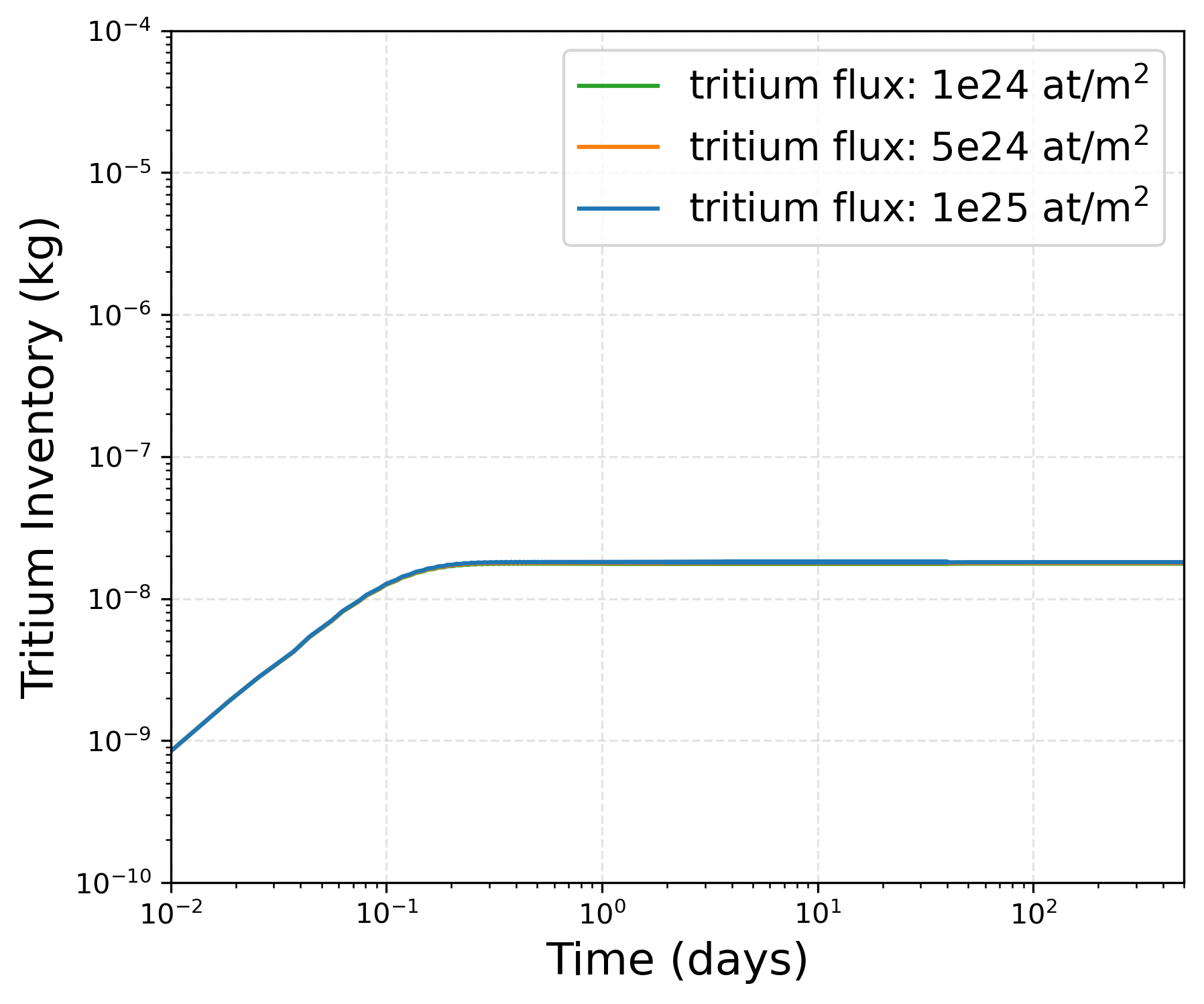}
    \caption{}
    \label{fig:fuel_cycle:multiple_changes_results:T}
\end{subfigure}
\begin{subfigure}{.43\linewidth}
  \centering
	\includegraphics[width = \linewidth]{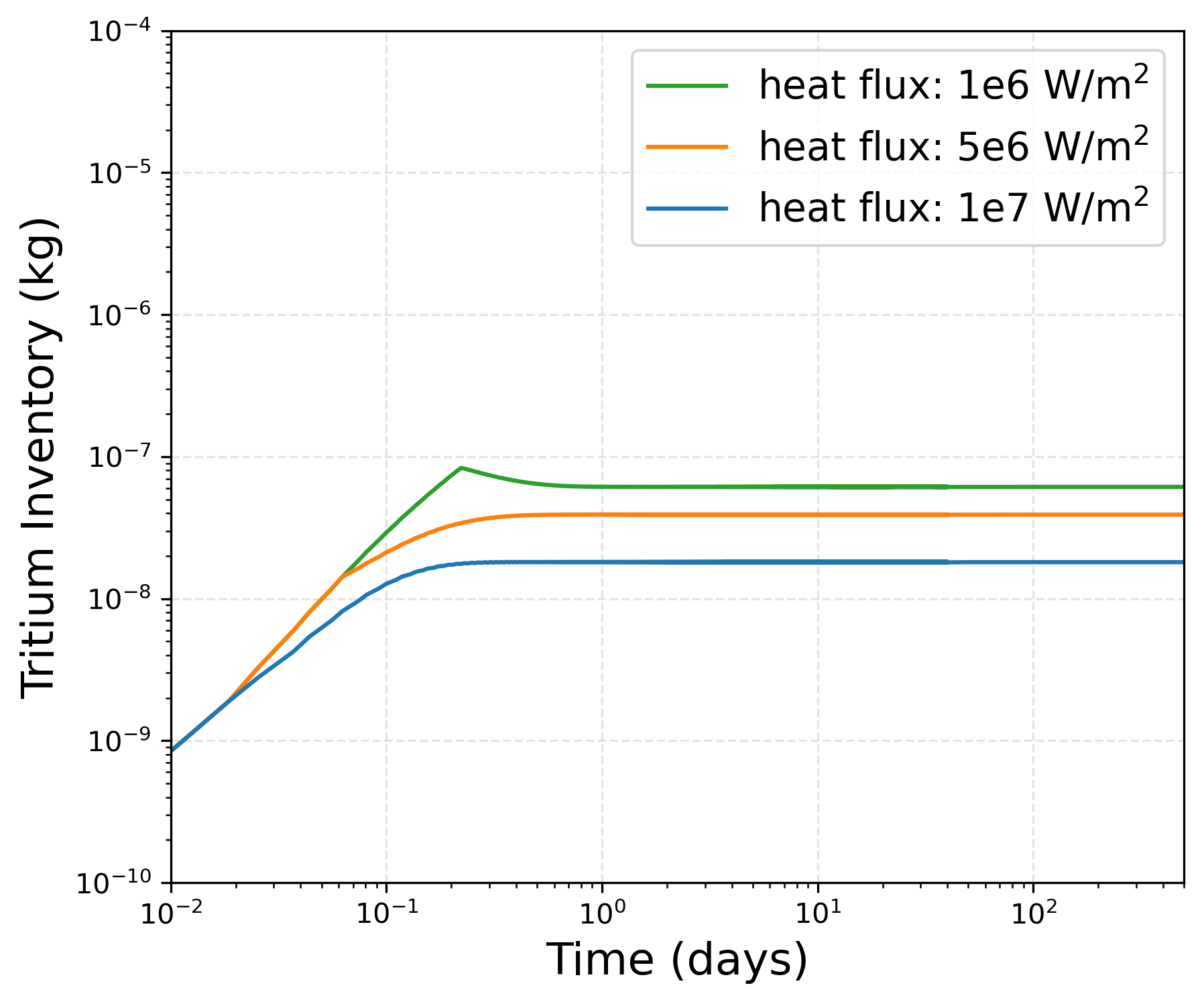}
    \caption{}
    \label{fig:fuel_cycle:multiple_changes_results:H}
\end{subfigure}
\begin{subfigure}{.43\linewidth}
  \centering
	\includegraphics[width = \linewidth]{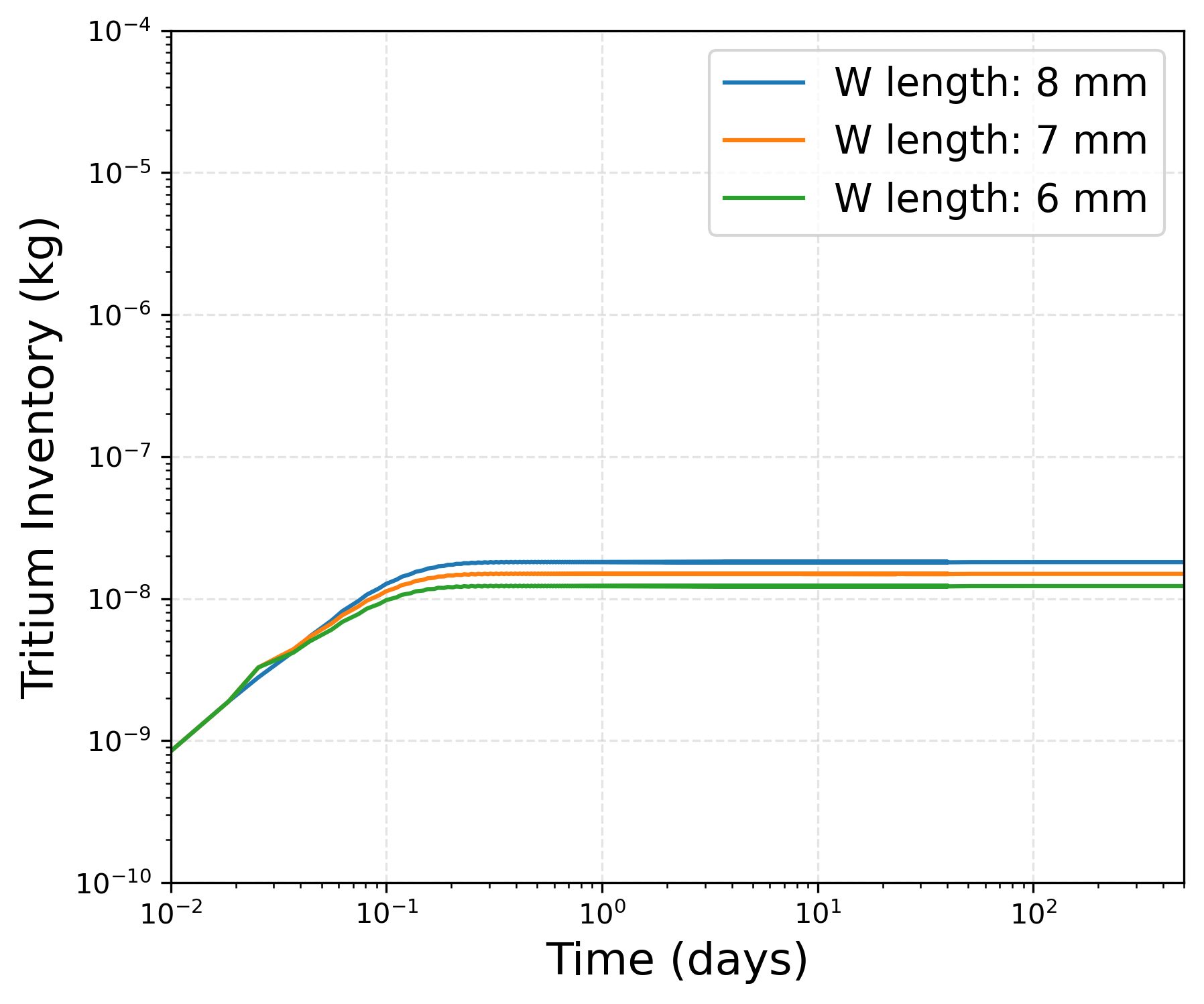}
    \caption{}
    \label{fig:fuel_cycle:multiple_changes_results:W}
\end{subfigure}
\begin{subfigure}{.43\linewidth}
  \centering
	\includegraphics[width = \linewidth]{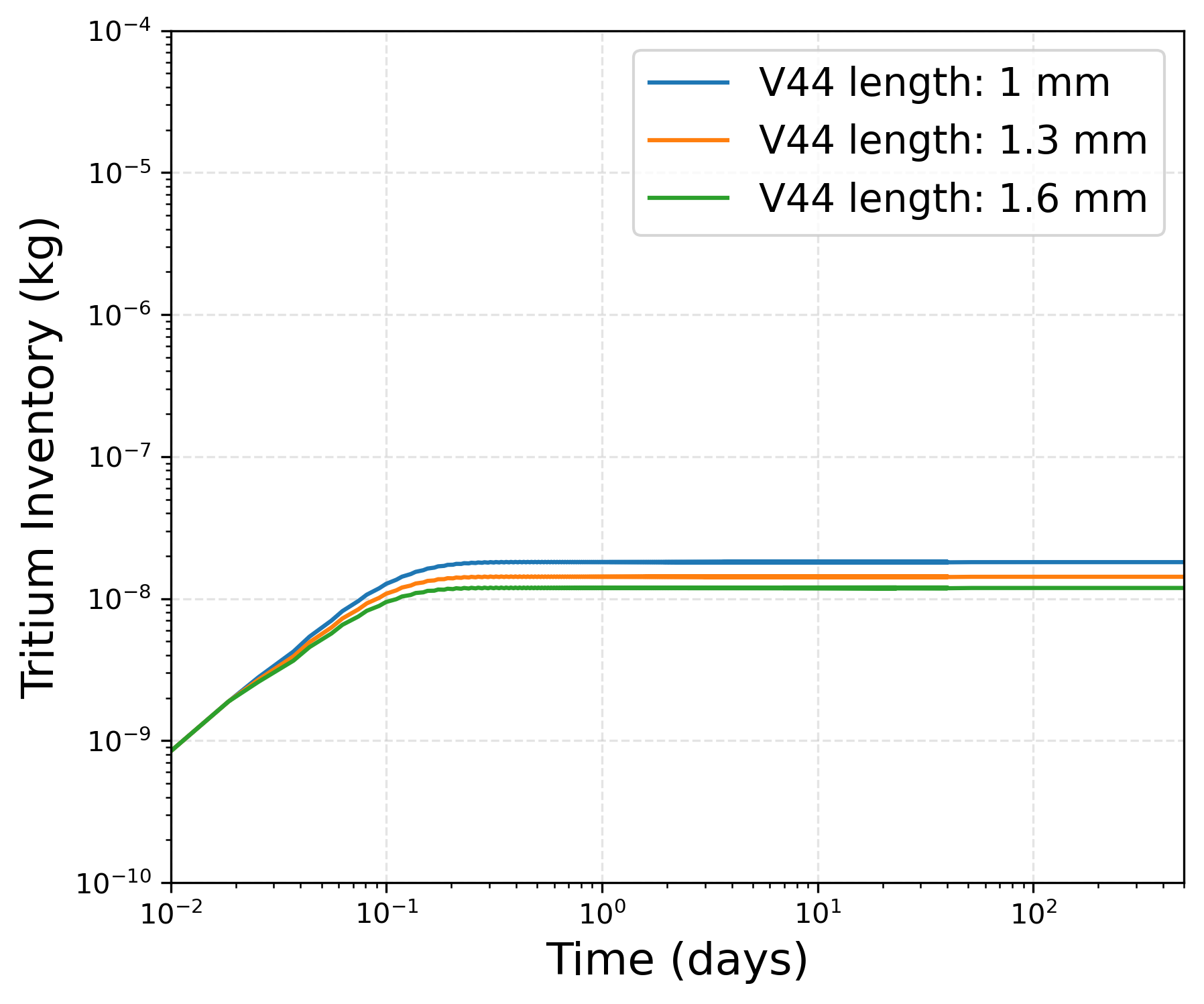}
    \caption{}
    \label{fig:fuel_cycle:multiple_changes_results:V44}
\end{subfigure}
\caption{The evolution of tritium inventory in the \acrshort{ccfw} with varying (a) tritium flux, (b) heat flux, (c) \acrshort{w} length, and (d) \acrshort{v44} length. The bumps at early times, particularly for the green curves in (b) and (c), arise from the change in tritium inventory rate before and after $\tau_0$. Before $\tau_0$, the \acrshort{ccfw} receives only an input tritium flux, whereas after $\tau_0$ it has both input and output fluxes.}
\label{fig:fuel_cycle:multiple_changes_results}
\end{figure}

We also vary the tritium flux, heat flux, \acrshort{w} length, and \acrshort{v44} length in the \acrshort{ccfw} component to compute their corresponding two-parameter residence times and to evaluate how these parameters influence the overall fuel cycle behavior. \Cref{fig:fuel_cycle:multiple_changes_input} shows the changes in these four input parameters and their corresponding residence time predictions from the surrogate models. \Cref{fig:fuel_cycle:multiple_changes_results} presents the resulting tritium inventory evolution in the \acrshort{ccfw} component. The heat flux has the strongest impact on tritium inventory evolution, whereas tritium flux exhibits the weakest influence, at least within the ranges considered in this study. The tritium inventories of varying \acrshort{w} length and \acrshort{v44} length have similar trends, with tritium inventory increasing as either thickness decreases.

\section{Discussion}
\label{sec:discussion}


In this work, we develop a tritium modeling framework that integrates component-level simulations, data-driven surrogate modeling, and system-level fuel cycle analysis. The component-level tritium transport models capture diffusion, trapping, and release behavior within the plasma-facing components and the \acrshort{vv}. Gaussian process surrogate models are then trained on these datasets to enable fast, low computational cost predictions of tritium transport behavior across a wide input parameter space. Finally, the residence time from the surrogate models are incorporated into a system-level tritium fuel cycle model, which provides fast and scalable feedback for tritium recycling design in fusion plants.

This integrated approach accelerates the iterative design for tritium management systems, reduces the computational cost of exploring impacts of component configurations, and supports optimization of tritium recycling strategies. However, several limitations and opportunities for future work remain and are discussed in this section.

\subsection{Component-Level Modeling}
\label{sec:discussion:component}

The current component-level simulations use material properties from the literature without explicitly considering parameter uncertainty or experimental error. In practice, diffusivity, solubility, and trapping parameters can vary significantly depending on microstructure, doping element, and surface condition. Moreover, the present component-level model assumes a single intrinsic trapping site for each plasma-facing material (\acrshort{w}, \acrshort{v44}, and \acrshort{ss}), while real materials exhibit multiple intrinsic trapping sites with distinct energies and densities \cite{hodille2021modelling}. Irradiation damage in the fusion environment will further introduce dynamic trapping properties, altering effective transport coefficients over time. Additionally, surface kinetics are represented by a simplified tritium-only reaction, neglecting isotopic competition and mixed tritium--deuterium that may modify recombination and dissociation kinetics. Finally, the model only considers one-dimensional geometry, which limits its ability to capture non-uniform temperature and flux distributions that occur in realistic plasma-facing component geometries, although the one-dimensional geometry can still provide comparable results on tritium release behavior under various configurations on plasma-facing components \cite{shimada2024toward,hodille2021modelling}.

Despite these simplifications, the component-level models provide valuable insight into how structural and operational parameters, such as armor thickness, pipe material, and heat flux, impact tritium retention and release. For instance, thinner \acrshort{w} armor reduces tritium retention due to smaller storage volume but increases coolant surface flux. The use of \acrshort{w} pipe instead of \acrshort{v44} has limited impact on overall retention but influences coolant surface flux due to differences in permeability. These findings assist the design of critical components on thermal and tritium transport performance.

\subsection{Surrogate Modeling}
\label{sec:discussion:surrogate}

The Gaussian process surrogate models effectively reproduce steady-state tritium flux predictions with a minimal RMSPE of approximately 9.92\%, while the residence time parameters ($\tau_0$ and $\tau_1$) exhibit higher uncertainty (33.58\% and 50.57\%) due to the greater variability in transient behavior and extra errors from the fitting process. Although the surrogates introduce some error relative to direct simulations, they dramatically reduce computational cost by a factor of 2.7$\times10^{-6}$.

This performance makes rapid parametric sweeps and uncertainty analyses feasible, especially when components are extended to higher-dimensional models. Future work will focus on improving surrogate accuracy through a larger input parameter space.

\subsection{Fuel Cycle Model}
\label{sec:discussion:fuel_cycle}

The system-level fuel cycle model demonstrates how residence time from the surrogate models can be coupled within system-level tritium recycling. In this study, surrogate data are integrated for the \acrshort{div}, \acrshort{ccfw}, and \acrshort{bkfw}, while other components still use baseline parameters from previous models \cite{TE_lead_paper}. 

The comparison between the one-parameter and two-parameter residence time fuel cycle models shows that considering an initial delay period reduces the apparent tritium inventory within plasma-facing components by more accurately tritium release behavior. This highlights the value of coupling component-level models to refine the system-level predictions of tritium recycling. Extending surrogate coupling to all components will further accelerate the design and analysis of tritium recycling systems in future fusion plants. Moreover, the comparison of tritium inventories to variations in the input parameters identify the dominant parameter governing fuel cycle performance, thereby guiding targeted optimization and enabling rapid design iteration for fusion systems. These outcomes therefore achieve our goals from \cref{sec:introduction}, demonstrating the feasibility of the integrated multiscale approach and enabling rapid design on key components.

\subsection{Discussion for Integrated Multiscale Approach}
\label{sec:discussion:integrated}


The current study demonstrates using a surrogate model to capture component-scale behavior at the scale of the fuel cycle. This enables the capture of important component-scale behavior while drastically reducing computational costs. Another method for doing this, if the computational framework enables it, is to perform integrated simulations at the component scale and feeding their results in memory into the overarching system-level simulation \cite{gaston2015physics}. This conventional approach obviates the need for surrogate model training and allows component-level models to be updated directly within the system-level models without the need to retrain surrogate models. Unfortunately, conventional approach can be time-consuming, especially when applied to detailed geometries and long operational timescales \cite{kuan1999new}. This can be prohibitive for iterative design and scoping studies. Therefore, the methodology developed in this study represents an initial step toward a fully integrated multiscale tritium transport framework. By combining component-level physics with fast, data-driven surrogate models and system-level integration, this approach balances computational efficiency with physical accuracy. The resulting workflow enables fast design iterations during early-stage fusion plant design, supports parametric sensitivity analyses, and reduces dependence on empirically calibrated or estimated and highly uncertain coefficients.

Future developments will focus on coupling the component-level and system-level models directly, allowing on-demand optimization of component predictions during system-level analysis. Incorporating additional physics such as radiation damage evolution, multi-trap kinetics, and surface chemistry under mixed-isotope conditions will further enhance fidelity. Ultimately, this integrated framework can enable the rapid design of self-sufficient, low-inventory fuel cycles to satisfy the tritium accountancy for fusion pilot plants.

\section{Conclusion}
\label{sec:conclusion}

In this work, we develop a component-level tritium transport model to evaluate tritium behavior in key fusion plant components during plasma operations. Using component-level simulations, we quantify tritium diffusion, trapping, and release on the plasma-facing components and the vacuum vessel. Based on the datasets from the tritium transport model, a Gaussian process surrogate model is trained to efficiently predict critical tritium transport parameters, including steady-state tritium flux at the coolant surface and residence time. The surrogate model is then applied to generate these parameters for a range of component configurations, and incorporated into a system-level tritium fuel cycle model to evaluate tritium recycling in a whole fusion plant.

This integrated multiscale approach provides a computationally efficient pathway for fast design iteration in fusion plant development. It reduces computational time significantly while maintaining high predictive accuracy. The approach enables the fast evaluation of tritium inventory evolution, assisting both industrial and research applications in optimizing tritium recycling strategies and minimizing overall tritium inventory requirements.

Future work will focus on extending the current one-dimensional model to two- and three-dimensional geometries to improve modeling accuracy in thermal and tritium transport \rtxt{by using a higher fidelity tritium transport model accounting for microstructural evolution and irradiation effects, for example}. In addition, component-level and system-level models will be directly coupled to assess trade-offs between the surrogate-based-iteration approach and high-fidelity coupled simulations. \rtxt{The integrated approach will also be applied to investigate mixed deuterium–tritium recycling and to assess the impact of neutron irradiation on tritium transport and fuel cycle performance.} These works will enhance predictive capabilities for tritium transport and accelerate the design of self-sufficient tritium-recycling fusion plants.


\section*{Data Availability}
\label{sec:data_availability}

The \acrshort{tmap8} code is open source and available at https://github.com/idaholab/TMAP8. When everything is cleared for release, the online \acrshort{tmap8} documentation will contain all the input files and Python scripts used to generate the simulation results and result figures in this manuscript. The complete TMAP8 documentation, including instructions on how to get started, the verification and validation cases, and descriptions of all of \acrshort{tmap8}’s capabilities, is available at https://mooseframework.inl.gov/TMAP8/index.html.

\section*{Acknowledgments}
\label{sec:acknowledgments}

This work was supported by the Tokamak Energy Ltd. under Contract No. SPP-24SP95. The United States Government retains, and the publisher, by accepting the article for publication, acknowledges that the United States Government retains a nonexclusive, paid-up, irrevocable, worldwide license to publish or reproduce the published form of this manuscript, or allow others to do so, for United States Government purposes.

This research made use of the resources of the High Performance Computing Center at Idaho National Laboratory, which is supported by the Office of Nuclear Energy of the U.S. Department of Energy and the Nuclear Science User Facilities, United States under Contract No. SPP-24SP95.

\bibliographystyle{ar-style3.bst}
\bibliography{sample}

\newpage
\appendix
\section{Surrogate Models Comparison}
\label{sec:appendix:surrogates}

Three types of surrogate models, polynomial regression, Gaussian process, and artificial neural network, are initially evaluated on tritium flux and one-parameter residence time using datasets generated from component-level tritium transport simulations with varied operating and material parameters. We use $k$-folds cross validation to validate the accuracy of different surrogate models with datasets of various sizes. 

$k$-folds cross validation is a widely used technique in machine-learning model validation \cite{stone1974cross,stone1977asymptotic,efron2004estimation}.
In this approach, the dataset is randomly partitioned into $k$ equally sized subsets (we choose $k=5$).
For each of the $k$ iterations, one fold is designated as the validation set, while the remaining $k - 1$ folds are used for training.
This process is repeated $k$ times, ensuring that each fold serves as both training and validation data exactly once.

To ensure statistical robustness, the entire $k$-fold cross-validation process is repeated for 10 trials, and the average \acrshort{rmspe} across all trials is reported as the final performance metric. The resulting surrogate models are integrated into the system-level model (see \cref{sec:methods:governing_system}) to provide rapid, accurate predictions for tritium inventory under varying operating conditions and with varied design parameters.

\begin{figure}[htb]
\centering
\begin{subfigure}{.43\linewidth}
  \centering
	\includegraphics[width = \linewidth]{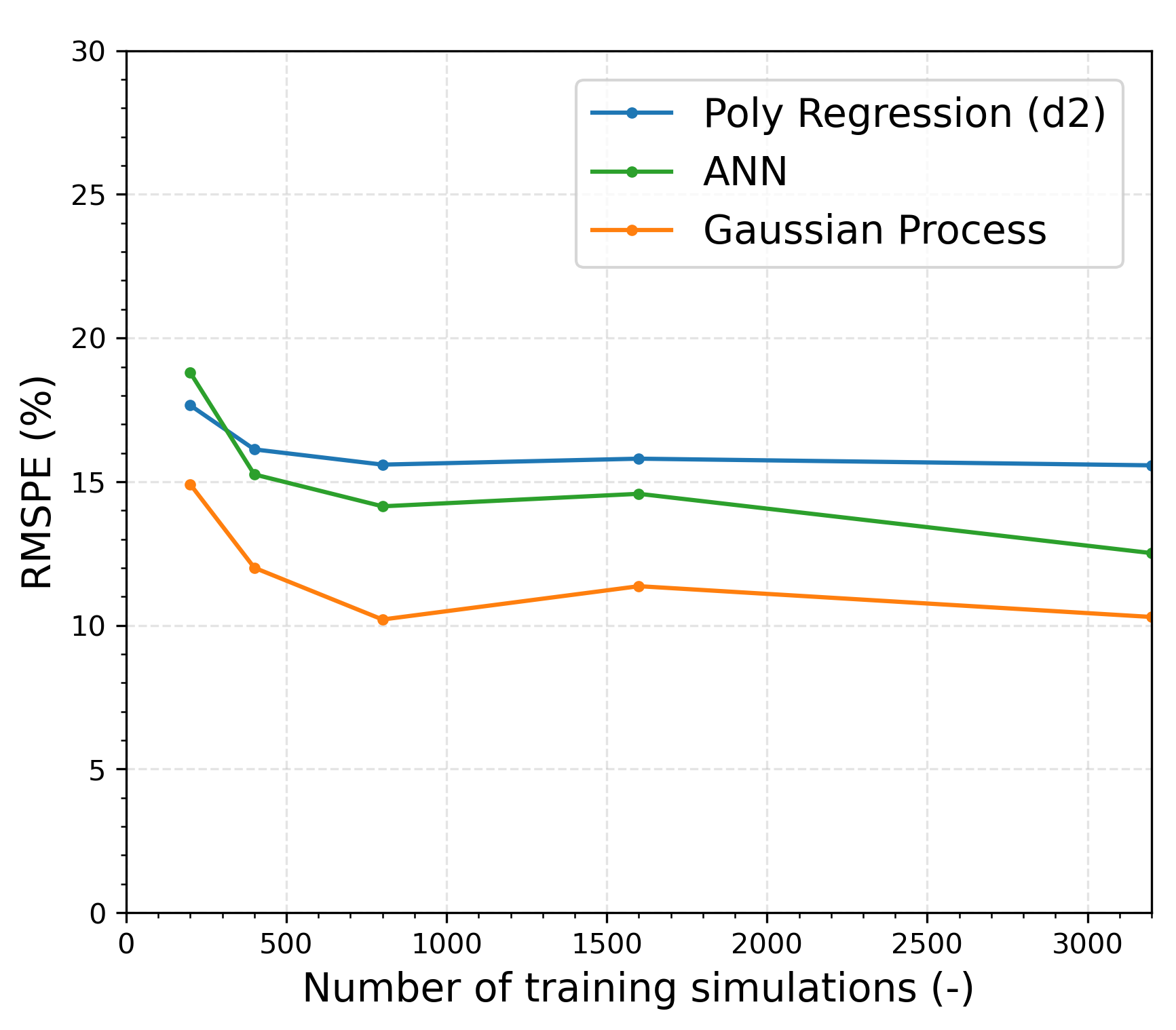}
    \caption{}
    \label{fig:surrogate:k_validation:flux}
\end{subfigure}
\begin{subfigure}{.43\linewidth}
  \centering
	\includegraphics[width = \linewidth]{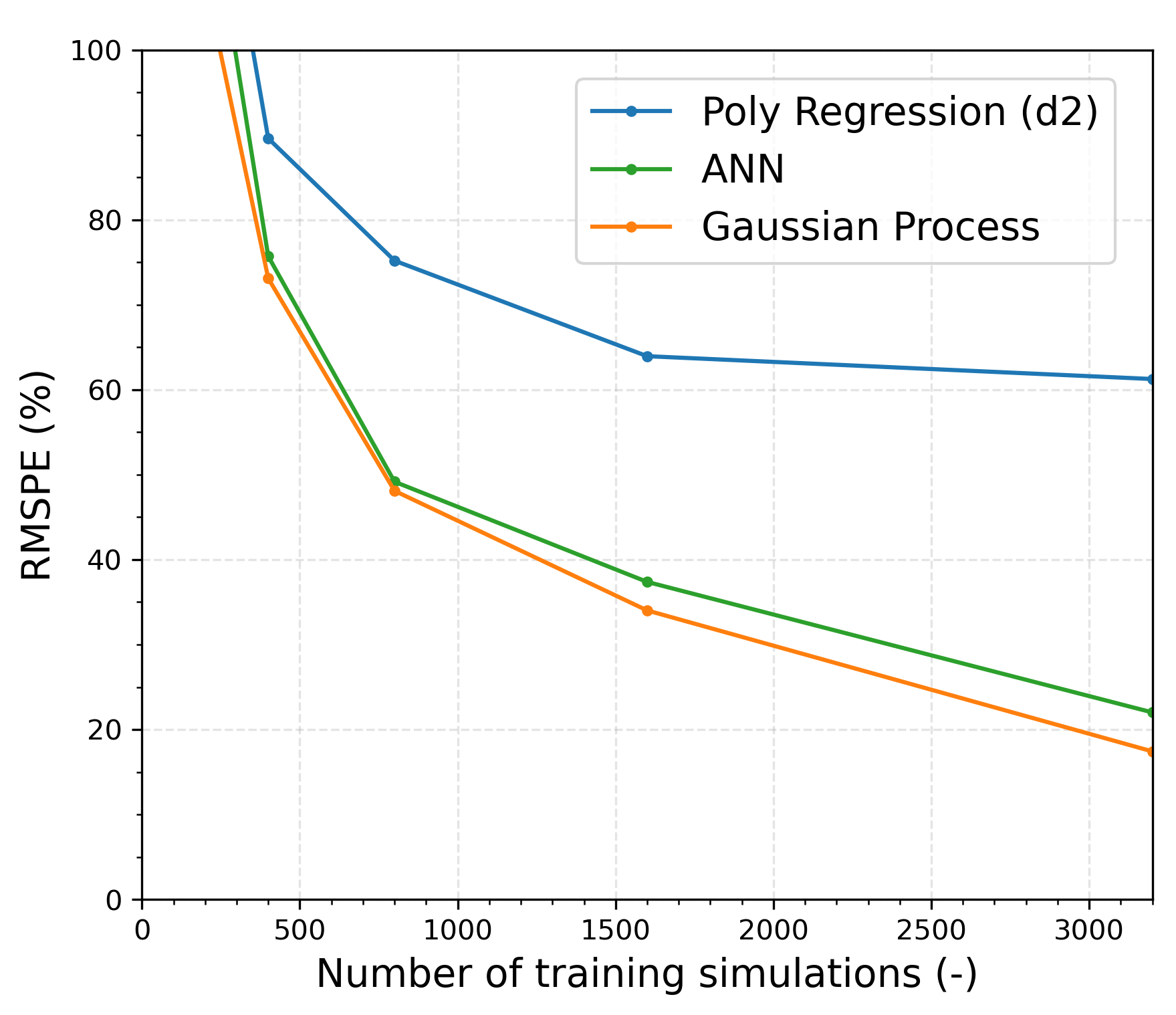}
    \caption{}
    \label{fig:surrogate:k_validation:tau}
\end{subfigure}
\caption{The $k$-fold cross-validation results evolve with the size of datasets from the surrogates models' predictions for (a) steady-state tritium flux at the coolant surface and (b) residence time.}
\label{fig:surrogate:k_validation}
\end{figure}

As shown in \cref{fig:surrogate:k_validation}, the RMSPE for all the surrogate models decreases with the number of simulations in the datasets, and the lowest RMSPEs are 10\% and 17\% for predicting tritium flux and one-parameter residence time, respectively, from the Gaussian process surrogate model. Therefore, we use the Gaussian process method for the surrogate model in this work since it is more accurate than the other methods we test.

\section{Numerical Instabilities from Component-Level Simulations}
\label{sec:appendix:flaw}

\begin{figure}[htb]
\centering
\begin{subfigure}{.43\linewidth}
  \centering
	\includegraphics[width = \linewidth]{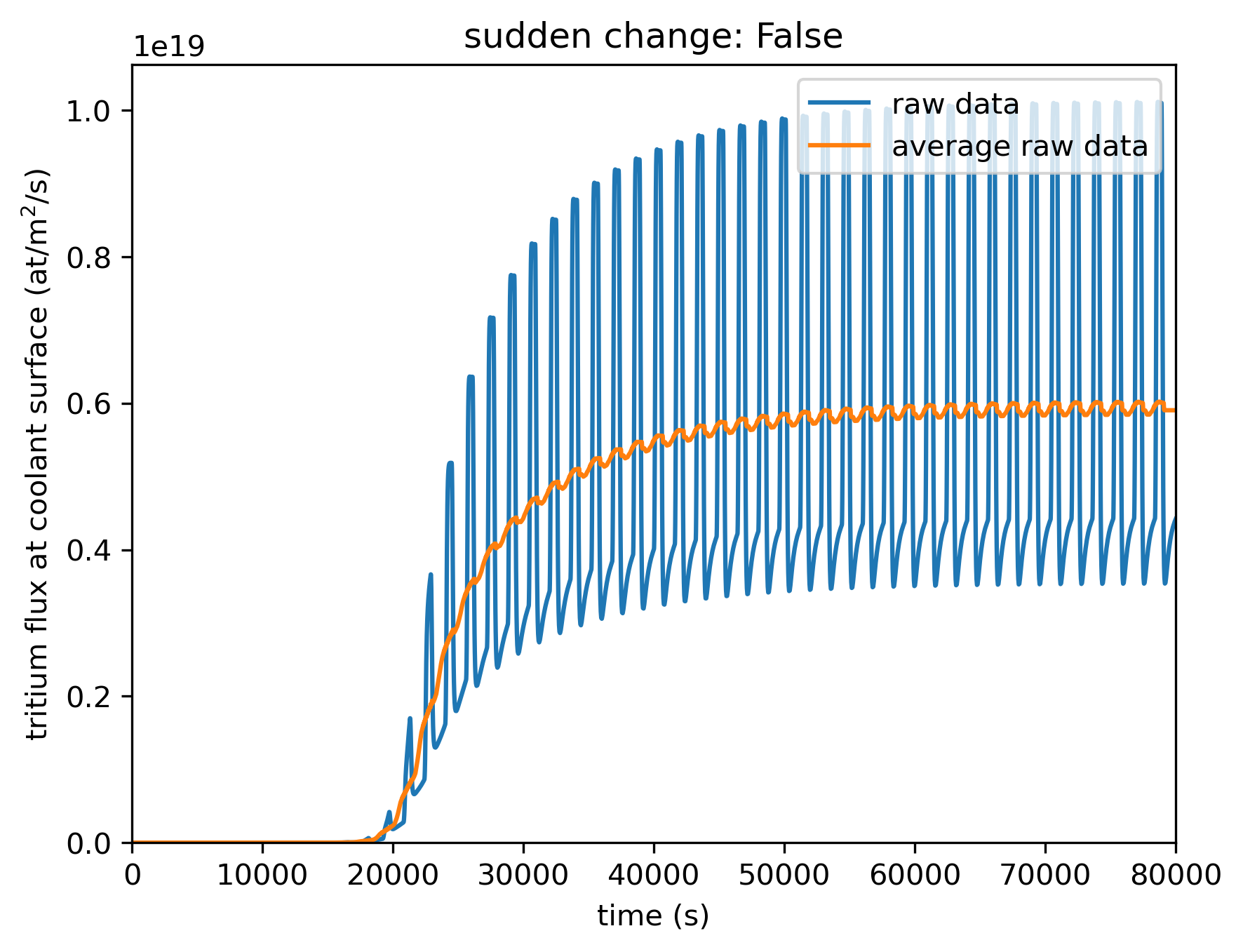}
    \caption{}
    \label{fig:appendex:flaw:false}
\end{subfigure}
\begin{subfigure}{.43\linewidth}
  \centering
	\includegraphics[width = \linewidth]{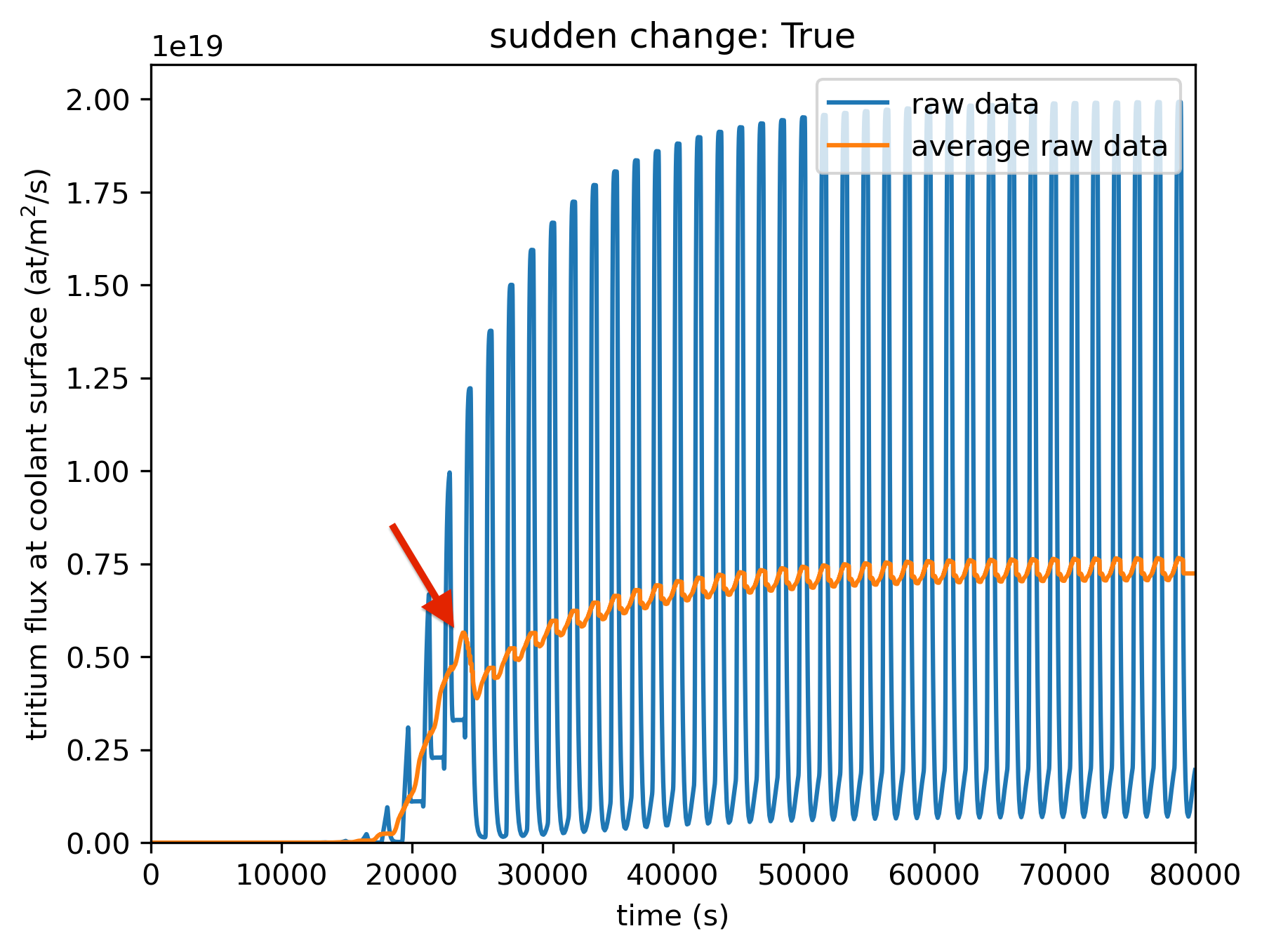}
    \caption{}
    \label{fig:appendex:flaw:true}
\end{subfigure}
\begin{subfigure}{.43\linewidth}
  \centering
	\includegraphics[width = \linewidth]{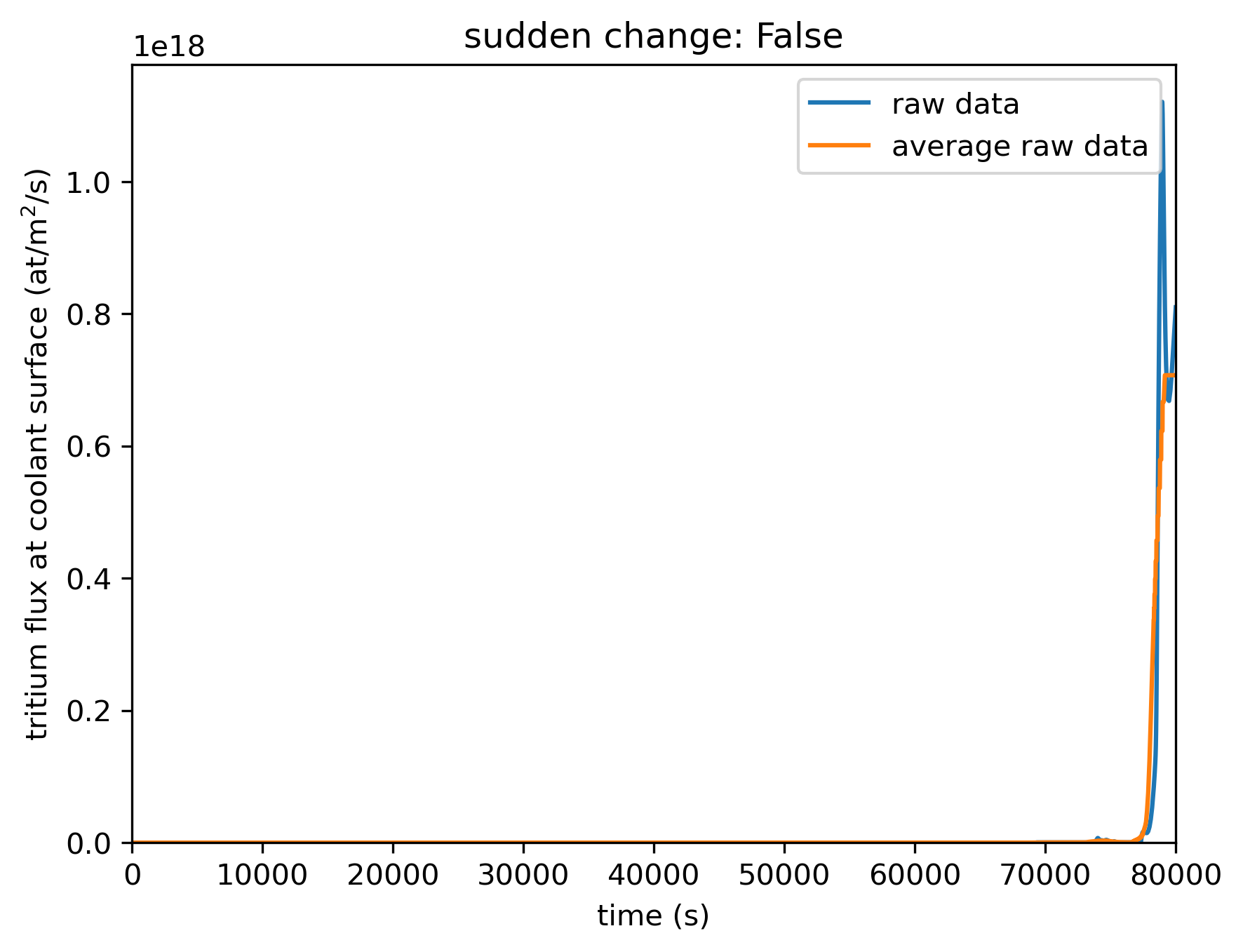}
    \caption{}
    \label{fig:appendex:flaw:beyond}
\end{subfigure}
\caption{Tritium flux evolution from three component-level simulations: (a) without an adverse drop, (b) with an adverse drop, and (c) with a residence time greater than 80,000. The blue curves represent the raw flux data that come directly from the simulations, while the orange curves show the flux averaged over a 1,600 s window. The presence of an adverse drop introduces numerical instabilities that can negatively affect surrogate model training.}
\label{fig:appendex:flaw}
\end{figure}

During the component-level simulations, less than 20\% of the simulations over all datasets exhibit numerical instabilities. As shown in \cref{fig:appendex:flaw:true}, the tritium flux evolution shows a sudden drop caused by the trade-off between computational time cost and numerical accuracy. These abrupt drops can reduce the accuracy of surrogate model training and the prediction of $J_\infty$, $\tau_0$, and $\tau_0$. We define an adverse drop as any decrease exceeding twice the normal flux fluctuation observed within a typical pulse cycle in this simulation, and such simulations are excluded from both the training and test datasets. After data cleaning, each dataset retains more than 80\% of its original size.

\section{Hyperparameters for Gaussian Process Models}
\label{sec:appendix:hyperparameters}

Here we present the hyperparameters of the optimal Gaussian process surrogate models for $J_\infty$ and $\tau$ on the \acrshort{div} with \acrshort{v44} and \acrshort{w} pipe in \cref{tab:appendix:hyperparameters_flux,tab:appendix:hyperparameters_tau}. $\vec{\ell}_i$ is the length factor for the corresponding input parameters (i.e, tritium flux, heat flux, \acrshort{w} armor thickness, pipe thickness, coolant temperature, trapping site fraction in the \acrshort{w} armor, and release energy in the \acrshort{w} armor), $\sigma_n$ is the noise variance, and $\sigma_f$ is the signal variance. $\pmb{A}_i$ and $\pmb{\lambda}_i$ are the weights variances in the multi-output Gaussian process model specific for $\tau$ with two residence time values. All the models and hyperparameters are from the Gaussian process model in \acrshort{stm} \cite{dhulipala2025moose,cheng2020sparse,liu2018remarks}.

\begin{table}[!ht]
\centering
\caption{The hyperparameters of the Gaussian process surrogate models for $J_\infty$.}
\label{tab:appendix:hyperparameters_flux}
\begin{tabular}{|c|c|c|c|c|}
\hline
\textbf{Parameter} & \textbf{Model for \acrshort{v44} pipe} & \textbf{Model for \acrshort{w} pipe} \\ \hline
$\vec{\ell}_1$ & 1.72 & 1.84 \\ \hline
$\vec{\ell}_2$ & 0.72 & 2.62 \\ \hline
$\vec{\ell}_3$ & 3.44 & 8.81 \\ \hline
$\vec{\ell}_4$ & 2.64 & 6.38 \\ \hline
$\vec{\ell}_5$ & 1.96 & 9.65 \\ \hline
$\vec{\ell}_6$ & 2.51 & 9.44 \\ \hline
$\vec{\ell}_7$ & 3.43 & 9.88 \\ \hline
$\sigma_n$ & 0.011 & 0.001 \\ \hline
$\sigma_f$ & 0.62 & 2.11 \\ \hline
\end{tabular}
\end{table}

\begin{table}[!ht]
\centering
\caption{The hyperparameters of the Gaussian process surrogate models for $\tau$.}
\label{tab:appendix:hyperparameters_tau}
\begin{tabular}{|c|c|c|c|c|}
\hline
\textbf{Parameter} & \textbf{Model for \acrshort{v44} pipe} & \textbf{Model for \acrshort{w} pipe} \\ \hline
$\pmb{A}_1$ & 1$\times10^{-9}$ & 0.87 \\ \hline
$\pmb{A}_2$ & 0.86 & 1$\times10^{-9}$ \\ \hline
$\pmb{\lambda}_1$ & 1.35 & 2.61 \\ \hline
$\pmb{\lambda}_2$ & 1.40 & 0.34 \\ \hline
$\vec{\ell}_1$ & 5.78 & 2.75 \\ \hline
$\vec{\ell}_2$ & 0.55 & 1.95 \\ \hline
$\vec{\ell}_3$ & 4.47 & 3.41 \\ \hline
$\vec{\ell}_4$ & 1.38 & 2.44 \\ \hline
$\vec{\ell}_5$ & 2.79 & 7.58 \\ \hline
$\vec{\ell}_6$ & 6.24 & 3.36 \\ \hline
$\vec{\ell}_7$ & 7.21 & 6.74 \\ \hline
$\sigma_n$ & 0.032 & 0.013 \\ \hline
$\sigma_f$ & 0.39 & 2.23 \\ \hline
\end{tabular}
\end{table}

\section{Tritium Transport Behavior During Plasma Operations}
\label{sec:appendix:tritium_behavior}

\begin{figure}[htb]
\centering
\begin{subfigure}{.45\linewidth}
  \centering
	\includegraphics[width = \linewidth]{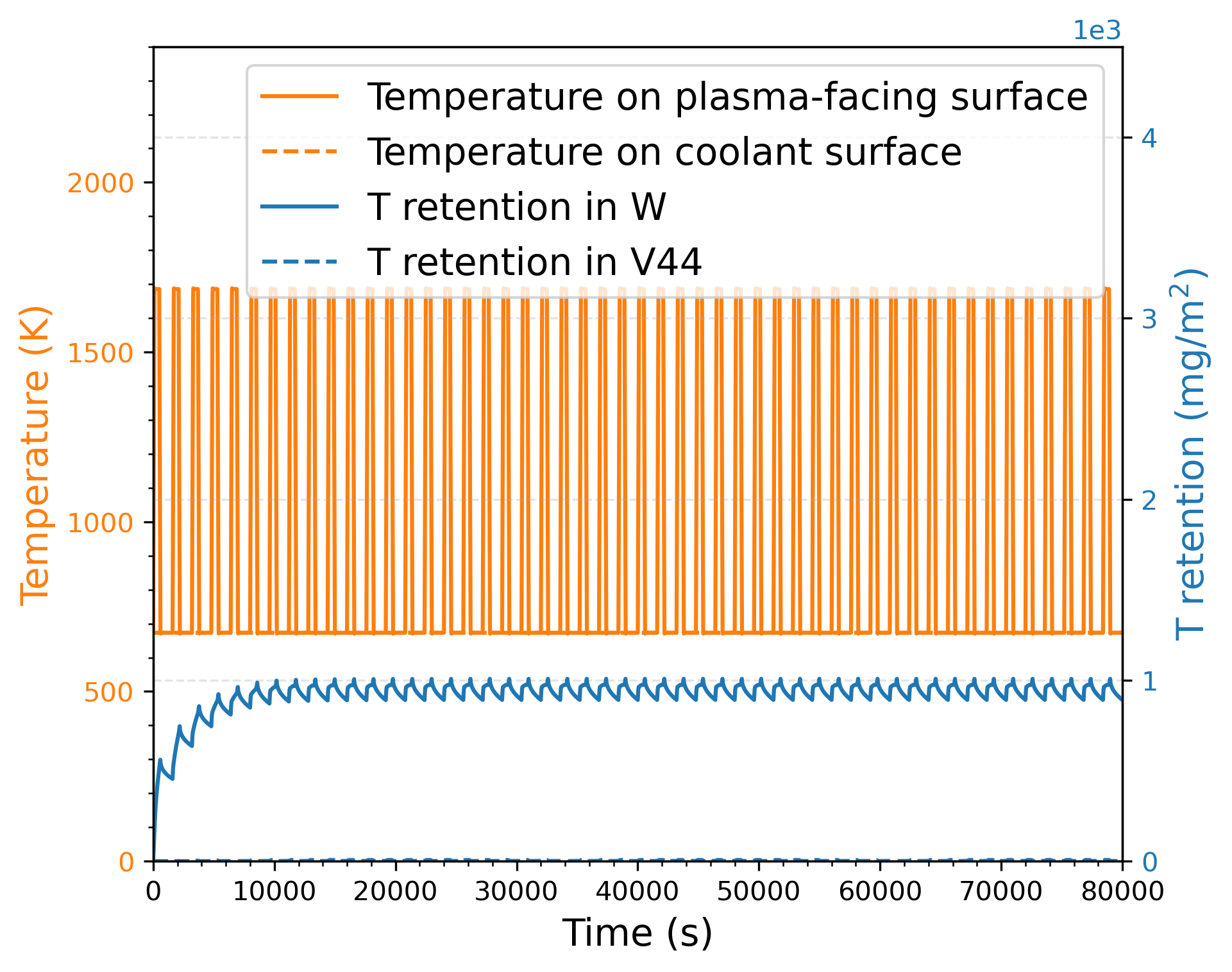}
    \caption{}
    \label{fig:1D_results:DIV_W_V44_v1:tritium_rentension}
\end{subfigure}
\begin{subfigure}{.45\linewidth}
  \centering
	\includegraphics[width = \linewidth]{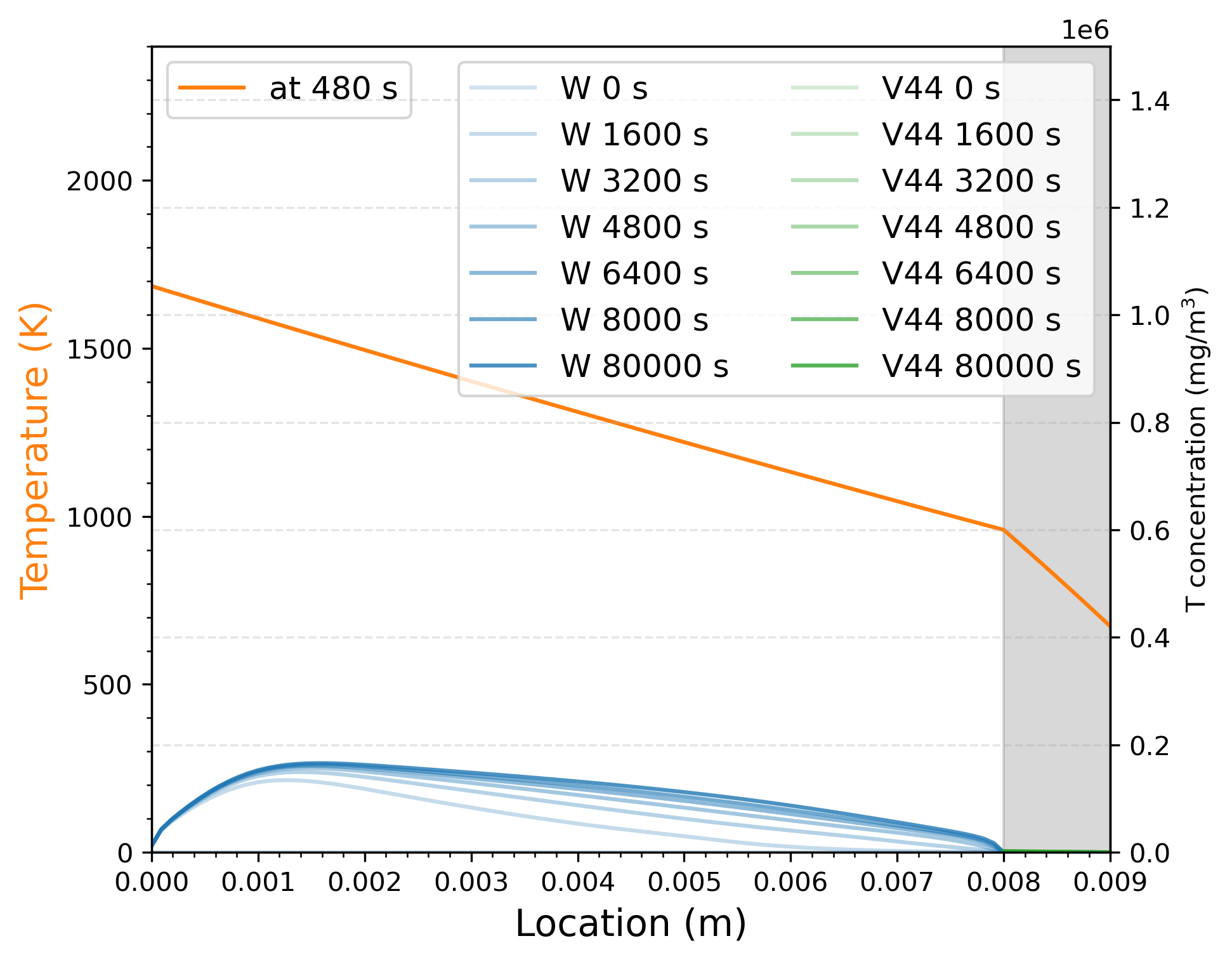}
    \caption{}
    \label{fig:1D_results:DIV_W_V44_v1:tritium_profile}
\end{subfigure}
\begin{subfigure}{.45\linewidth}
  \centering
	\includegraphics[width = \linewidth]{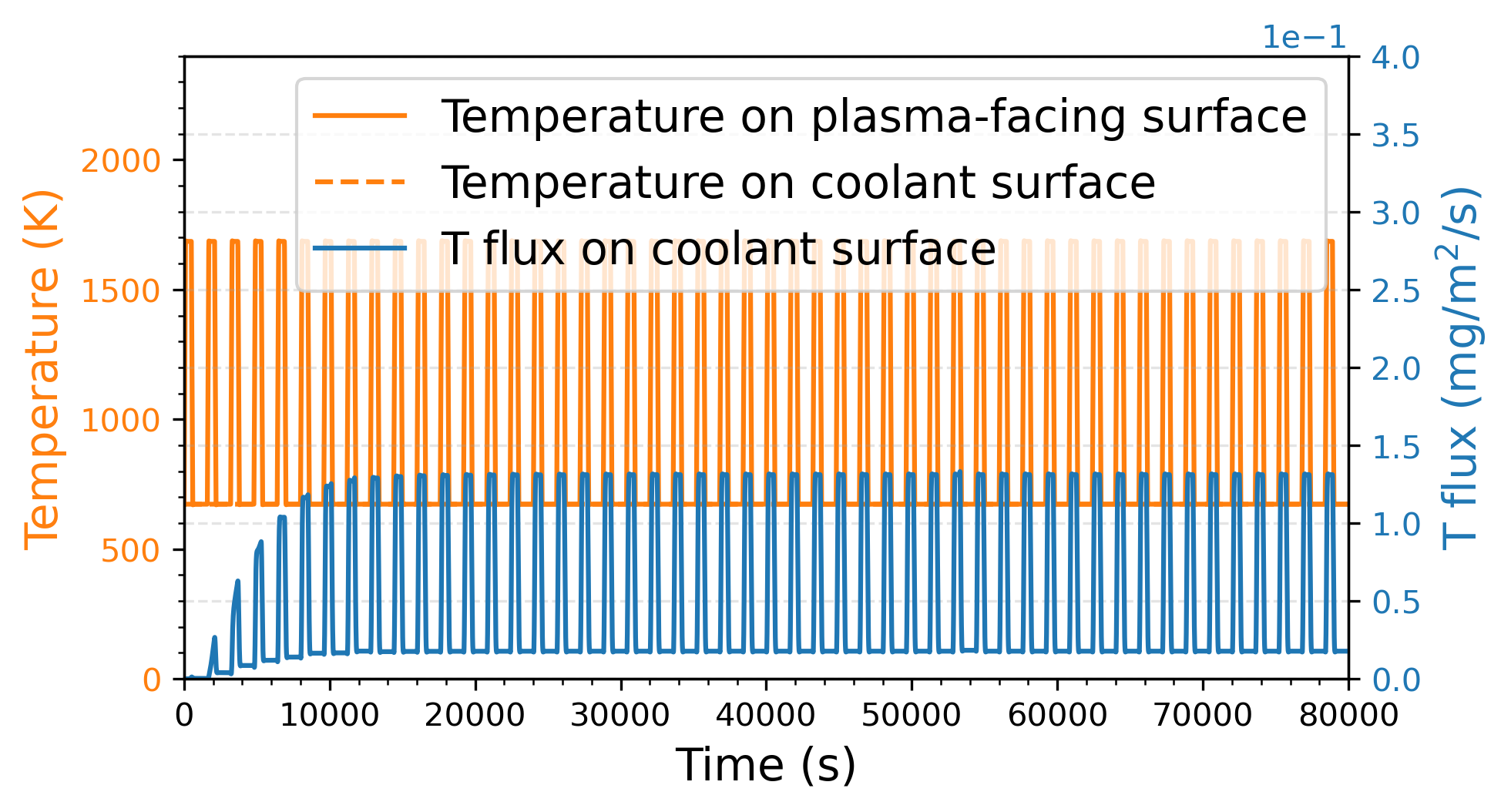}
    \caption{}
    \label{fig:1D_results:DIV_W_V44_v1:tritium_flux}
\end{subfigure}
\caption{The temperature and tritium evolution and corresponding profile for the \acrshort{div} with 8 mm \acrshort{w} armor and 1 mm \acrshort{v44} pipe during 50 pulses. Panel (a) is the temperature and tritium retention evolution during 50 pulses, (b) is the maximum temperature profile and tritium profile at the end of the first five pulses and the final pulse, and (c) is the temperature and tritium flux evolution at the coolant side during 50 pulses. The orange curves represent temperature and the blue curves the tritium.}
\label{fig:1D_results:DIV_W_V44_v1}
\end{figure}

\begin{figure}[htb]
\centering
\begin{subfigure}{.45\linewidth}
  \centering
	\includegraphics[width = \linewidth]{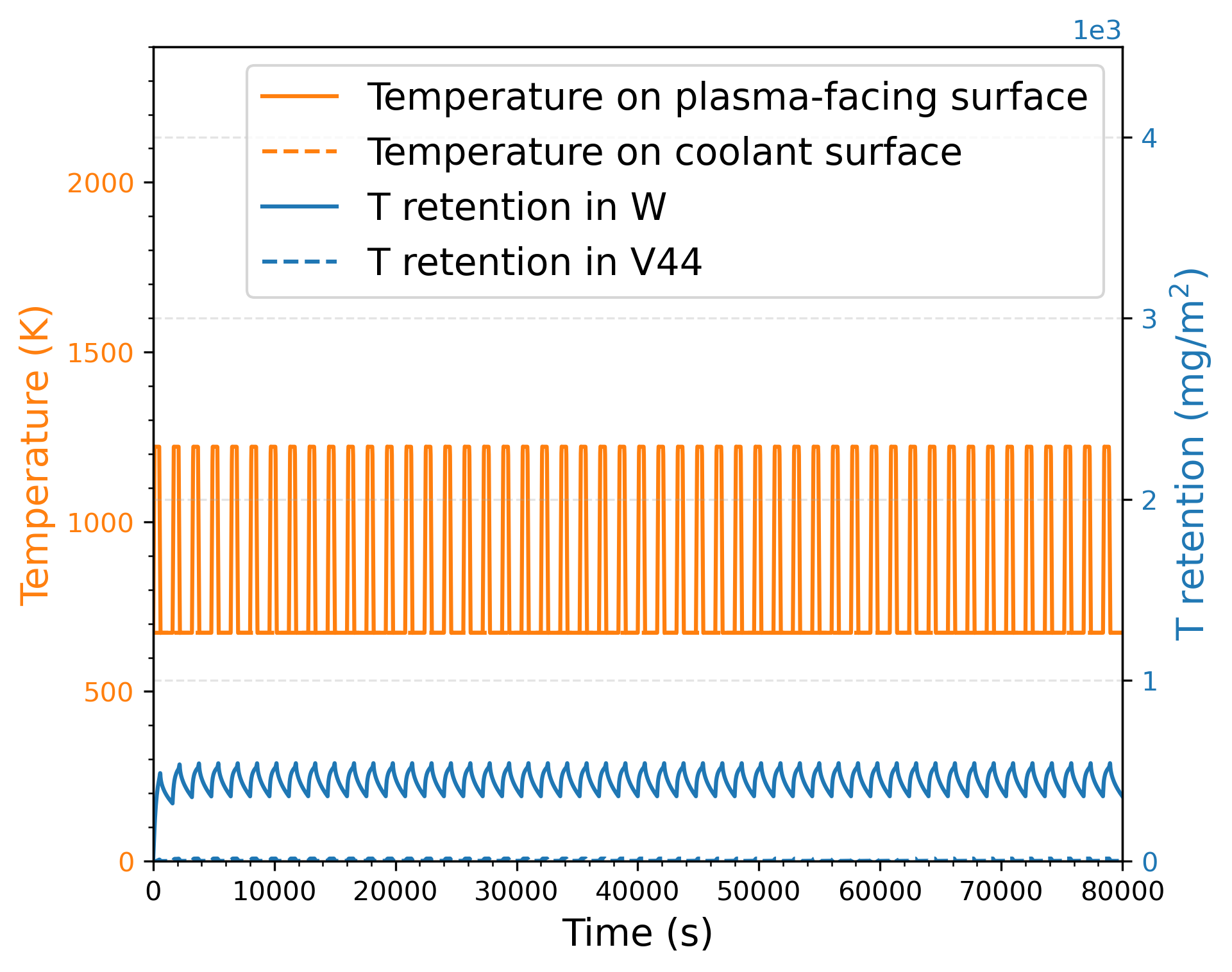}
    \caption{}
    \label{fig:1D_results:DIV_W_V44_v2:tritium_rentension}
\end{subfigure}
\begin{subfigure}{.45\linewidth}
  \centering
	\includegraphics[width = \linewidth]{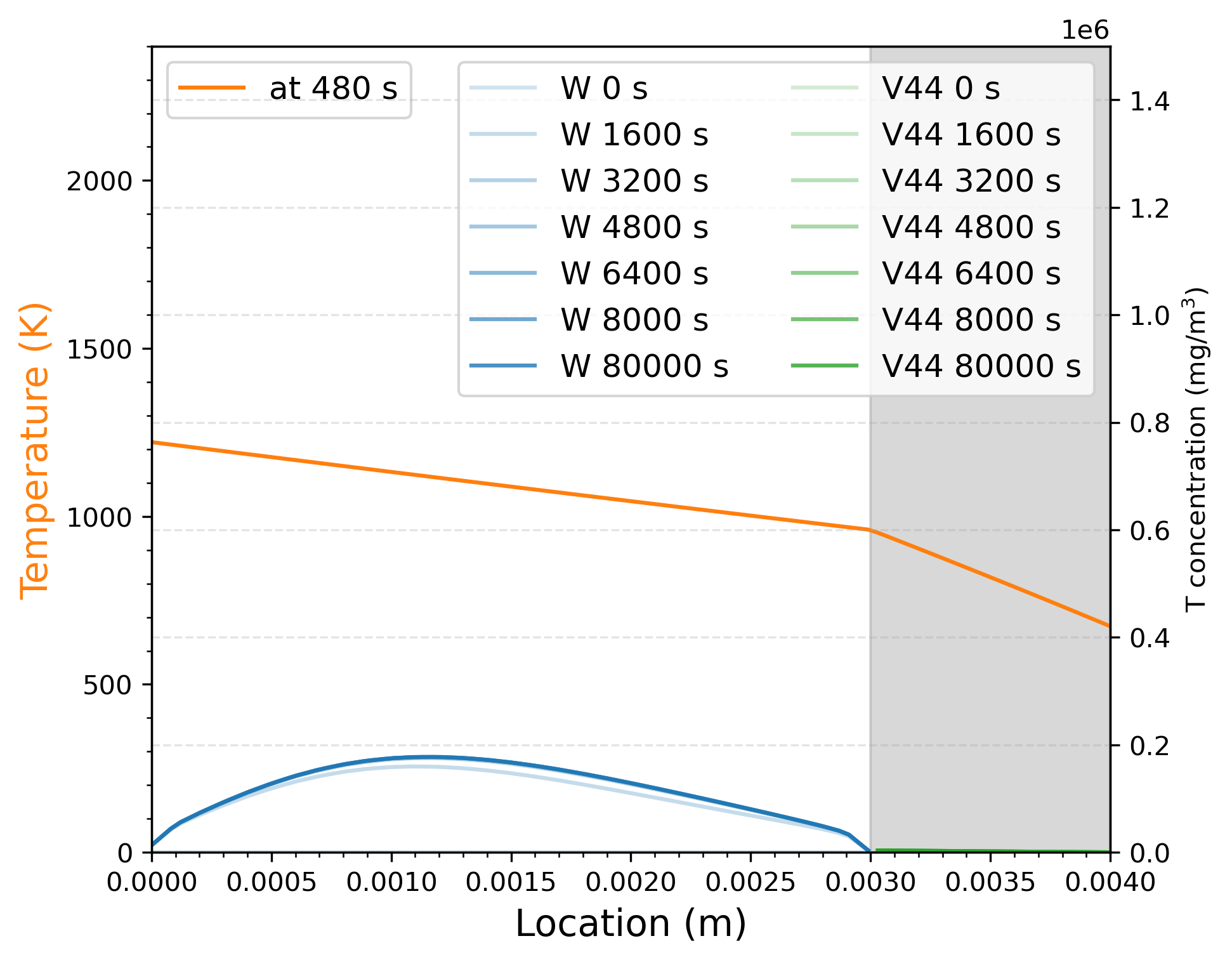}
    \caption{}
    \label{fig:1D_results:DIV_W_V44_v2:tritium_profile}
\end{subfigure}
\begin{subfigure}{.45\linewidth}
  \centering
	\includegraphics[width = \linewidth]{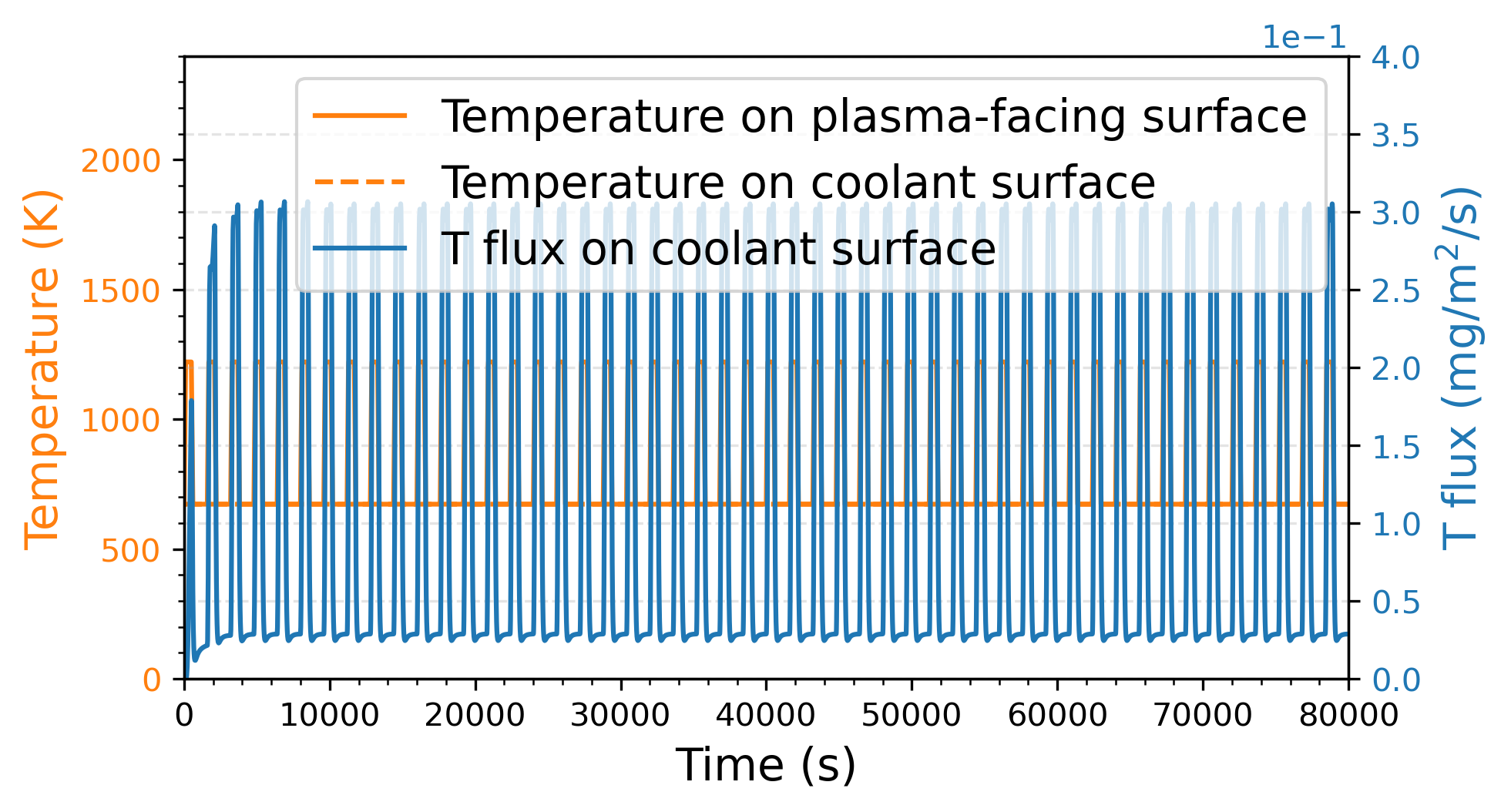}
    \caption{}
    \label{fig:1D_results:DIV_W_V44_v2:tritium_flux}
\end{subfigure}
\caption{The temperature and tritium evolution and corresponding profile for the \acrshort{div} with 3 mm \acrshort{w} armor and 1 mm \acrshort{v44} pipe during 50 pulses. Panel (a) is the temperature and tritium retention evolution during 50 pulses, (b) is the maximum temperature profile and tritium profile at the end of the first five pulses and the final pulse, and (c) is the temperature and tritium flux evolution at the coolant side during 50 pulses. The orange curves represent temperature and the blue curves the tritium.}
\label{fig:1D_results:DIV_W_V44_v2}
\end{figure}

\begin{figure}[htb]
\centering
\begin{subfigure}{.45\linewidth}
  \centering
	\includegraphics[width = \linewidth]{figures/Experiment_W_W_trapping_DIVv1_50_pulses_Temperature_and_Integrated_H_evolution.png}
    \caption{}
    \label{fig:1D_results:DIV_W_W_v1:tritium_rentension}
\end{subfigure}
\begin{subfigure}{.45\linewidth}
  \centering
	\includegraphics[width = \linewidth]{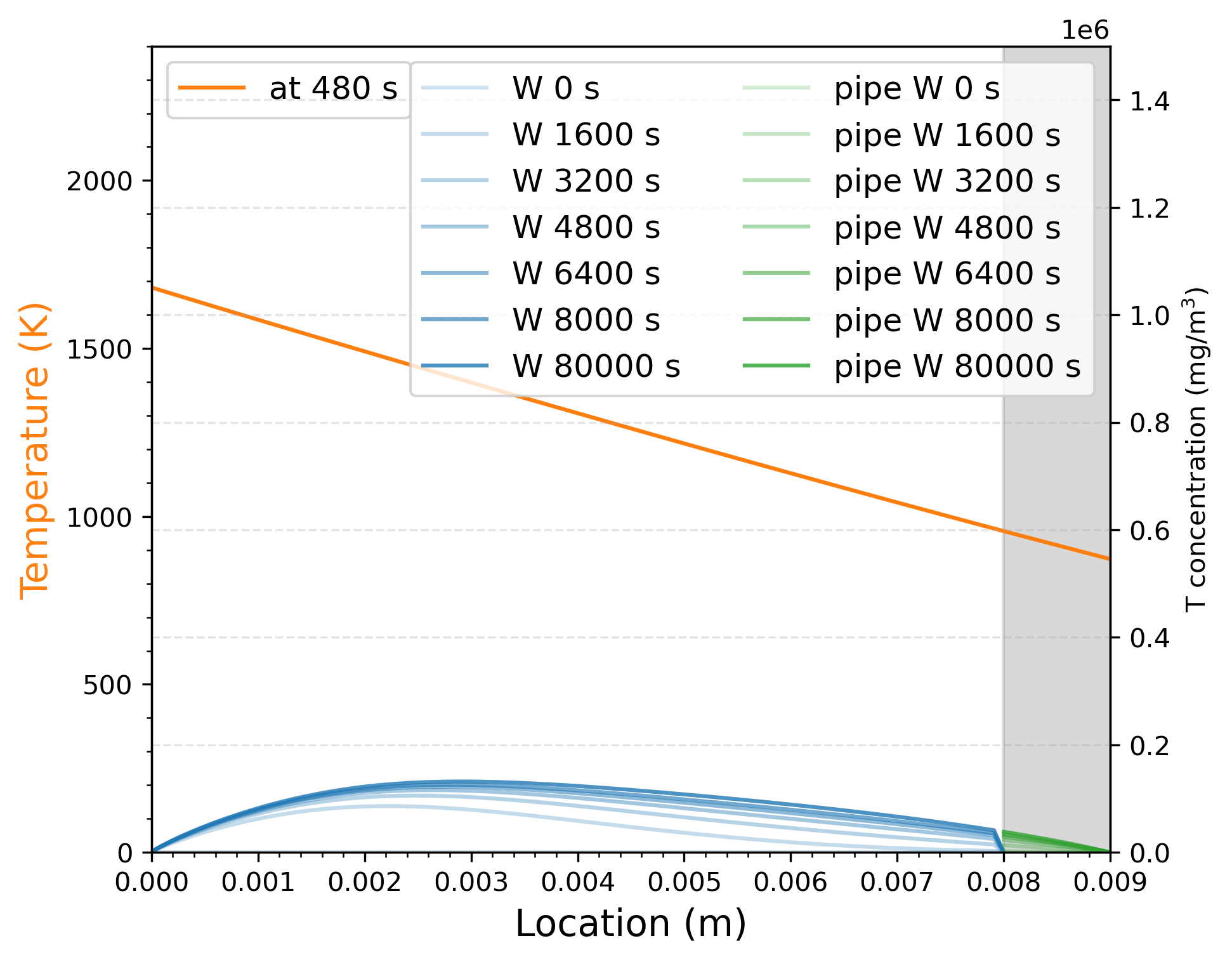}
    \caption{}
    \label{fig:1D_results:DIV_W_W_v1:tritium_profile}
\end{subfigure}
\begin{subfigure}{.45\linewidth}
  \centering
	\includegraphics[width = \linewidth]{figures/Experiment_W_W_trapping_DIVv1_50_pulses_Flux.png}
    \caption{}
    \label{fig:1D_results:DIV_W_W_v1:tritium_flux}
\end{subfigure}
\caption{The temperature and tritium evolution and corresponding profile for the \acrshort{div} with 8 mm \acrshort{w} armor and 1 mm \acrshort{w} pipe during 50 pulses. Panel (a) is the temperature and tritium retention evolution during 50 pulses, (b) is the maximum temperature profile and tritium profile at the end of the first five pulses and the final pulse, and (c) is the temperature and tritium flux evolution at the coolant side during 50 pulses. The orange curves represent temperature and the blue curves the tritium.}
\label{fig:1D_results:DIV_W_W_v1}
\end{figure}

\begin{figure}[htb]
\centering
\begin{subfigure}{.45\linewidth}
  \centering
	\includegraphics[width = \linewidth]{figures/Experiment_W_W_trapping_DIVv2_50_pulses_Temperature_and_Integrated_H_evolution.png}
    \caption{}
    \label{fig:1D_results:DIV_W_W_v2:tritium_rentension}
\end{subfigure}
\begin{subfigure}{.45\linewidth}
  \centering
	\includegraphics[width = \linewidth]{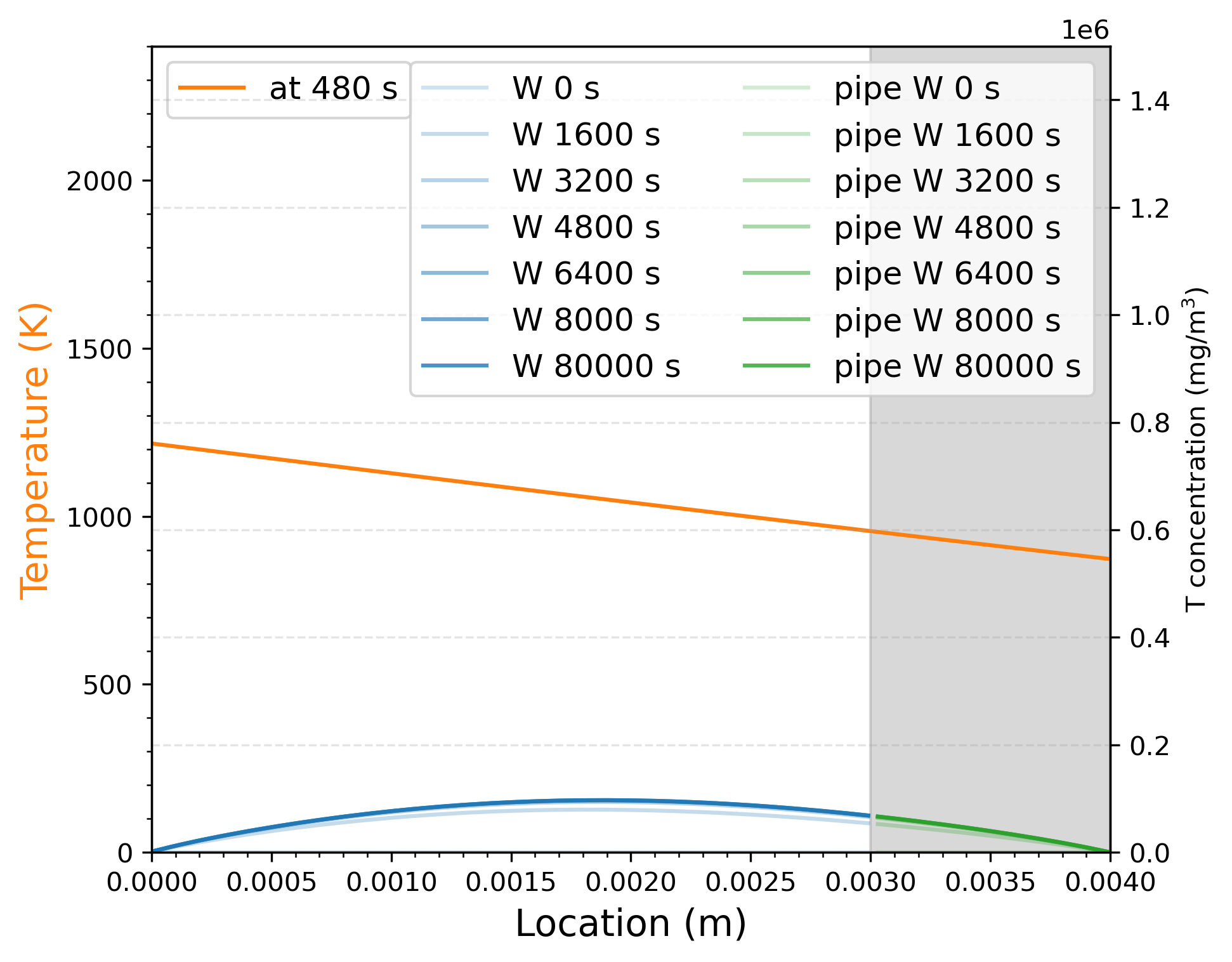}
    \caption{}
    \label{fig:1D_results:DIV_W_W_v2:tritium_profile}
\end{subfigure}
\begin{subfigure}{.45\linewidth}
  \centering
	\includegraphics[width = \linewidth]{figures/Experiment_W_W_trapping_DIVv2_50_pulses_Flux.png}
    \caption{}
    \label{fig:1D_results:DIV_W_W_v2:tritium_flux}
\end{subfigure}
\caption{The temperature and tritium evolution and corresponding profile for the \acrshort{div} with 3 mm \acrshort{w} armor and 1 mm \acrshort{w} pipe during 50 pulses. Panel (a) is the temperature and tritium retention evolution during 50 pulses, (b) is the maximum temperature profile and tritium profile at the end of the first five pulses and the final pulse, and (c) is the temperature and tritium flux evolution at the coolant side during 50 pulses. The orange curves are for the temperature and the blue curves are for tritium.}
\label{fig:1D_results:DIV_W_W_v2}
\end{figure}

\begin{figure}[htb]
\centering
\begin{subfigure}{.45\linewidth}
  \centering
	\includegraphics[width = \linewidth]{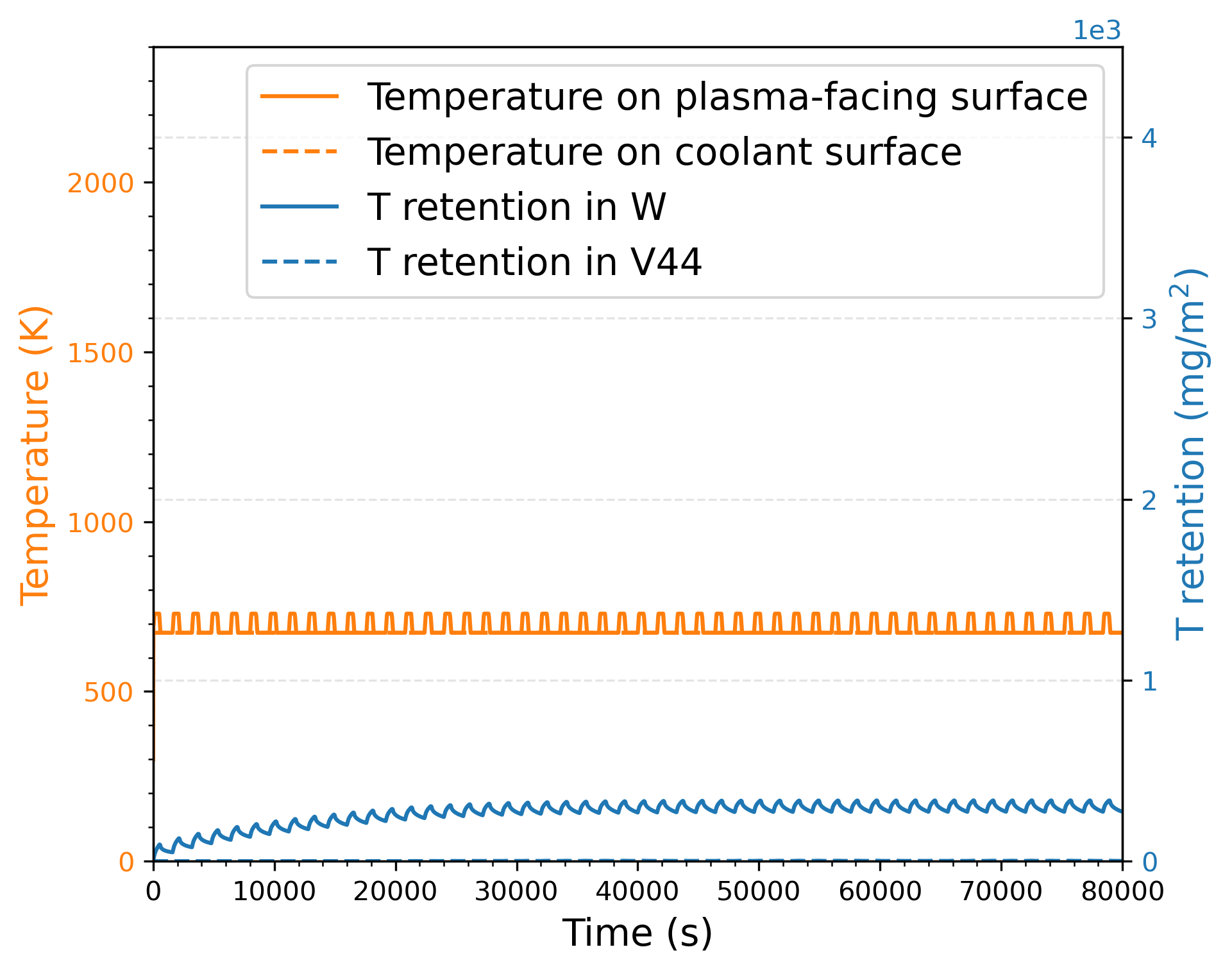}
    \caption{}
    \label{fig:1D_results:CCFW_W_V44:tritium_rentension}
\end{subfigure}
\begin{subfigure}{.45\linewidth}
  \centering
	\includegraphics[width = \linewidth]{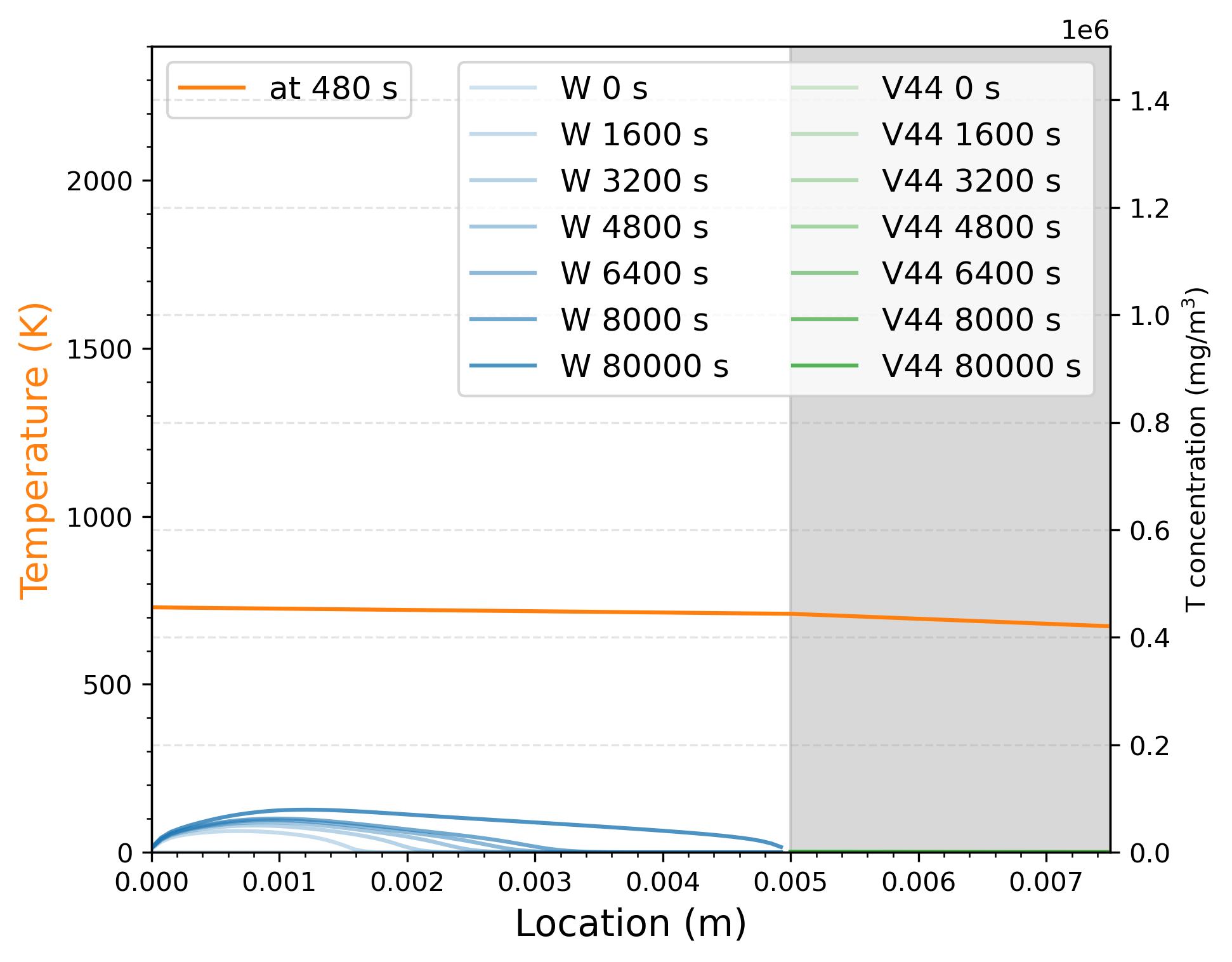}
    \caption{}
    \label{fig:1D_results:CCFW_W_V44:tritium_profile}
\end{subfigure}
\begin{subfigure}{.45\linewidth}
  \centering
	\includegraphics[width = \linewidth]{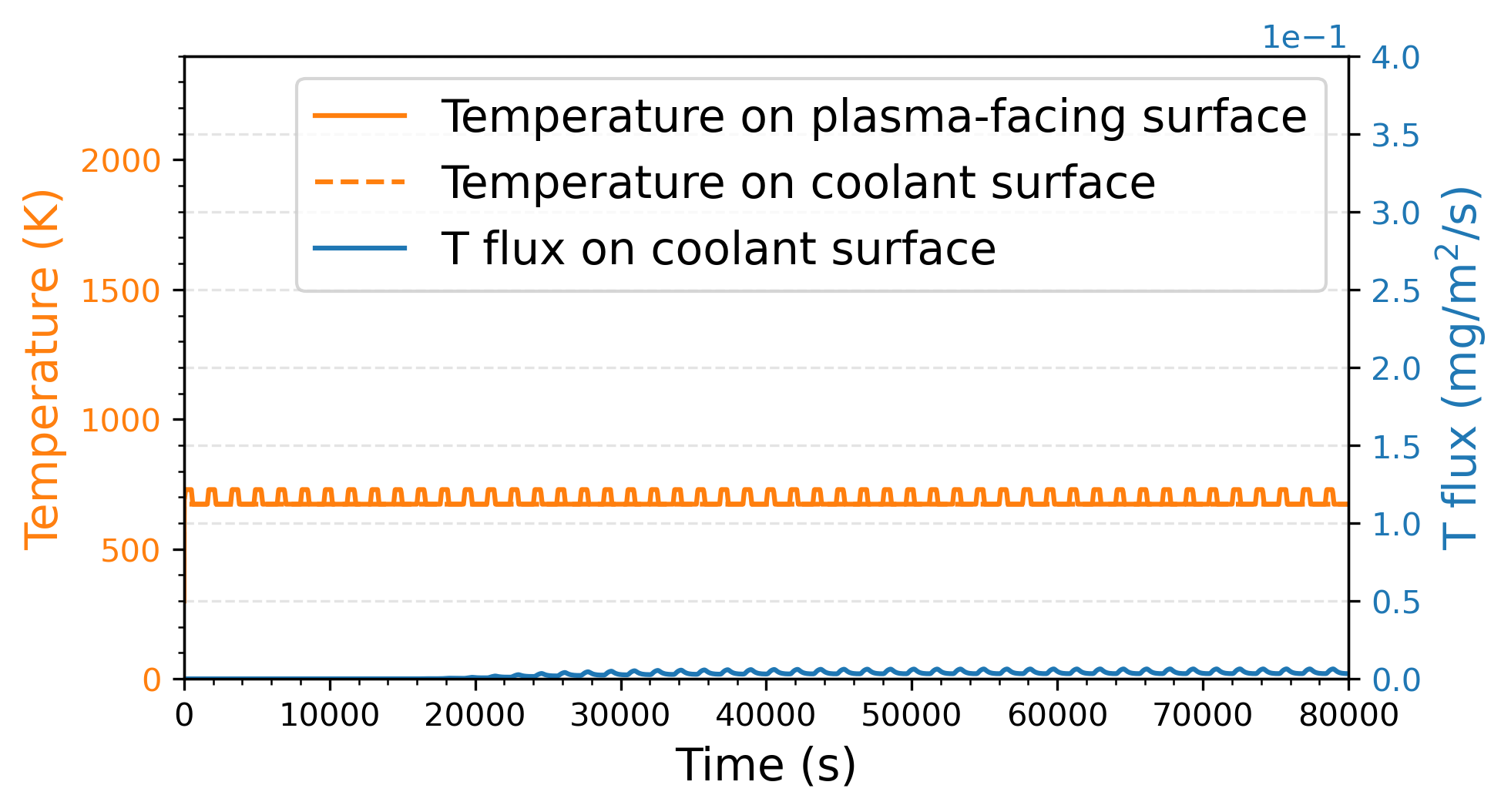}
    \caption{}
    \label{fig:1D_results:CCFW_W_V44:tritium_flux}
\end{subfigure}
\caption{The temperature and tritium evolution and corresponding profile for the \acrshort{ccfw} with 5 mm \acrshort{w} armor and 2.5 mm \acrshort{v44} pipe during 50 pulses. Panel (a) is the temperature and tritium retention evolution during 50 pulses, (b) is the maximum temperature profile and tritium profile at the end of the first five pulses and the final pulse, and (c) is the temperature and tritium flux evolution at the coolant side during 50 pulses. The orange curves represent temperature and the blue curves the tritium.}
\label{fig:1D_results:CCFW_W_V44}
\end{figure}

\begin{figure}[htb]
\centering
\begin{subfigure}{.45\linewidth}
  \centering
	\includegraphics[width = \linewidth]{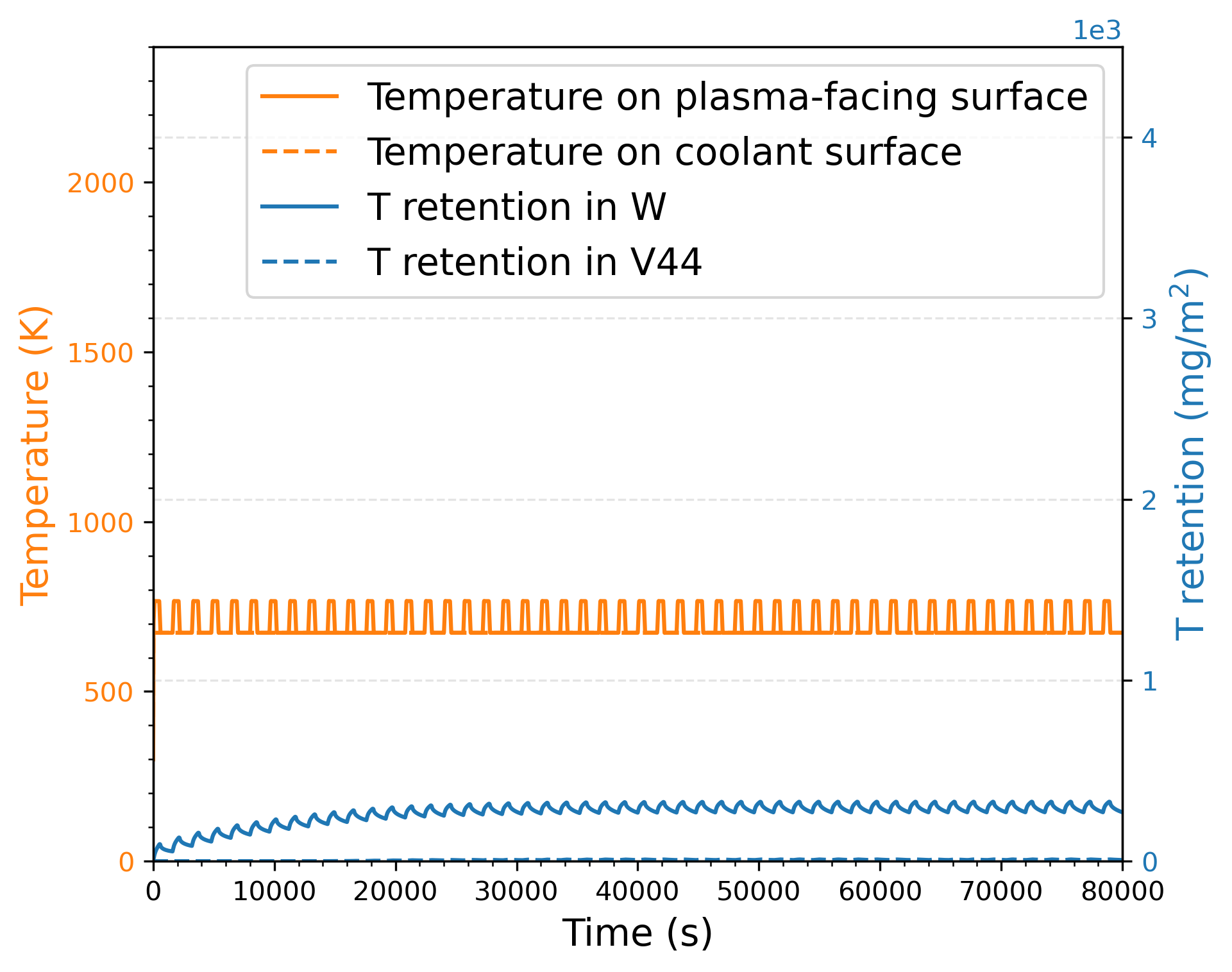}
    \caption{}
    \label{fig:1D_results:BKFW_W_V44_v1:tritium_rentension}
\end{subfigure}
\begin{subfigure}{.45\linewidth}
  \centering
	\includegraphics[width = \linewidth]{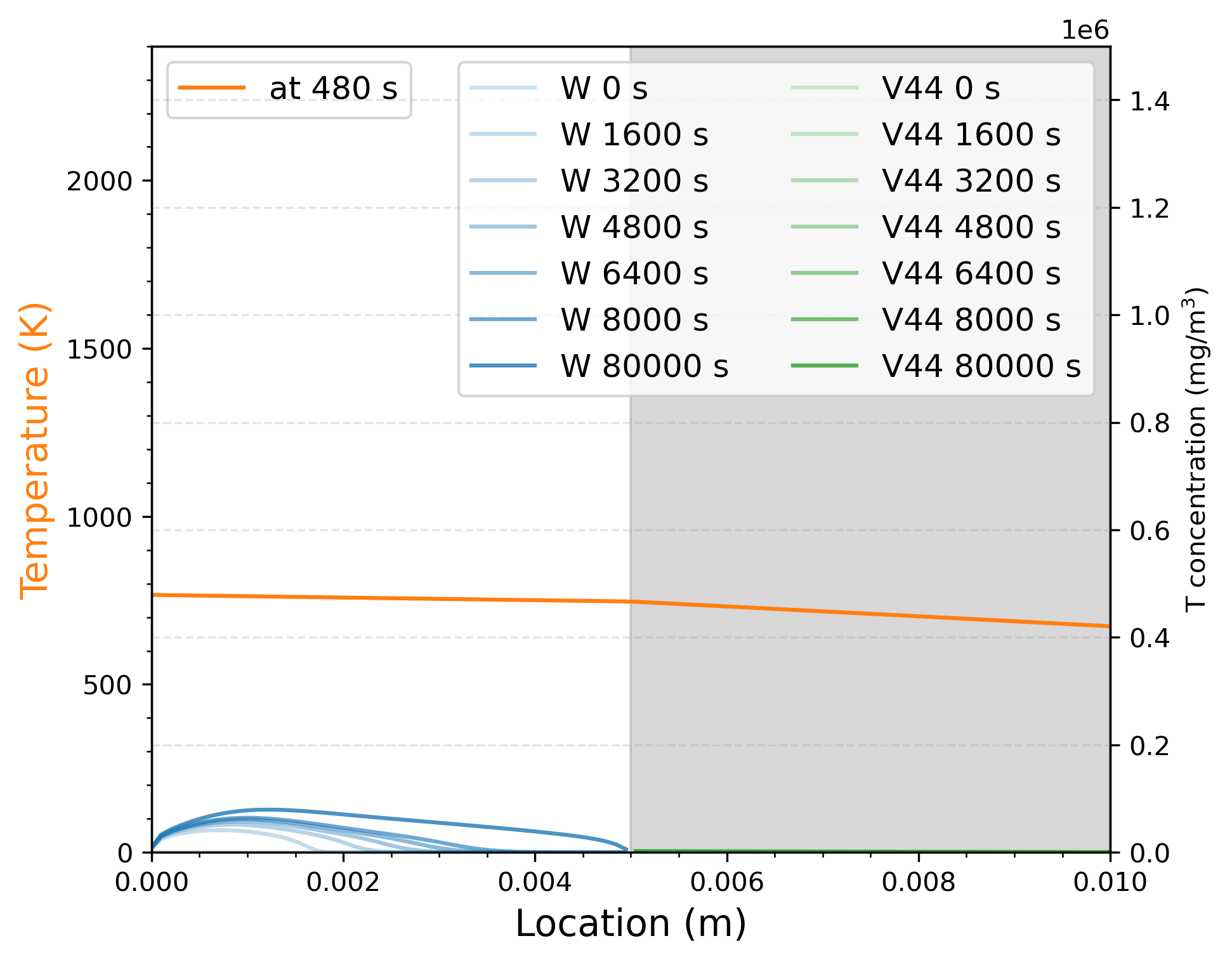}
    \caption{}
    \label{fig:1D_results:BKFW_W_V44_v1:tritium_profile}
\end{subfigure}
\begin{subfigure}{.45\linewidth}
  \centering
	\includegraphics[width = \linewidth]{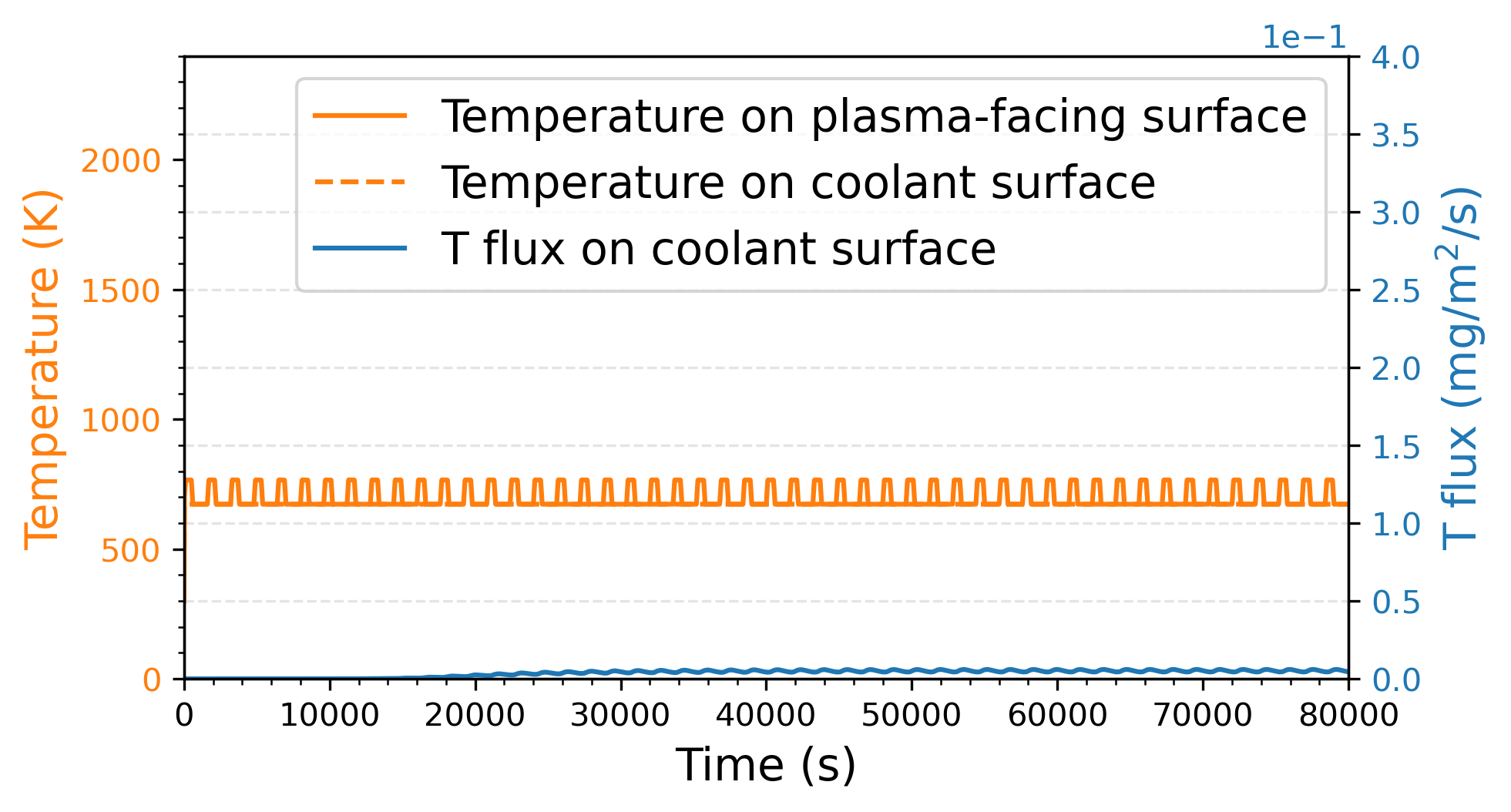}
    \caption{}
    \label{fig:1D_results:BKFW_W_V44_v1:tritium_flux}
\end{subfigure}
\caption{The temperature and tritium evolution and corresponding profile for the \acrshort{bkfw} with 5 mm \acrshort{w} armor and 5 mm \acrshort{v44} pipe during 50 pulses. Panel (a) is the temperature and tritium retention evolution during 50 pulses, (b) is the maximum temperature profile and tritium profile at the end of the first five pulses and the final pulse, and (c) is the temperature and tritium flux evolution at the coolant side during 50 pulses. The orange curves represent temperature and the blue curves the tritium.}
\label{fig:1D_results:BKFW_W_V44_v1}
\end{figure}

\begin{figure}[htb]
\centering
\begin{subfigure}{.45\linewidth}
  \centering
	\includegraphics[width = \linewidth]{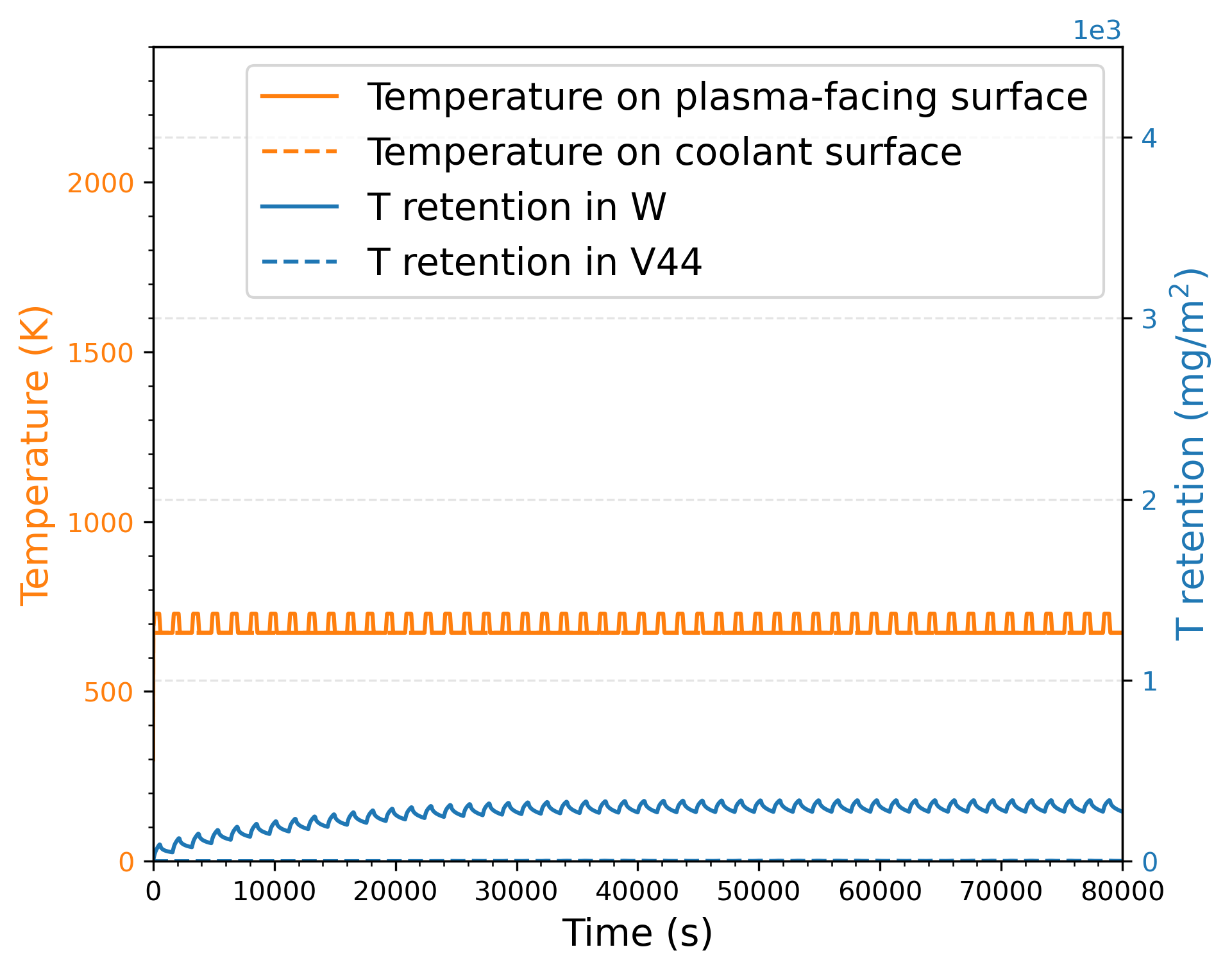}
    \caption{}
    \label{fig:1D_results:BKFW_W_V44_v2:tritium_rentension}
\end{subfigure}
\begin{subfigure}{.45\linewidth}
  \centering
	\includegraphics[width = \linewidth]{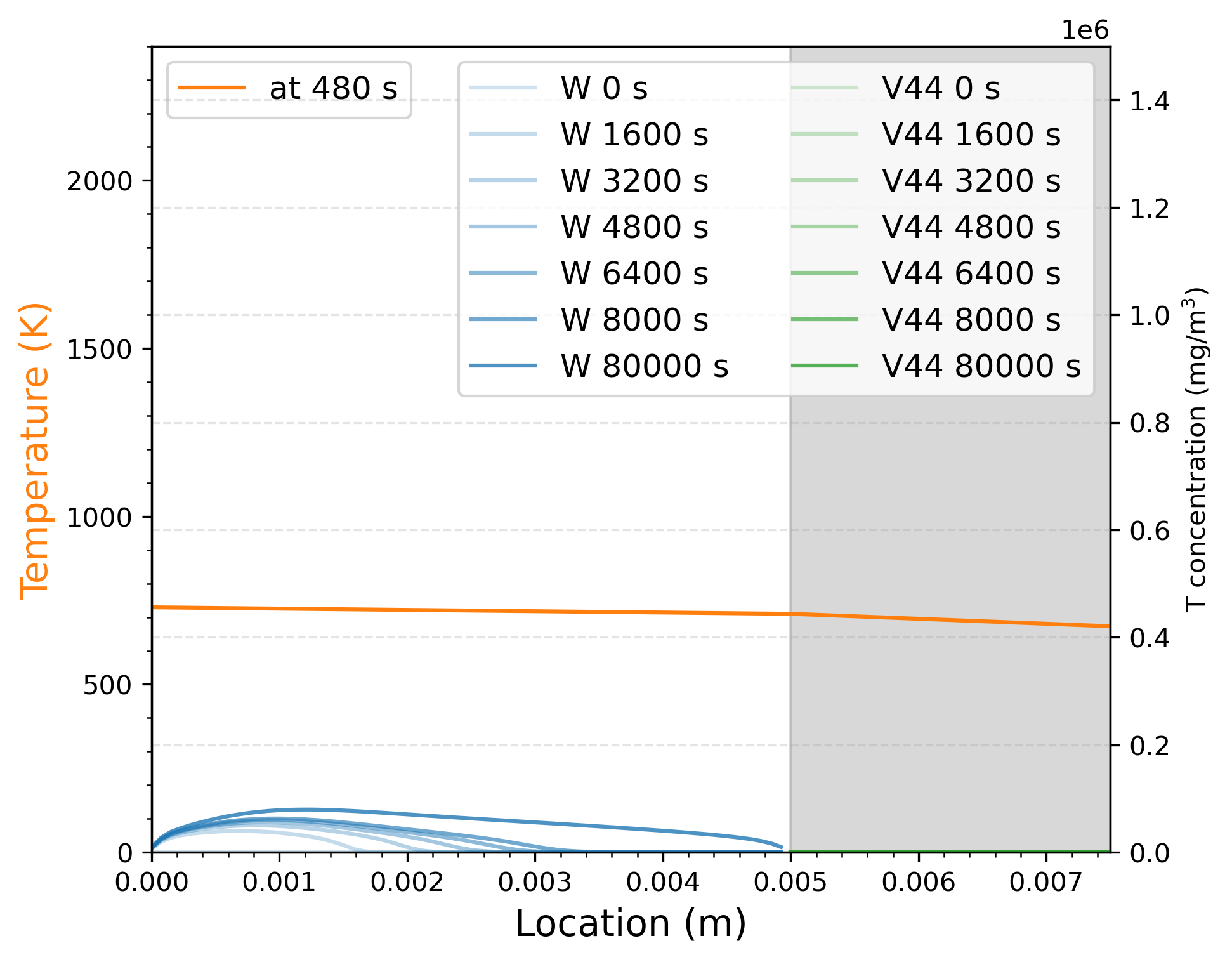}
    \caption{}
    \label{fig:1D_results:BKFW_W_V44_v2:tritium_profile}
\end{subfigure}
\begin{subfigure}{.45\linewidth}
  \centering
	\includegraphics[width = \linewidth]{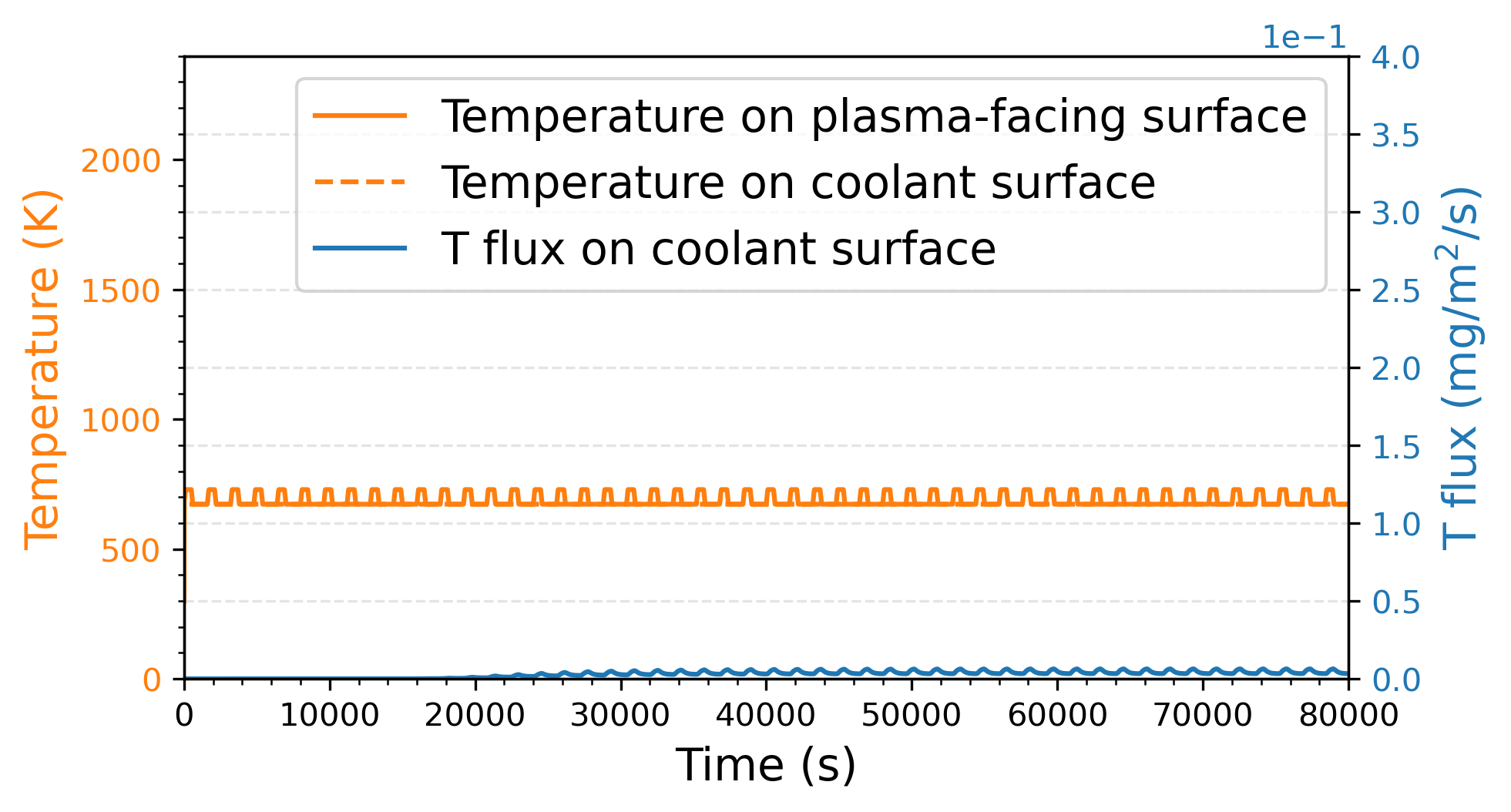}
    \caption{}
    \label{fig:1D_results:BKFW_W_V44_v2:tritium_flux}
\end{subfigure}
\caption{The temperature and tritium evolution and corresponding profile for the \acrshort{bkfw} with 5 mm \acrshort{w} armor and 2.5 mm \acrshort{v44} pipe during 50 pulses. Panel (a) is the temperature and tritium retention evolution during 50 pulses, (b) is the maximum temperature profile and tritium profile at the end of the first five pulses and the final pulse, and (c) is the temperature and tritium flux evolution at the coolant side during 50 pulses. The orange curves represent temperature and the blue curves the tritium.}
\label{fig:1D_results:BKFW_W_V44_v2}
\end{figure}

\begin{figure}[htb]
\centering
\begin{subfigure}{.45\linewidth}
  \centering
	\includegraphics[width = \linewidth]{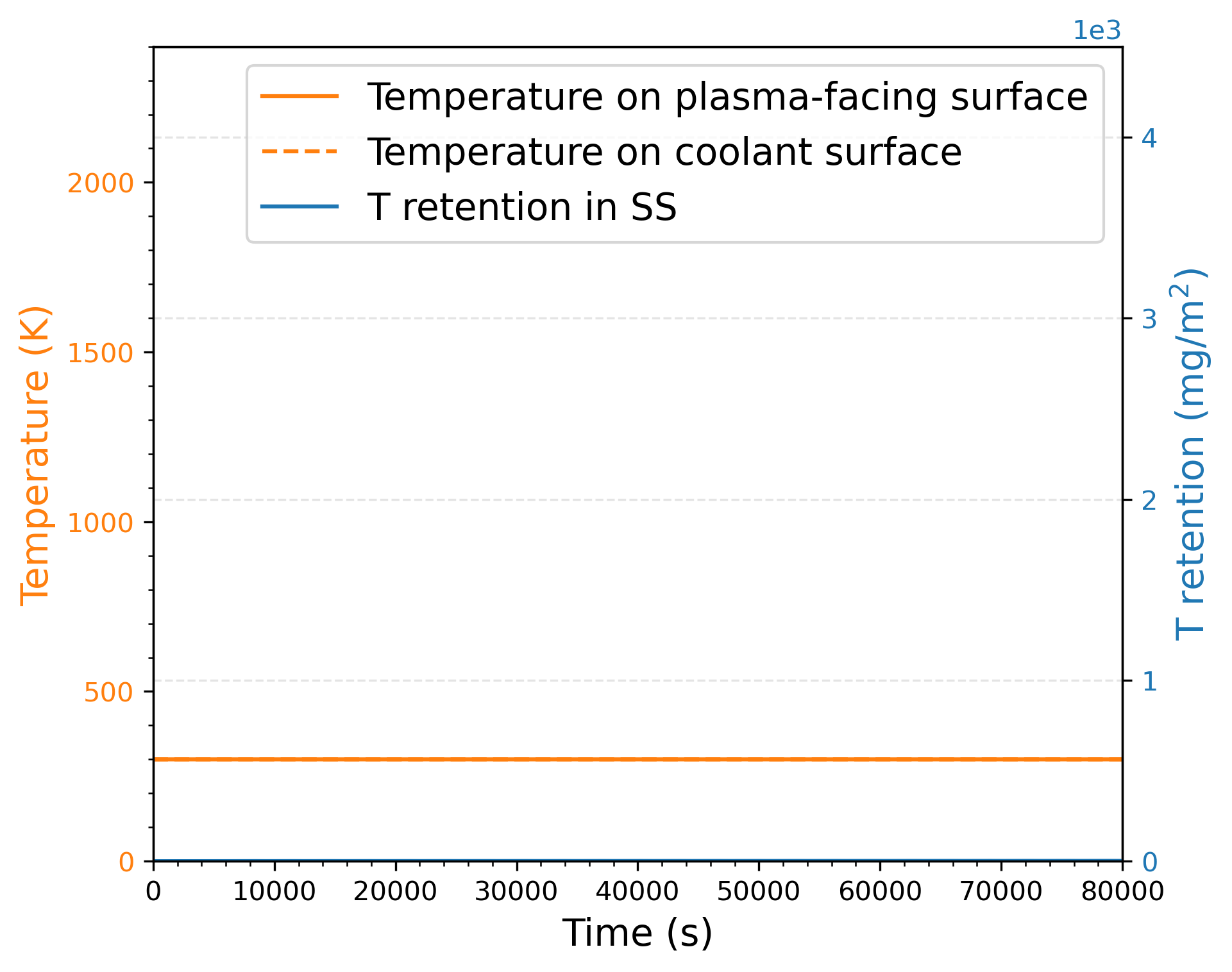}
    \caption{}
    \label{fig:1D_results:VV_SS:tritium_rentension}
\end{subfigure}
\begin{subfigure}{.45\linewidth}
  \centering
	\includegraphics[width = \linewidth]{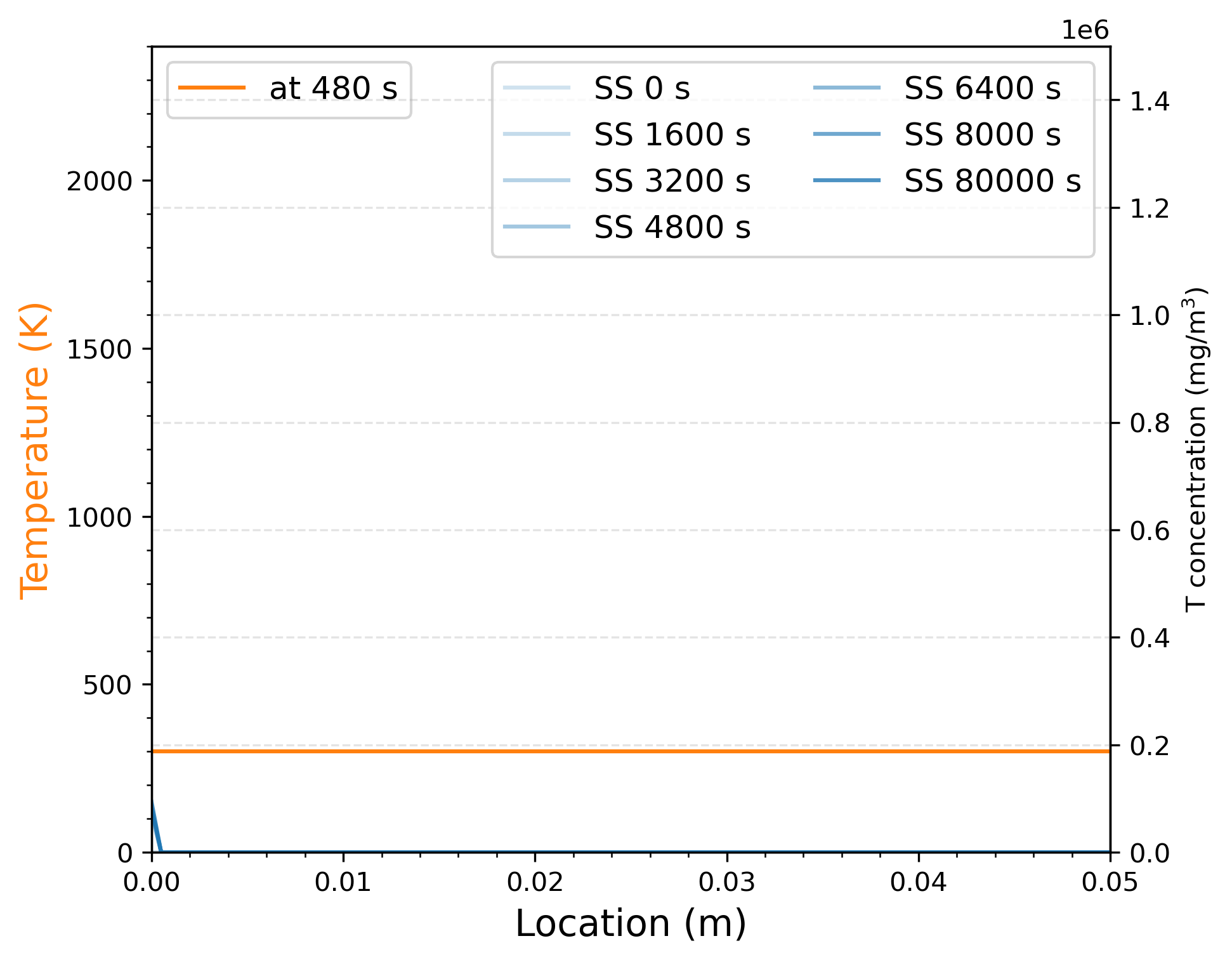}
    \caption{}
    \label{fig:1D_results:VV_SS:tritium_profile}
\end{subfigure}
\begin{subfigure}{.45\linewidth}
  \centering
	\includegraphics[width = \linewidth]{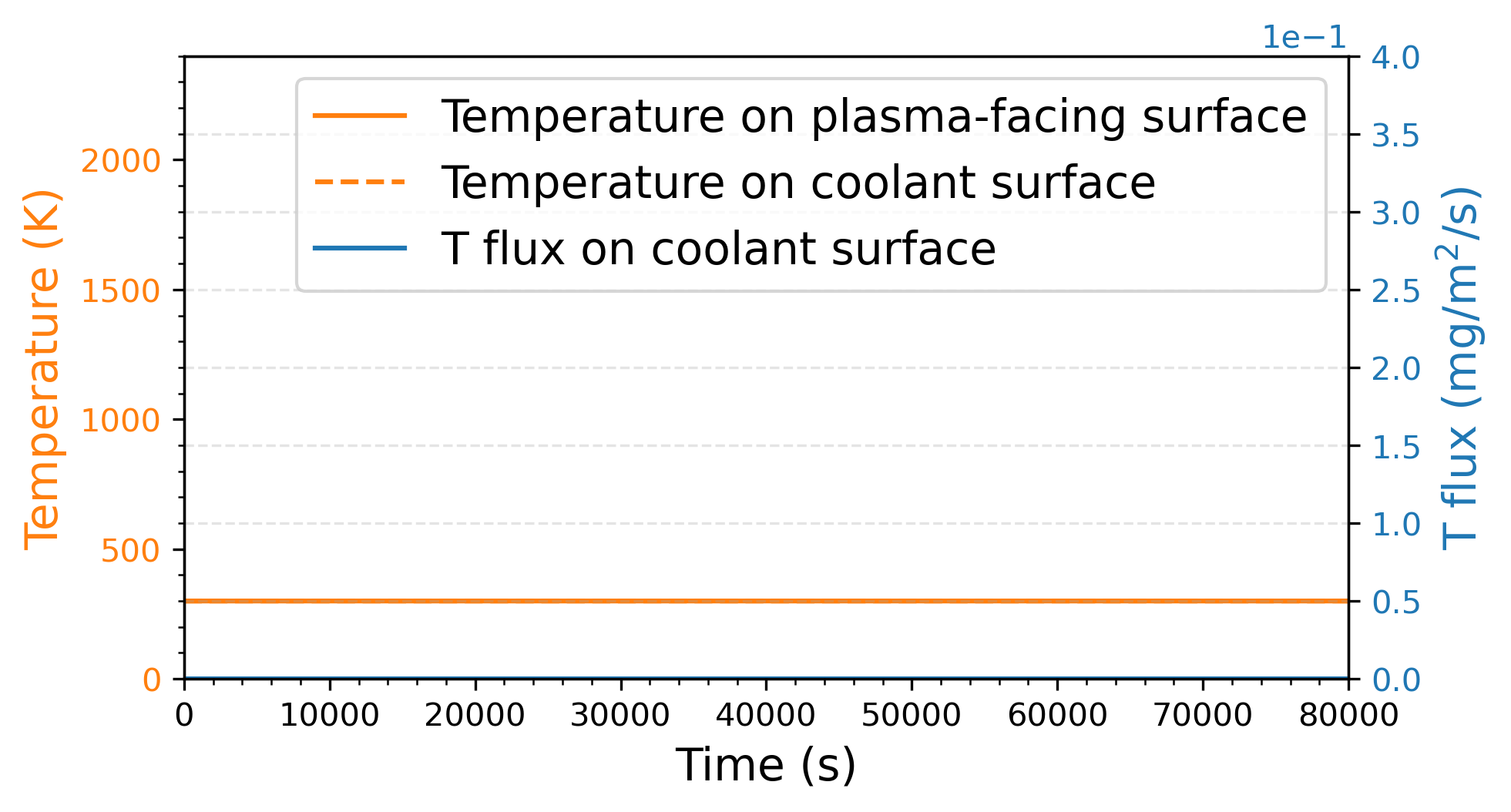}
    \caption{}
    \label{fig:1D_results:VV_SS:tritium_flux}
\end{subfigure}
\caption{The temperature and tritium evolution and corresponding profile for the \acrshort{vv} with 50 mm \acrshort{ss} during 50 pulses. Panel (a) is the temperature and tritium retention evolution during 50 pulses, (b) is the maximum temperature profile and tritium profile at the end of the first five pulses and the final pulse, and (c) is the temperature and tritium flux evolution at the coolant side during 50 pulses. The orange curves represent temperature and the blue curves the tritium.}
\label{fig:1D_results:VV_SS}
\end{figure}

Here we present the simulated tritium transport behavior for all components introduced in \cref{sec:components_information:components}, as shown in \cref{fig:1D_results:DIV_W_V44_v1,fig:1D_results:DIV_W_V44_v2,fig:1D_results:DIV_W_W_v1,fig:1D_results:DIV_W_W_v2,fig:1D_results:CCFW_W_V44,fig:1D_results:BKFW_W_V44_v1,fig:1D_results:BKFW_W_V44_v2,fig:1D_results:VV_SS}. Across all component configurations, both the tritium retention and the tritium flux at the coolant interface arrive at a quasi steady-state condition after sufficient pulsed operation. This quasi steady-state behavior indicates that the tritium retention and the release flux approach saturation under the applied periodic thermal and particle loading.

\end{document}